\documentclass[apj]{emulateapj}
\usepackage{amsmath, amsthm, amssymb}
\usepackage[figuresright]{rotating}
\usepackage{graphicx}

\usepackage{xcolor}

\newcommand\T{\rule{0pt}{2.6ex}}       
\newcommand\B{\rule[-1.2ex]{0pt}{0pt}} 

\slugcomment{Accepted to the Astrophysical Journal Supplement Series}

\shorttitle{Robust Chauvenet Rejection}
\shortauthors{Maples et al.}

\begin{document}
	
	
	\title{Robust Chauvenet Outlier Rejection}
	
	
	\author{M.~P.~Maples, D.~E.~Reichart, {\color{black}N.~C.~Konz,} T.~A.~Berger, A.~S.~Trotter, J.~R.~Martin, D.~A.~Dutton, M.~L.~Paggen, R.~E.~Joyner, C.~P.~Salemi}
	\affil{Department of Physics and Astronomy, University of North Carolina at Chapel Hill, Chapel Hill, NC 27599}
	\email{reichart@unc.edu}
	
	
	
	\begin{abstract}
		Sigma clipping is commonly used in astronomy for outlier rejection, but the number of standard deviations beyond which one should clip data from a sample ultimately depends on the size of the sample.  Chauvenet rejection {\color{black}is one of the oldest, and simplest, ways to account for this, but, like sigma clipping, depends on the sample's mean and standard deviation, neither of which are robust quantities:  Both are easily contaminated by the very outliers they are being used to reject.  Many, more robust measures of central tendency, and of sample deviation, exist, but each has a tradeoff with precision.  Here, we demonstrate that outlier rejection can be both very robust \textit{and} very precise if decreasingly robust but increasingly precise techniques are applied in sequence.}  To this end, we present a variation on Chauvenet rejection that we call ``robust'' Chauvenet rejection {\color{black}(RCR), which uses three decreasingly robust/increasingly precise measures of central tendency, and four decreasingly robust/increasingly precise measures of sample deviation.  We show this sequential approach to be very} effective for a wide variety of contaminant types, even when a significant {\color{black}--} even dominant {\color{black}--} fraction of the sample is contaminated, and especially when the contaminants are strong.  Furthermore, we have developed a bulk-rejection variant, to significantly decrease computing times, and {\color{black}RCR} can be applied both to weighted data{\color{black}, and when fitting parameterized models to data}.  We present aperture photometry in a contaminated, crowded field as an {\color{black}example.  RCR} may be used by anyone at https://skynet.unc.edu/rcr, and source code is available there as well.  
	\end{abstract}
	
	
	
	
	
	\section{Introduction}
	
	{\color{black}Consider a sample of outlying and non-outlying measurements, where the non-outlying measurements are drawn from a given statistical distribution, due to a given physical process, and the outlying measurements are sample contaminants, drawn from a different statistical distribution, due to a different, or additional, physical process, or due to non-statistical errors in measurement.  Furthermore, the statistical distribution from which the outlying measurements are drawn is often unknown.  Whether (1)~combining this sample of measurements into a single value, or (2)~fitting a parameterized model to these data, outliers} can result in incorrect {\color{black}inferences.}
	
	{\color{black}There are a great many methods for identifying and either down-weighting (see \textsection2.1) or outright rejecting outliers.  The most ubiquitous, particularly in astronomy, is sigma clipping.  Here, measurements are identified as outlying and rejected if they are} more than a certain number of standard deviations from {\color{black}the mean}, assuming that the sample is otherwise distributed normally.  {\color{black}Sigma clipping, for example, is a staple of aperture photometry, where it is used to reject signal above the noise (e.g., other sources, Airy rings, diffraction spikes, cosmic rays, and hot pixels),} as well as overly negative deviations (e.g., cold pixels), when measuring the background level in a surrounding annulus.  
	
	Sigma clipping, however, is crude in a number of ways, the first being where to set the {\color{black}rejection} threshold.  For example, if working with $\approx$100 data points, 2-sigma {\color{black}deviations from the mean} are expected but 4-sigma {\color{black}deviations} are not, so one might choose to set the threshold between 2 and 4.  However, if working with $\approx$10$^4$ points, 3-sigma {\color{black}deviations} are expected but 5-sigma {\color{black}deviations} are not, in which case a greater threshold should be applied.
	
	Chauvenet rejection is {\color{black}one of the oldest, and also the most straightforward, improvement to sigma clipping, in that it quantifies this rejection threshold, and does so very simply (Chauvenet 1863).  Chauvenet's criterion for rejecting a measurement} is:
	
	\begin{equation}
	NP(\rm{>}|z|)<0.5,
	\end{equation}
	
	\noindent where $N$ is the number of {\color{black}measurements in the sample,} and $P(\rm{>}|z|)$ is the cumulative probability of {\color{black}the measurement} being more than $z$ standard deviations from the mean, assuming a Gaussian distribution.  We apply {\color{black}Chauvenet's} criterion iteratively, rejecting only one {\color{black}measurement} at a time for increased stability, but consider the case of (bulk) rejecting all {\color{black}measurements} that meet {\color{black}Chauvenet's} criterion each iteration in \textsection5.{\color{black}\footnote{{\color{black}Some care must be taken here:  Since both the mean and the standard deviation change each iteration, measurements that were outlying can become not outlying (though the opposite is usually the case).}}}  In either case, after each iteration (1)~we lower $N$ by the number of points that we rejected, and (2)~we re-estimate the mean and standard deviation, which are used to compute each {\color{black}measurement's} $z$ value, from the remaining{\color{black}, non-rejected measurements}.
	
	However, {\color{black}both traditional Chauvenet rejection, as well as its more-general, less-defined version, sigma clipping, suffer} from neither the mean nor the standard deviation being ``robust'' {\color{black}quantities:  Both are easily contaminated by the very outliers they} are being used to reject.  {\color{black}In \textsection2, we consider increasingly robust (but decreasingly precise; see below) replacements for the mean and standard deviation, settling on three measures of central tendency (\textsection2.1) and four measures of sample deviation (\textsection2.2).  We calibrate seven pairings of these, using uncontaminated data, in \textsection2.3.}
	
	{\color{black}In \textsection3, we evaluate these increasingly robust improvements to traditional Chauvenet rejection against different contaminant types:}  In \textsection3.1, we consider the case of two-sided contaminants, meaning that outliers are as likely to be {\color{black}high as they are to be low; and in} \textsection3.2, we consider the {\color{black}(}more challenging{\color{black})} case of one-sided contaminants, {\color{black}where all or almost all of the outliers are high (or low; we also consider in-between cases here).}  In \textsection3.3, we consider the case of rejecting outliers from mildly non-Gaussian distributions.  
	
	In \textsection3, {\color{black}we show that these increasingly robust improvements to traditional Chauvenet rejection do indeed result in increasingly accurate measurements, and they do so in the face of increasingly high contaminant fractions and contaminant strengths.  But at the same time, these measurements are decreasingly precise.  However, in \textsection4, we show that one can make measurements that are both very accurate \textit{and} very precise, by applying these techniques in sequence, with more-robust techniques applied before more-precise techniques.}
	
	In \textsection5, we evaluate the effectiveness of bulk rejection, which can be significantly less demanding computationally.  In \textsection6, we consider the case of weighted data.  In \textsection7, we exercise both of these techniques with an astronomical example.  
	
	In \textsection8, we show how {\color{black}RCR} can be applied to model fitting{\color{black}, which first requires a generalization of this, traditionally, non-robust process}.  In \textsection9, we compare {\color{black}RCR} to Peirce rejection {\color{black}(Peirce 1852; Gould 1855), which is perhaps the next-most commonly used outlier-rejection technique (Ross 2003).  Peirce rejection is a} non-iterative {\color{black}alternative to traditional Chauvenet rejection,} that can also be applied to model fitting, and that has a reputation of being superior to {\color{black}traditional} Chauvenet rejection.  
	
	We summarize our findings in \textsection10.
	
	{\color{black}\section{Robust Techniques}
		
		\subsection{Measures of Central Tendency}
		
		There are a wide variety of, increasingly robust, ways to measure central tendency.  For example, instead of the mean, one could use the Windsorized mean, in which the values in each tail of a distribution are replaced by the most extreme value remaining, before calculating the mean.  Or, one could use the truncated mean, in which these values are instead simply discarded.  In either case, such measures are a tradeoff, or a compromise, between robustness and precision, depending on what fraction of each side of the distribution is replaced or discarded:  If 0\% is replaced or discarded, these measures are just the mean, which is not robust, but is precise; in the limit that all but one value (or two, if there are an even number of values in the distribution) are replaced or discarded, these measures are equivalent to the median, which is more robust than the mean, but less precise.  
		
		In this paper, we are not trying to introduce a compromise between robustness and precision.  Rather, we are attempting to have both by applying measures with differing properties in sequence.  Consequently, we limit this investigation to the three most-common measures of central tendency, which already have a wide range of properties:  the mean, the median, and the mode, which are increasingly robust, but decreasingly precise.
		
		The mean and median are calculated in the usual ways:  The mean is given by summing a data set's values, and dividing by its number of values, $N$; the median is given by instead sorting these values, and taking the middle value if $N$ is odd, and the mean of the} two middle values if $N$ is even.
	
	{\color{black}Given continuous data, the mode, however, can be defined in a variety of ways.  We adopt an iterative half-sample} approach (e.g., Bickel \& Fr\"uhwirth 2005), {\color{black}and calculate} the mode as follows.  Sort the data, $x_i$, and for every index $j$ in the first half of the data set, including the middle value if $N$ is odd, let $k$ be the largest integer such that:
	
	\begin{equation}
	k \leq j + 0.5N.
	\end{equation}
	
	\noindent Of these $(j,k)$ combinations, select the one for which $|x_k-x_j|$ is smallest.  If multiple combinations meet this criterion, let $j$ be the smallest of their $j$ values and $k$ be the largest of their $k$ values.  Restricting oneself to only the $k-j+1$ values between and including $j$ and $k$, repeat this procedure, iterating to completion.  Take the median of the final $k-j+1$ (typically two) values.
	
	{\color{black}\subsection{Measures of Sample Deviation}
		
		As with central tendency, there are a wide variety of measures of sample deviation.  When we use the mean to measure central tendency, we use the standard deviation to measure sample deviation:  Neither are robust, but both are precise.  
		
		However, when using more-robust measures of central tendency, like the median or the mode, we need to pair these with more-robust measures of sample deviation.  For this, we use what we will call the 68.3-percentile deviation, which we define here in three increasingly robust ways.
		
		The first way is to sort the absolute values of the individual deviations from the measure of central tendency (either the median or the mode), and then} to simply take the 68.3-percentile value from {\color{black}this} sorted distribution.  This is {\color{black}analogous to the ``median absolute deviation'' measure of sample deviation, but with the 68.3-percentile value instead of the 50-percentile value (which we do to remain analogous to the standard deviation, in the limit of a Gaussian distribution).  
		
		This technique} works well as long as less than 40\% -- 70\% of the measurements are contaminated (see \textsection3.1 and \textsection3.2).  However, sometimes a greater fraction of the sample may be contaminated.  In this case, we model the 68.3-percentile deviation from the lower-deviation measurements.  
	
	Consider the case of $N$ measurements, distributed normally and sorted by the absolute value of their deviations from $\mu$ (equal to either the median or the mode).  If weighted uniformly (however, see \textsection6), the percentile of the $i$th element is given by:
	
	\begin{equation}
	\frac{i-1+\Delta i}{N} = P\left(\rm{<}\left|\frac{\delta_i}{\sigma}\right|\right),
	\end{equation}
	
	\noindent where $P(\rm{<}|\delta_i/\sigma|)$ is the cumulative probability of being within $|\delta_i/\sigma|$ standard deviations of the mean, $\delta_i$ is the $i$th sorted deviation, $\sigma$ is the 68.3-percentile deviation, and $0<\Delta i<1$ is the bin center.  We set $\Delta i=0.683$ to yield intuitive results in the limit that $N\rightarrow1$ and $\mu$ is known a priori (\textsection6).  Solving for $\delta_i$ yields:
	
	\begin{equation}
	\delta_i = \sigma\left[\sqrt{2}\mathrm{erf}^{-1}\left(\frac{i-0.317}{N}\right)\right].
	\end{equation}
	
	\noindent Consequently, if plotted $\delta_i$ vs.\@ $\sqrt{2}\mathrm{erf}^{-1}[(i-0.317)/N]$, the distribution is linear, and the slope of this line yields $\sigma$ (see Figure~1).
	
	\begin{figure}
		\plotone{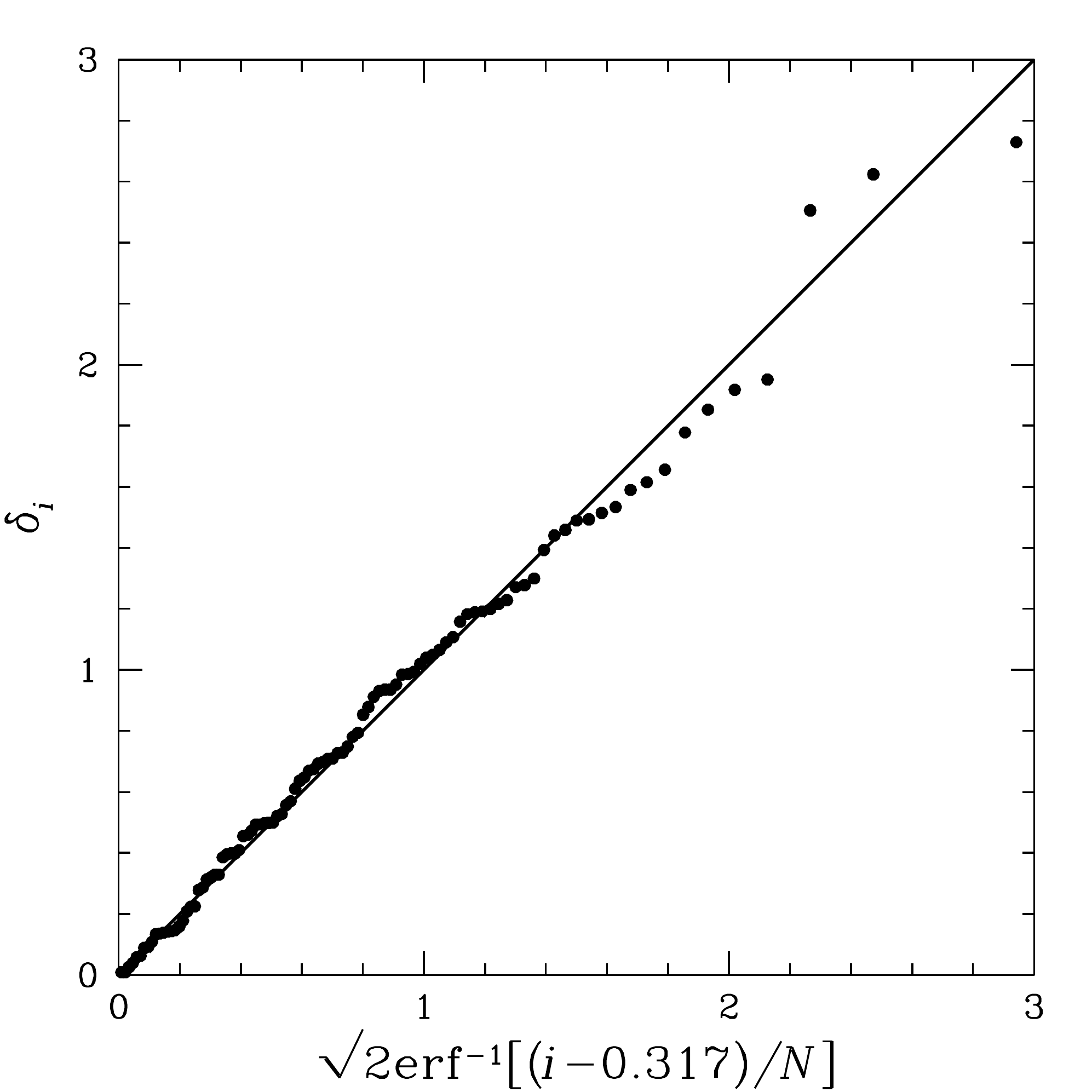}
		\caption{100 sorted deviations from the median, all drawn from a Gaussian distribution of standard deviation $\sigma=1$.  The measured 68.3-percentile deviation is also $\approx$1.}
	\end{figure}
	
	However, if a fraction of the sample is contaminated, the shape of the distribution changes:  The slope steepens, and (1)~if the value from which the deviations are measured (the median or the mode) still approximates that of the uncontaminated measurements, and (2)~if the contaminants are drawn from a sufficiently broader distribution, the curve breaks upward (see Figure~2, upper left).\footnote{If the median or the mode no longer approximates that of the uncontaminated measurements, the curve can instead break downward, making the following three 68.3-percentile deviation measurement techniques decreasingly robust, instead of increasingly robust (see \textsection3.2, Figure~17).}  Consequently, we model the 68.3-percentile deviation of the uncontaminated measurements in three, increasingly accurate ways:  (1)~by simply using the 68.3-percentile value, as described above (e.g., Figure~2, upper right); (2)~by fitting a zero-intercept line to the $\sqrt{2}\mathrm{erf}^{-1}[(i-0.317)/N]<\sqrt{2}\mathrm{erf}^{-1}(0.683)=1$ data{\color{black},} and using the fitted slope (e.g., Figure~2, lower left){\color{black};} and (3)~by fitting a broken line of intercept zero (see Appendix A for fitting details) to the same data{\color{black},} and using the fitted slope of the first component (e.g., Figure~2, lower right).
	
	\begin{figure*}
		\plotone{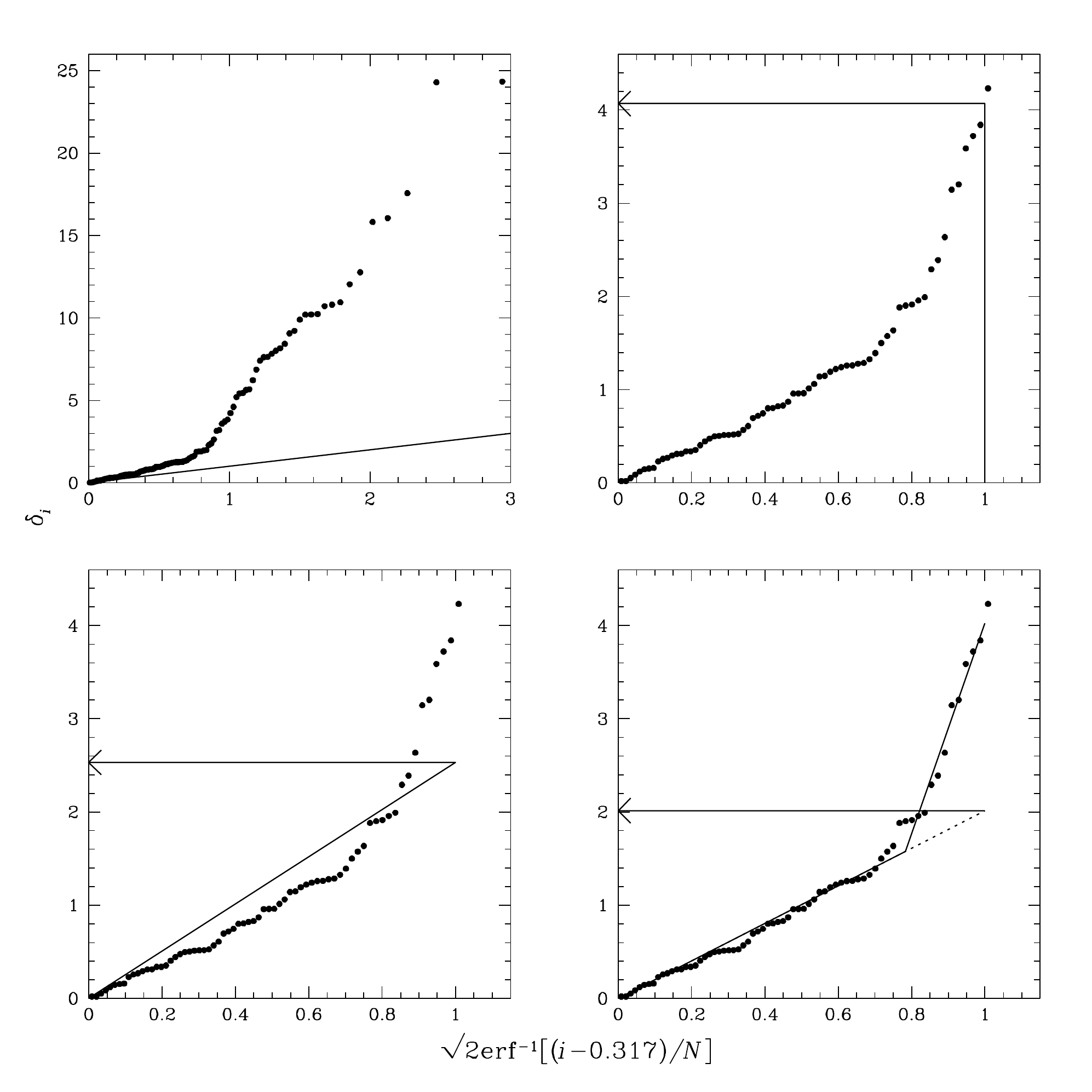}
		\caption{\textbf{Upper left:}  100 sorted deviations from the median, with fraction $f_1=0.5$ drawn from a Gaussian distribution of standard deviation $\sigma_1=1$, and fraction $f_2=0.5$, representing contaminated measurments, drawn from a Gaussian distribution of standard deviation $\sigma_2=10$.  \textbf{Upper right:}  Zoom-in of the upper-left panel, with the 68.3-percentile deviation measured using technique~1, yielding a pre-rejection value of $\sigma_1=4.07$.  \textbf{Lower left:}  Zoom-in of the upper-left panel, with the 68.3-percentile deviation measured using technique~2, yielding a pre-rejection value of $\sigma_1=2.53$.  \textbf{Lower right:}  Zoom-in of the upper-left panel, with the 68.3-percentile deviation measured using technique~3, yielding a pre-rejection value of $\sigma_1=2.01$.  See Figure~3 for post-rejection versions and measured values.}
	\end{figure*}
	
	We then iteratively Chauvenet{\color{black}-}reject the greatest outlier (\textsection1), using either (1)~the median or (2)~the mode instead of the mean, and the 68.3-percentile deviation instead of the standard deviation.\footnote{With the following exception:  We never reject down to a sample of identical measurements.  In the standard case of producing a single measurement from multiple, we always leave at least two distinct measurements.  In the more general case of fitting a multiple-parameter model to multiple measurements (see \textsection8), we always leave at least $M+1$ distinct measurements, where $M$ is the number of model parameters.}  The effect of this on the data presented in Figure~2 can be seen in Figure~3, for each of our three, increasingly robust, 68.3-percentile deviation measurement techniques.
	
	\begin{figure*}
		\centering
		\includegraphics[height=0.9\textheight]{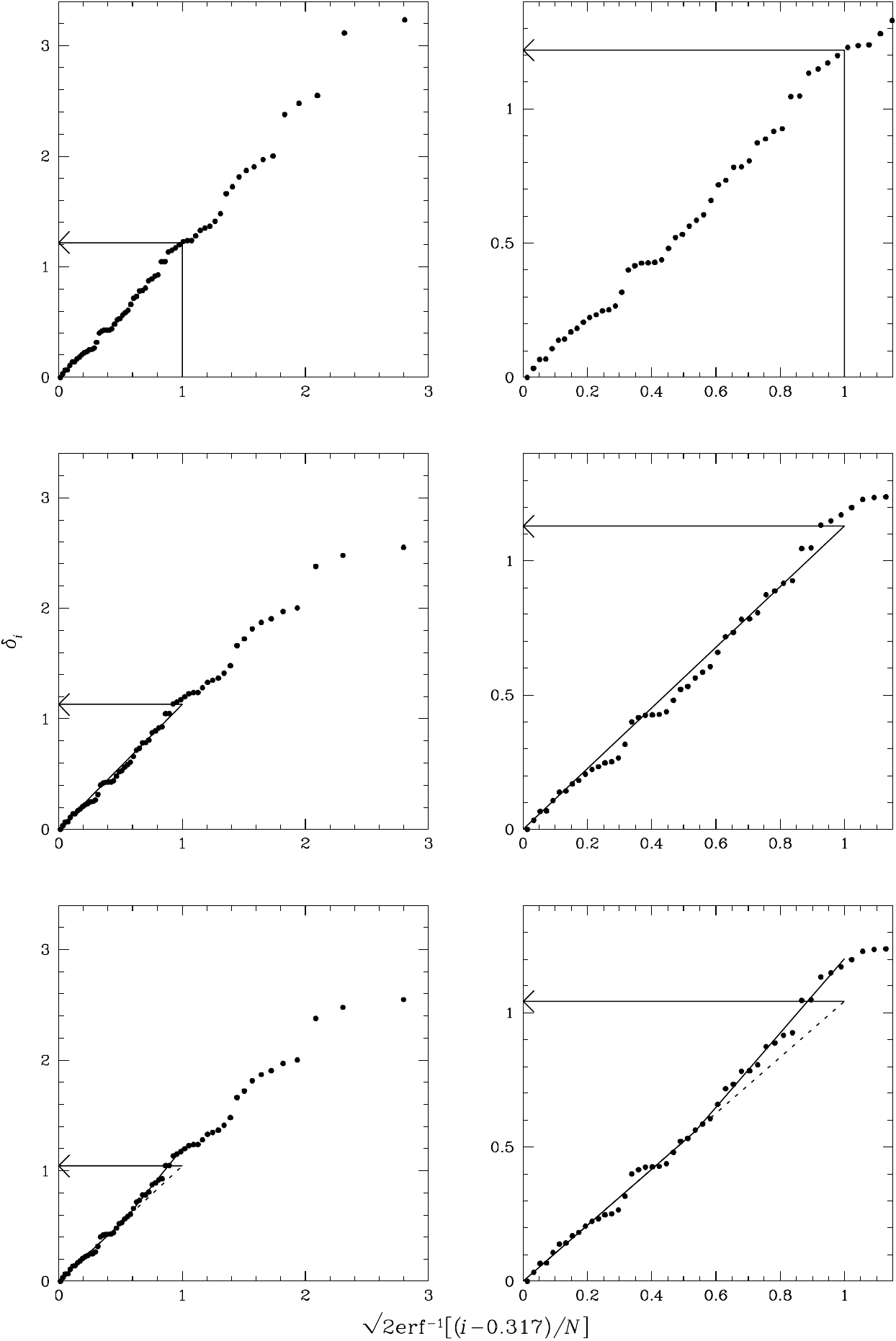}
		\caption{Figure~2, after iterated Chauvenet rejection.  \textbf{Upper left:}  Using the 68.3-percentile deviation from technique~1, yielding a final measured value of $\sigma_1=1.22$.  \textbf{Upper right:}  Zoom-in of the upper-left panel.  \textbf{Middle left:}  Using the 68.3-percentile deviation from technique~2, yielding a final measured value of $\sigma_1=1.13$.  \textbf{Middle right:}  Zoom-in of the middle-left panel.  \textbf{Lower left:}  Using the 68.3-percentile deviation from technique~3, yielding a final measured value of $\sigma_1=1.04$.  \textbf{Lower right:}  Zoom-in of the lower-left panel.}
	\end{figure*}
	
	{\color{black}\subsection{Calibration}
		
		Before further using these two more-robust measures of central tendency (\textsection2.1) and three more-robust measures of sample deviation (\textsection2.2) to Chauvenet-reject outliers, we calibrate these $2 \times 3 = 6$ more-robust techniques, using uncontaminated data.  We also calibrate two less-robust, comparison techniques, using} the mean and standard deviation (1)~without and (2)~with iterated Chauvenet rejection.
	
	For each sample size $2\leq N\leq100$, as well as for $N=1000$, we drew 100,000 samples from a Gaussian distribution of mean $\mu=0$ and standard deviation $\sigma=1$, and then recovered $\mu$ and $\sigma$ using each technique.  Averaged over the 100,000 samples, the recovered value of $\mu$ was always $\approx$0, and the recovered value of $\sigma$ was $\approx$1 in the limit of large $N$.  However, all of the techniques, including the traditional, comparison techniques,\footnote{It is well known that although the variance can be computed without bias using Bessel's correction, the standard deviation cannot, and the correction depends on the shape of the distribution.  For a normal distribution, without rejection of outliers, the correction is given by $\sqrt{\frac{N-1}{2}}\frac{\Gamma \left(\frac{N-1}{2}\right)}{\Gamma \left(\frac{N}{2}\right)}$, which matches what we determined empirically, and plot in the upper-left panel of Figure~4 (solid black curve).} underestimated $\sigma$ in the limit of small $N$ (see Figure~4).  
	\begin{figure*}
		\plotone{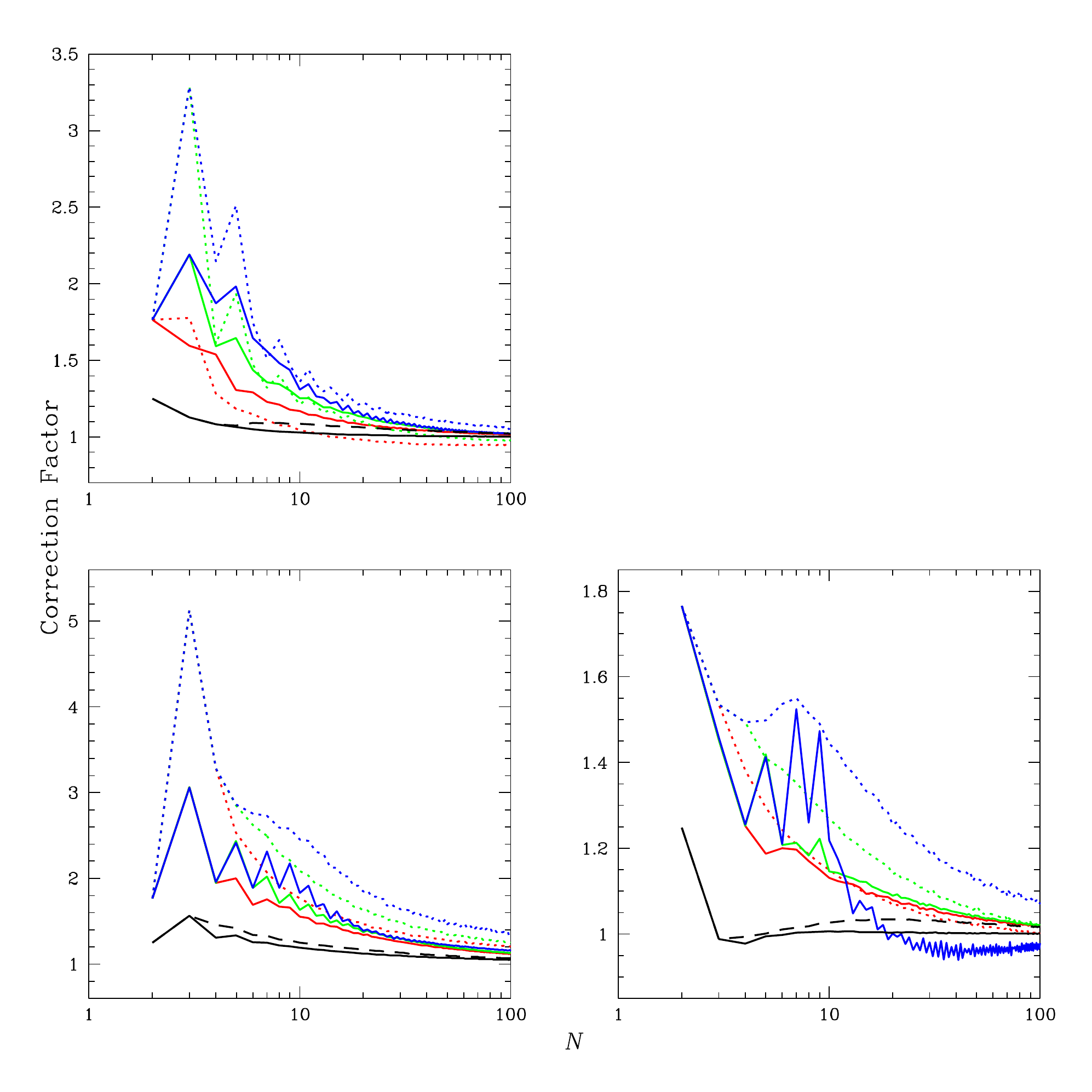}
		\caption{Correction factors by which standard and 68.3-percentile deviations, measured from uncontaminated data, need to be multiplied to yield the correct result, on average, and to avoid overaggressive rejection {\color{black}(although this can still happen in sufficiently small samples; see \textsection3.3.1)}, (1)~for the case of no rejection, using the mean and standard deviation (solid black curves; see Footnote 3); (2)~for the case of Chauvenet rejection, using the mean and standard deviation (dashed black curves); (3)~for the case of Chauvenet rejection, using the median and 68.3-percentile deviation as measured by technique~1 from \textsection2.2 (solid red curves), as measured by technique~2 from \textsection2.2 (solid green curves), and as measured by technique~3 from \textsection2.2 (solid blue curves); and (4)~for the case of Chauvenet rejection, using the mode and 68.3-percentile deviation as measured by technique~1 (dotted red curves), technique~2 (dotted green curves), and technique~3 (dotted blue curves).  \textbf{Upper left:}  For the simplest case of computing a single $\sigma$ (standard or 68.3-percentile deviation), using the deviations both below and above $\mu$ (the mean, the median, or the mode; see \textsection3.1).  \textbf{Lower left:}  For the case of computing separate $\sigma$ below and above $\mu$ ($\sigma_-$ and $\sigma_+$, respectively) and using the smaller of the two when rejecting outliers (see \textsection3.2). \textbf{Lower right:}  For the same case, but using $\sigma_-$ to reject outliers below $\mu$ and $\sigma_+$ to reject outliers above $\mu$ (see \textsection3.3.1).  Note that technique~3 defaults to technique~2 when the two are statistically equivalent (see Appendix A), or when fitting to fewer than three points (e.g., when $N<4$ for the cases in the top row and when $N<7$ for the median cases in the bottom row).  Similarly, technique~2 defaults to technique~1 when fitting to fewer than two points (e.g., when $N<3$ for the cases in the top row and when $N<5$ for the median cases in the bottom row).  Oscillations are not noise, but odd-even effects (e.g., with equally weighted data, when $N$ is odd, use of the median always results in at least one zero deviation, requiring a larger correction factor).  We use look-up tables for $N\leq100$ and power-law approximations for $N>100$ (see Appendix B).}
	\end{figure*}
	
	In Figure~4, we plot correction factors by which measured standard and 68.3-percentile deviations need to be multiplied to yield the correct result, on average.  We make use of these correction factors throughout this paper, to {\color{black}avoid} overaggressive rejection {\color{black}(although this can still happen in sufficiently small samples; see \textsection3.3.1)}.
	
	{\color{black}\section{Robust Techniques Applied to Contaminated Distributions}}
	
	We now evaluate the effectiveness {\color{black}(1)~of the two traditional, less-robust techniques, and (2)~of the $2 \times 3 = 6$ more-robust techniques, that we introduced in \textsection2 at rejecting outliers from Gaussian (see \textsection3.1 and \textsection3.2) and mildly non-Gaussian (see \textsection3.3) distributions.  In \textsection3.1, we consider the case of two-sided contaminants, meaning that outliers are as likely to be high as they are to be low.  In \textsection3.2, we consider the (more challenging) case of one-sided contaminants, where all or almost all of the outliers are high (or low); we also consider in-between cases here.\\  
		\\
		
		\subsection{Normally Distributed Uncontaminated Measurements with Two-Sided Contaminants}}
	
	For sample sizes $N=1000$, 100, and 10, we draw $f_1 N$ uncontaminated measurements from a Gaussian distribution of mean $\mu_1=0$ and standard deviation $\sigma_1=1$, and $f_2 N$ contaminated measurements, where $f_2=1-f_1$.  In this section, we model the contaminants as two-sided, meaning that outliers are as likely to be {\color{black}high as they are to be low}.  We draw {\color{black}contaminants} from a Gaussian distribution of mean $\mu_2=0$ and standard deviation $\sigma_2$, and add them to uncontaminated measurements, drawn as above.{\color{black}\footnote{{\color{black}In this paper, we draw our contaminants from Gaussian distributions}{\color{black}, which ensures that we include some of the worst-case scenarios for rejecting outliers (from Gaussian uncontaminated-}{\color{black}measurement distributions) in our analyses.  For example, with two-sided contaminants, the contaminated measurements are then also distributed as a Gaussian, of standard deviation $\sqrt{\sigma_1^2+\sigma_2^2}$, which becomes increasingly difficult to distinguish from the uncontaminated-measurement distribution as $\sigma_2\rightarrow\sigma_1$ (and of course as $\sigma_2\rightarrow0$).  Furthermore, this contaminated-measurement distribution becomes increasingly difficult to distinguish from any, Gaussian uncontaminated-measurement distribution as $f_2\rightarrow1$.  We have also experimented with non-Gaussian contaminant distributions, but always to similar, or greater, effect:  While these outlier-rejection techniques do depend on the assumed shape of the uncontaminated-measurement distribution, they do not depend on the assumed shape of the contaminant distribution (other than strongly contaminated measurements are of course easier to identify than weakly contaminated measurements).}}}  In the case of two-sided contaminants, the mean, median, and mode are all three, on average, insensitive to outliers, even in the limit of a large fraction of the sample being contaminated ($f_2\rightarrow1$; see Figure~6).  Consequently, this is a good case to evaluate the effectiveness of our three, increasingly robust, 68.3-percentile deviation techniques.  (We explore the more challenging case of one-sided contaminants in \textsection3.2.)
	
	For each technique and sample size, we draw 100 samples for each combination of $f_2=0$, 0.1, 0.2, 0.3, 0.4, 0.5, 0.6, 0.7, 0.8, 0.9, 1 and $\sigma_2=1$, 1.6, 2.5, 4.0, 6.3, 10, 16, 25, 40, 63, 100 (see Figure~5), and plot the average recovered $\mu_1$ in Figure~6, the uncertainty in the recovered $\mu_1$ in Figure~7, the average recovered $\sigma_1$ in Figure~8, and the uncertainty in the recovered $\sigma_1$ in Figure~9.
	
	\begin{figure}
		\plotone{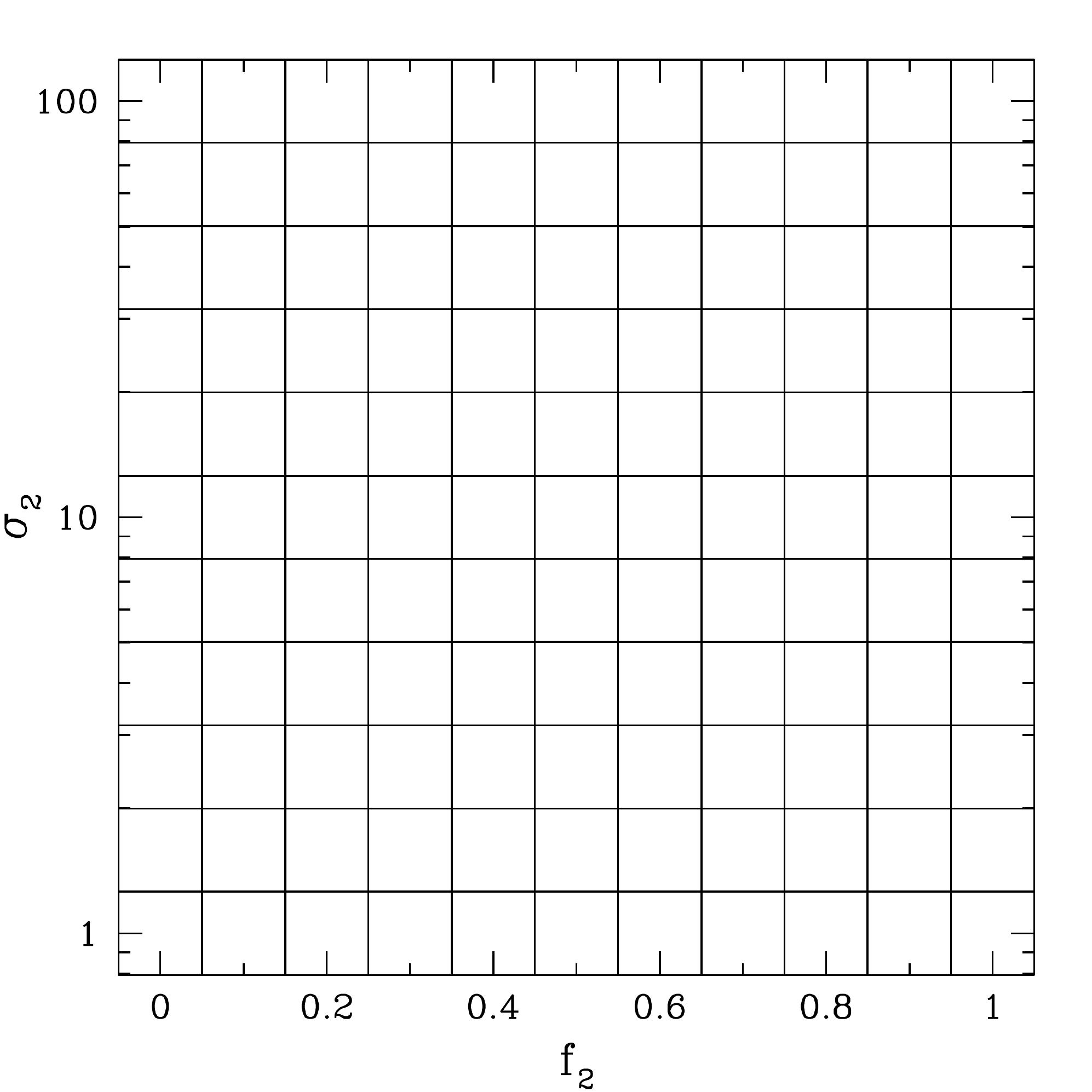}
		\caption{Blank contaminant strength ($\sigma_2$) vs.\@ fraction of sample ($f_2$) figure.  Each pixel corresponds to either a recovered quantity ($\mu_1$ or $\sigma_1$) or the uncertainty in a recovered quantity ($\Delta\mu_1$ or $\Delta\sigma_1$), measured from 100 samples with contaminants modeled by $f_2$ and $\sigma_2$.  This figure is provided as reference, as axis information would be too small to be easily readable in upcoming figures.}
	\end{figure}
	
	\begin{sidewaysfigure*}
		\centering
		\vspace{3.75in}
		\includegraphics[width=0.9\textwidth]{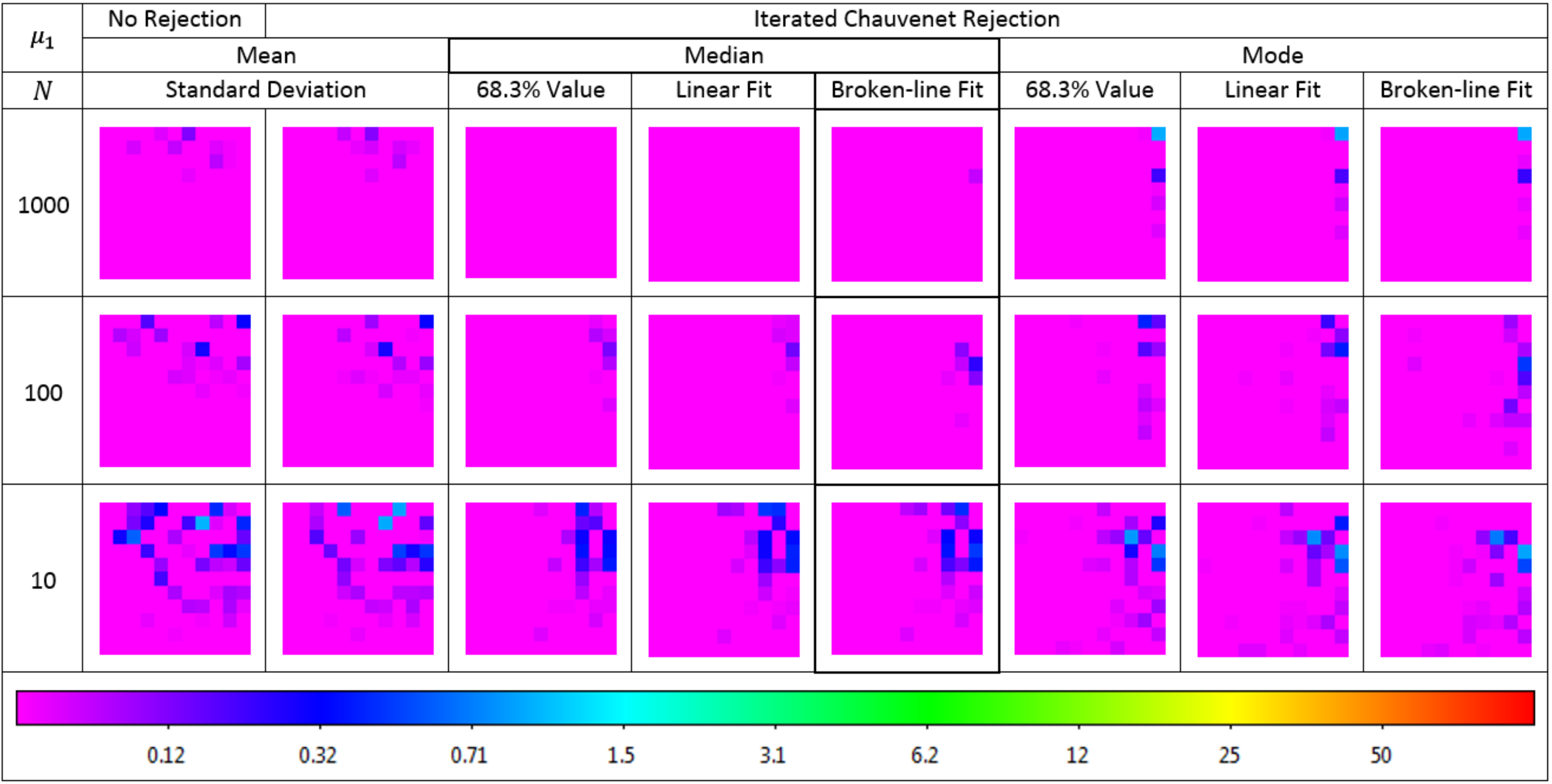}
		\caption{Average recovered $\mu_1$ for increasingly robust measurement techniques and decreasing sample sizes ($N$), for two-sided contaminants.  See Figure~5 for contaminant strength ($\sigma_2$) vs.\@ fraction of sample ($f_2$) axis information.  As expected with two-sided contaminants, the recovered values are $\approx$0, independent of contaminant fraction or strength.  Variation about zero is due to drawing only 100 samples, and is larger for larger values of $f_2$ and $\sigma_2$, and for smaller values of $N$ (see Figure~7).  {\color{black}All things considered (Figures 6 -- 9; \textsection3.1), the best-performing technique for Chauvenet-rejecting two-sided contaminants is highlighted with a bold outline.}  The colors are scaled logarithmically, and cut off at 0.02, to match the upcoming figures, permitting direct comparison of colors between figures.}
	\end{sidewaysfigure*}
	
	\begin{sidewaysfigure*}
		\centering
		\vspace{3.75in}
		\includegraphics[width=0.9\textwidth]{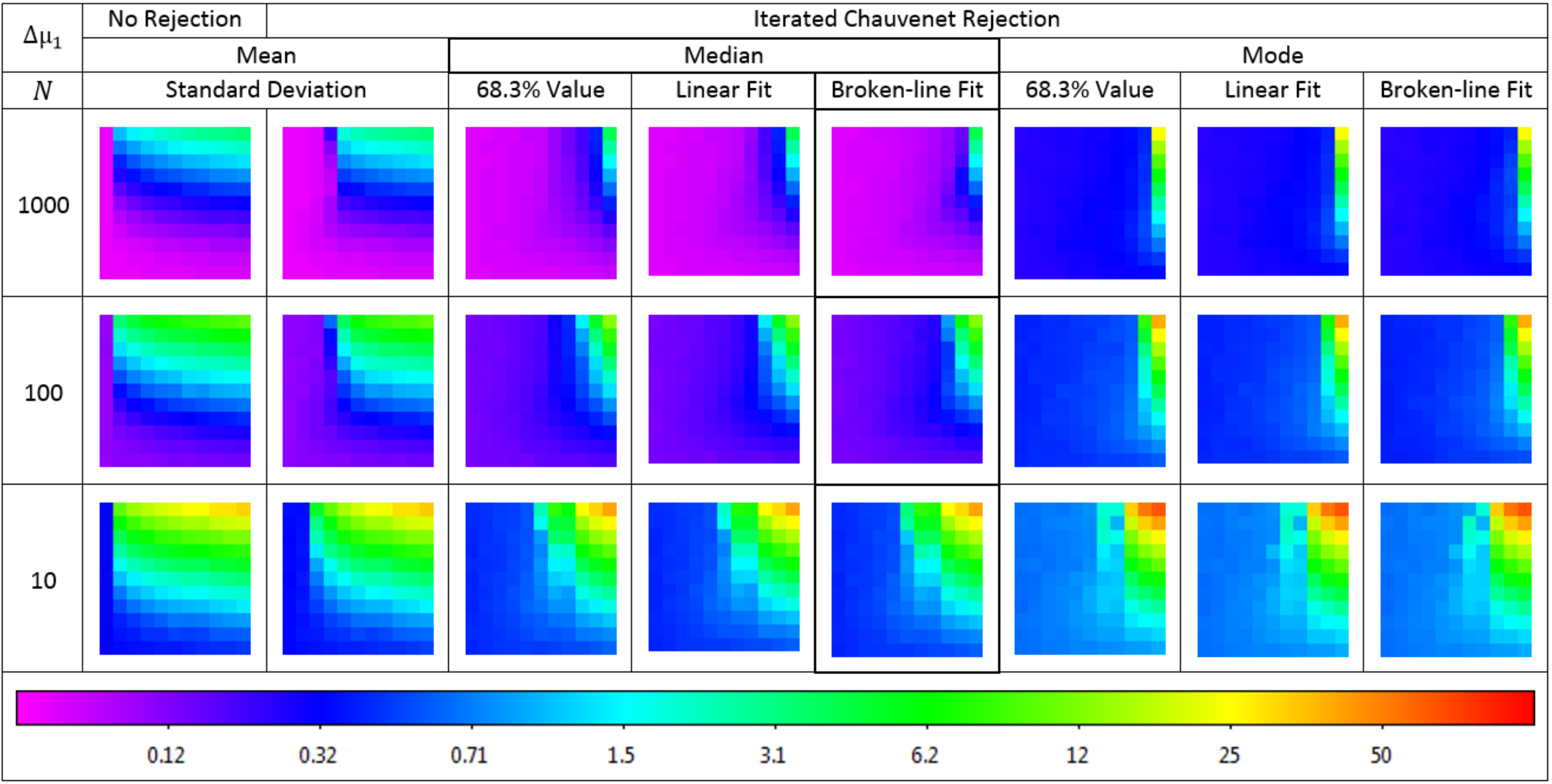}
		\caption{Uncertainty in the recovered $\mu_1$ for increasingly robust measurement techniques and decreasing sample sizes ($N$), for two-sided contaminants.  See Figure~5 for contaminant strength ($\sigma_2$) vs.\@ fraction of sample ($f_2$) axis information.  The effect of the contaminants, without rejection, can be seen in the first column:  Larger contaminant fractions and strengths, as well as smaller sample sizes, result in less precise recovered values of $\mu_1$.  However, increasingly robust measurement techniques are increasingly effective at rejecting outliers in large-$f_2$ samples, allowing $\mu_1$ to be measured significantly more precisely.  Note that this is at a marginal cost:  When applied to uncontaminated samples ($f_2=0$), these techniques recover $\mu_1$ with degrading precisions, of $\Delta\mu_1/\sigma_1\approx1.0N^{-1/2}$, $1.0N^{-1/2}$, $1.3N^{-1/2}$, $1.3N^{-1/2}$, $1.3N^{-1/2}$, $7.6N^{-1/2}$, $7.6N^{-1/2}$, and $7.6N^{-1/2}$, respectively.  {\color{black}However, all things considered (Figures 6 -- 9; \textsection3.1), the best-performing technique for Chauvenet-rejecting two-sided contaminants is highlighted with a bold outline.}  The colors are scaled logarithmically, between 0.02 and 100.}
	\end{sidewaysfigure*}
	
	\begin{sidewaysfigure*}
		\centering
		\vspace{3.75in}
		\includegraphics[width=0.9\textwidth]{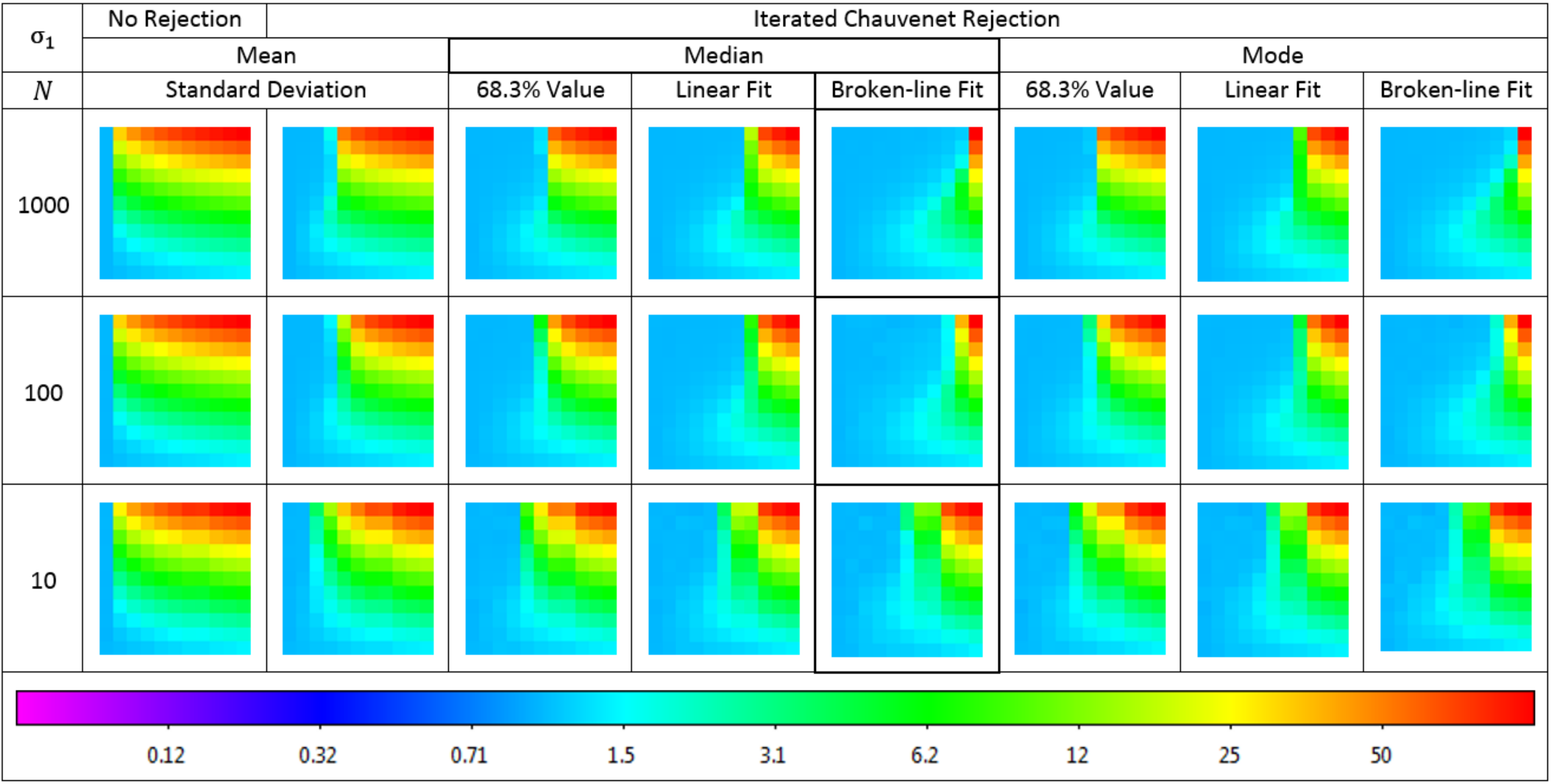}
		\caption{Average recovered $\sigma_1$ for increasingly robust measurement techniques and decreasing sample sizes ($N$), for two-sided contaminants.  See Figure~5 for contaminant strength ($\sigma_2$) vs.\@ fraction of sample ($f_2$) axis information.  The effect of the contaminants, without rejection, can be seen in the first column:  Larger contaminant fractions and strengths result in larger recovered values of $\sigma_1$.  However, increasingly robust measurement techniques are increasingly effective at rejecting outliers in large-$f_2$ samples, allowing $\sigma_1$ to be measured significantly more accurately.  {\color{black}All things considered (Figures 6 -- 9; \textsection3.1), the best-performing technique for Chauvenet-rejecting two-sided contaminants is highlighted with a bold outline.}  The colors are scaled logarithmically, between 0.02 and 100.}
	\end{sidewaysfigure*}
	
	\begin{sidewaysfigure*}
		\centering
		\vspace{3.75in}
		\includegraphics[width=0.9\textwidth]{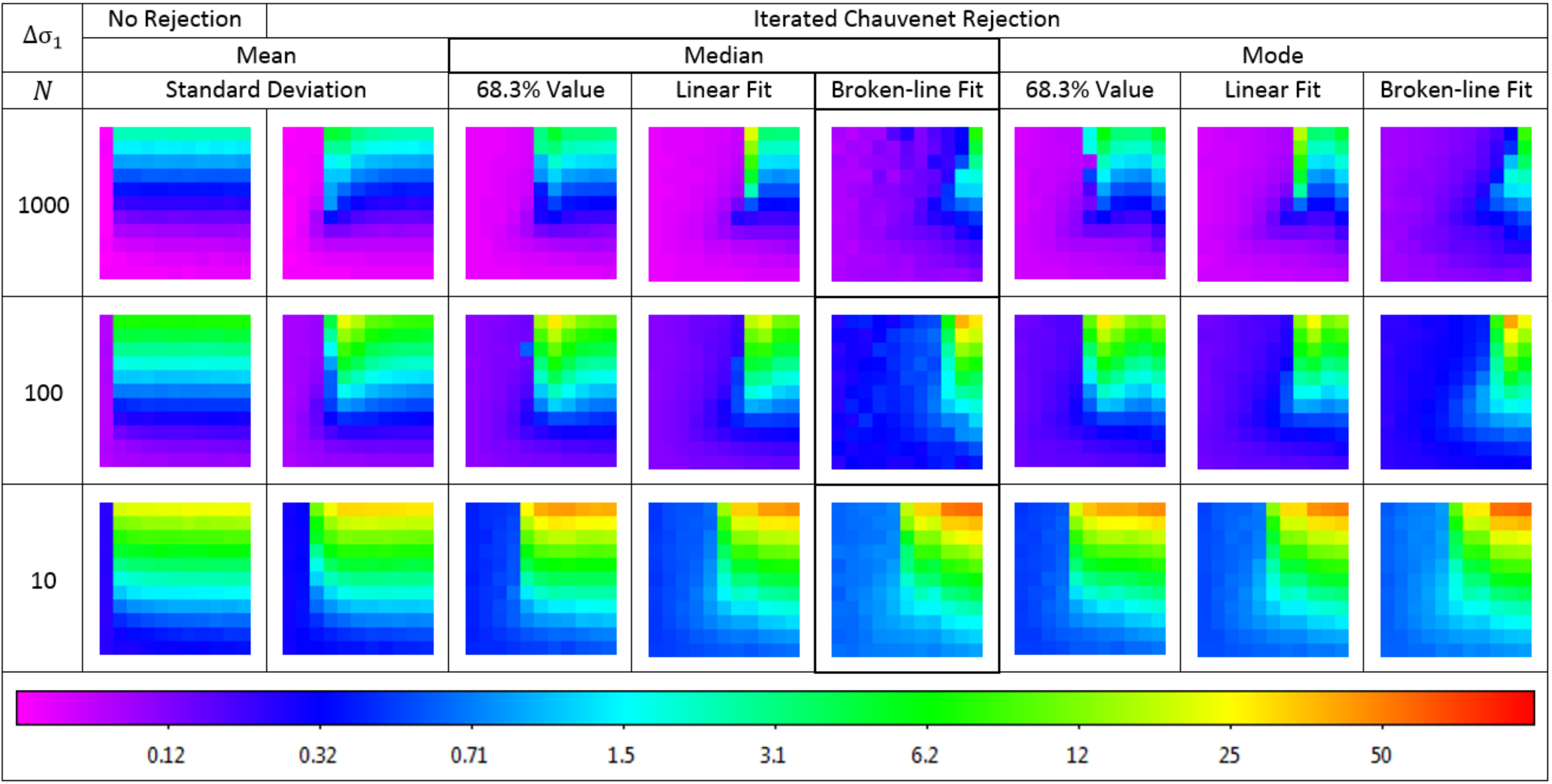}
		\caption{Uncertainty in the recovered $\sigma_1$ for increasingly robust measurement techniques and decreasing sample sizes ($N$), for two-sided contaminants.  See Figure~5 for contaminant strength ($\sigma_2$) vs.\@ fraction of sample ($f_2$) axis information.  The effect of the contaminants, without rejection, can be seen in the first column:  Larger contaminant fractions and strengths, as well as smaller sample sizes, result in less precise recovered values of $\sigma_1$.  However, increasingly robust measurement techniques are increasingly effective at rejecting outliers in large-$f_2$ samples, allowing $\sigma_1$ to be measured significantly more precisely.  Note that this is at a marginal cost:  When applied to uncontaminated samples ($f_2=0$), these techniques recover $\sigma_1$ with degrading precisions of $\Delta\sigma_1/\sigma_1\approx0.7N^{-1/2}$, $0.8N^{-1/2}$, $1.0N^{-1/2}$, $1.0N^{-1/2}$, $2.3N^{-1/2}$, $1.5N^{-1/2}$, $1.5N^{-1/2}$, and $2.8N^{-1/2}$, respectively.  {\color{black}However, all things considered (Figures 6 -- 9; \textsection3.1), the best-performing technique for Chauvenet-rejecting two-sided contaminants is highlighted with a bold outline.}  The colors are scaled logarithmically, between 0.02 and 100.}
	\end{sidewaysfigure*}
	
	As expected with two-sided contaminants, the average recovered $\mu_1$ is always $\approx$0.  However, the uncertainty in the recovered $\mu_1$, the average recovered $\sigma_1$, and the uncertainty in the recovered $\sigma_1$ are all susceptible to contamination, especially when $f_2$ and $\sigma_2$ are large.  However, our increasingly robust 68.3-percentile deviation measurement techniques are increasingly effective at rejecting outliers in large-$f_2$ samples, allowing $\sigma_1$ to be measured significantly more accurately, and both $\mu_1$ and $\sigma_1$ to be measured significantly more precisely.  Note that this is at a marginal cost:  When applied to uncontaminated samples, our increasingly robust measurement techniques recover $\mu_1$ and $\sigma_1$ with degrading precisions (Figures 7 and 9).  This suggests that one can reach a point of diminishing returns; however, this is a drawback that we largely eliminate in \textsection4.  Given this, when {\color{black}Chauvenet-rejecting} two-sided contaminants, we recommend using (1)~the median (because it is just as accurate as the mode (in this case), more precise, and computationally faster) and (2)~the 68.3-percentile deviation as measured by technique~3 from \textsection2.2 (the broken-line fit).  {\color{black}This technique is highlighted in Figures 6 -- 9 with a bold outline.\\
		\\
		
		\subsection{Normally Distributed Uncontaminated Measurements with One-Sided Contaminants}}
	
	We now repeat the analysis of \textsection3.1, but for the more challenging case of one-sided contaminants, which we model by drawing values from only the positive side of a Gaussian distribution of mean $\mu_2=0$ and standard deviation $\sigma_2$.  This case is more challenging because even though the median is more robust than the mean, and the mode is more robust than the median, even the mode will be biased in the direction of the contaminants (see Figure~10), and increasingly so as the fraction of the sample that is contaminated increases (see Figures 11 and 12).
	
	\begin{figure*}
		\plottwo{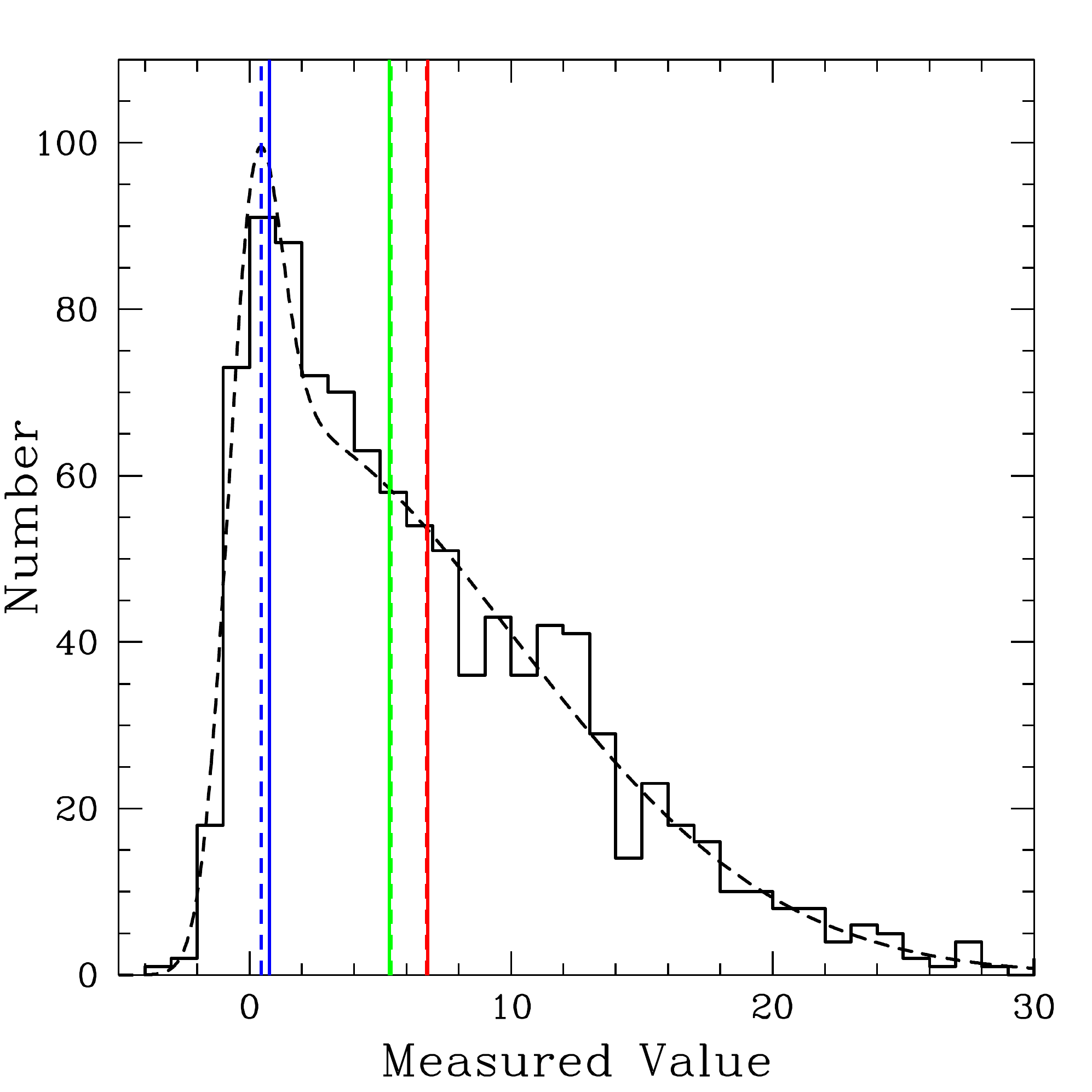}{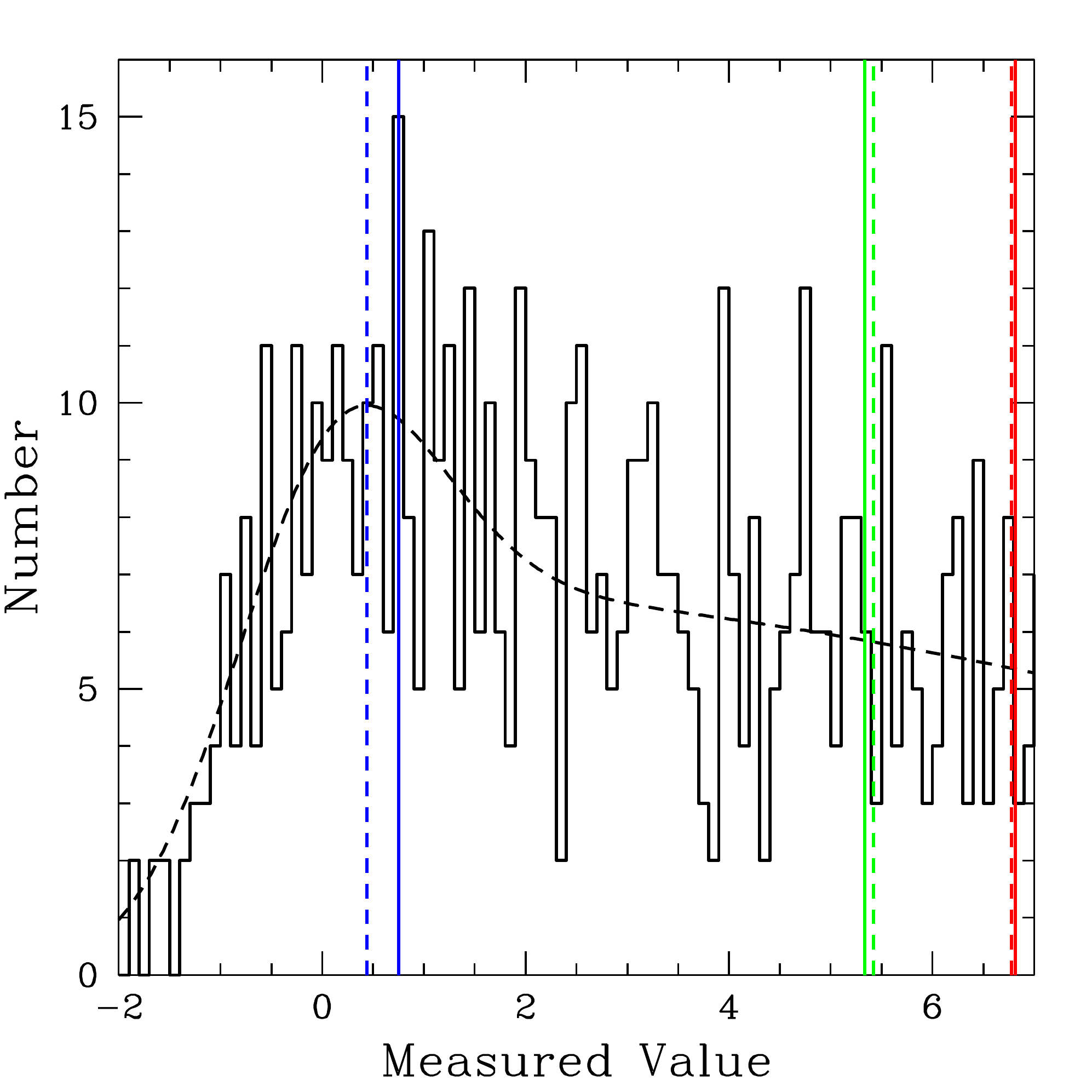}
		\caption{\textbf{Left:}  1000 measurements, with fraction $f_1=0.15$ drawn from a Gaussian distribution of mean $\mu_1=0$ and standard deviation $\sigma_1=1$, and fraction $f_2=0.85$, representing contaminated measurements, drawn from the positive side of a Gaussian distribution of mean $\mu_2=0$ and standard deviation $\sigma_2=10$, and added to uncontaminated measurements, drawn as above.  The measurements have been binned, and the mean (solid red line), median (solid green line), and mode (solid blue line) have been marked.  The dashed black curve marks the theoretical, or large-N, distribution, and for this the mean, median, and mode have also been marked, with dashed lines.  \textbf{Right:}  Zoom-in of the left panel, with smaller bins.  A large $f_2$ was chosen to more clearly demonstrate that the mode is biased in the direction of the contaminants, albeit only marginally.  Also, the sample mode differs from the theoretical mode more than the sample median and mean differ from the theoretical median and mean, due to ``noise'' peaks, caused by random sampling.  This is typical, and why the mode, although significantly more accurate, is less precise.}
	\end{figure*}
	
	\begin{figure*}
		\centering
		\includegraphics[height=0.9\textheight]{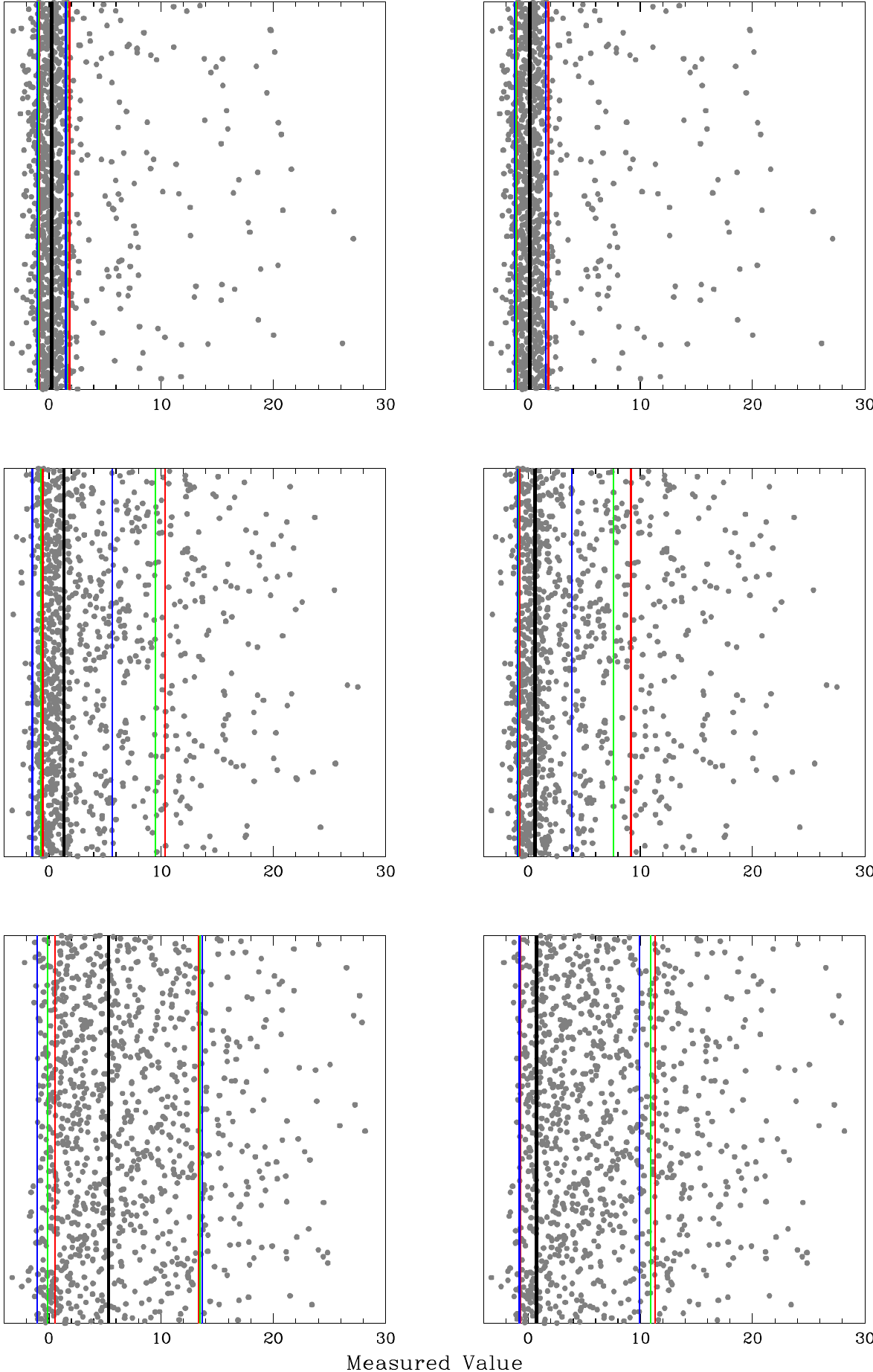}
		\caption{1000 measurements, with fraction $f_1=1-f_2$ drawn from a Gaussian distribution of mean $\mu_1=0$ and standard deviation $\sigma_1=1$, and fraction $f_2=0.15$ (top row), 0.5 (middle row), and 0.85 (bottom row), representing contaminated measurements, drawn from the positive side of a Gaussian distribution of mean $\mu_2=0$ and standard deviation $\sigma_2=10$, and added to uncontaminated measurements, drawn as above.  \textbf{Left column:}  Median (black line) and 68.3-percentile deviations, measured both below and above the median, using technique~1 from \textsection2.2 (red lines), using technique~2 from \textsection2.2 (green lines), and using technique~3 from \textsection2.2 (blue lines).  \textbf{Right column:}  Same as the left column, except using the mode instead of the median.  The mode performs better, especially in the limit of large $f_2$.  The 68.3-percentile deviation performs better when paired with the mode, and when measured in the opposite direction as the contaminants.  See Figure~12 for post-rejection versions.}
	\end{figure*}
	
	\begin{figure*}
		\centering
		\includegraphics[height=0.9\textheight]{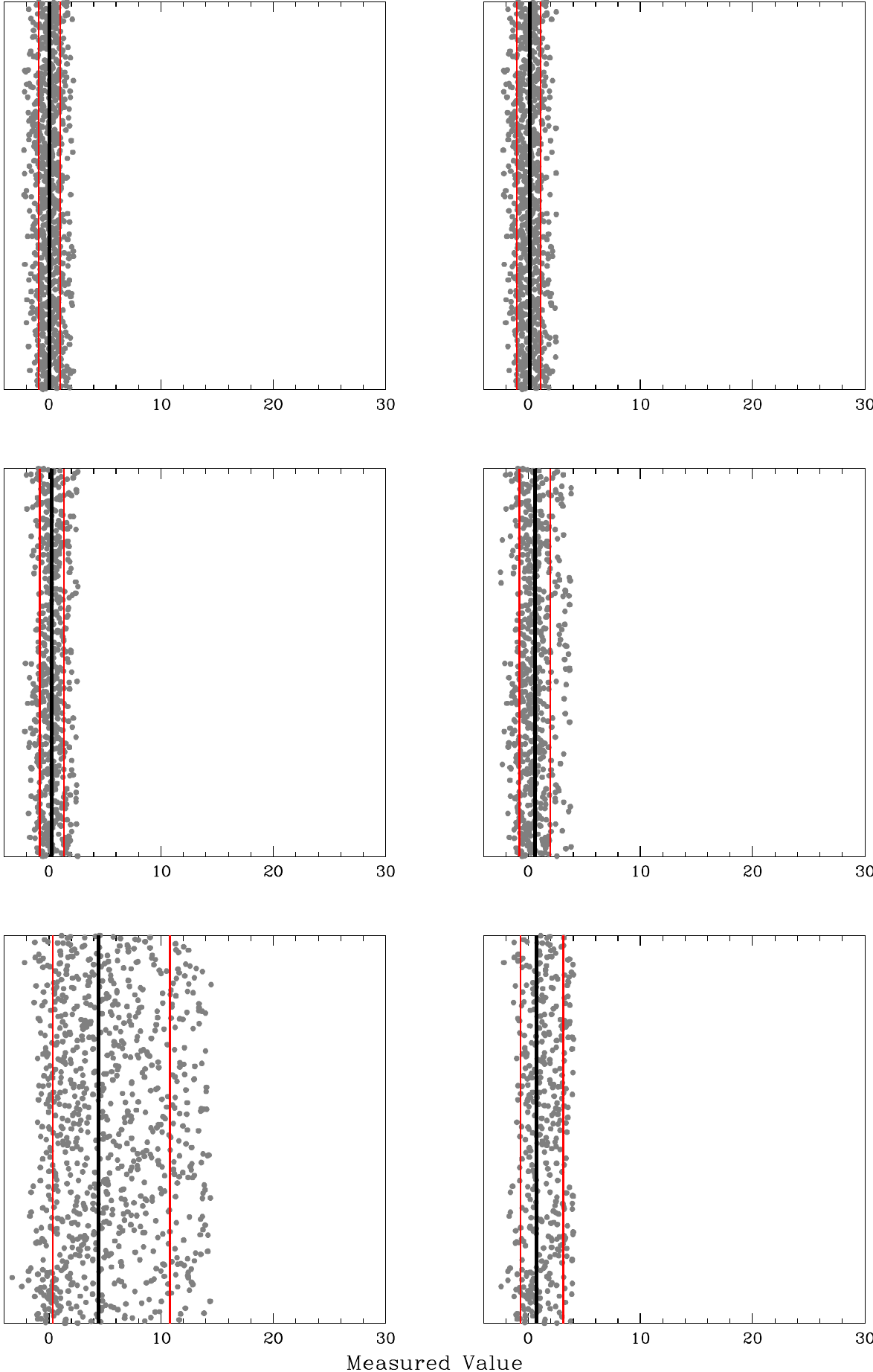}
		\caption{Figure~11, after iterated Chauvenet rejection, using the smaller of the below- and above-measured 68.3-percentile deviations, in this case as measured by technique~1 from \textsection2.2.  Techniques 2 and 3 from \textsection2.2 yield similar post-rejection samples and $\mu$ and $\sigma$ measurements.  The mode continues to perform better in the limit of large $f_2$.}
	\end{figure*}
	
	Furthermore, as $\mu$ (equal to the mean, the median, or the mode) becomes more biased in the direction of the one-sided contaminants, $\sigma$ (equal to the standard deviation or the 68.3-percentile deviation, as measured by any of the techniques presented in \textsection2.2) becomes more biased as well, (1)~because of the contaminants, and (2)~because it is measured from $\mu$.  However, $\sigma$ can be measured with less bias, if measured using only the deviations {\color{black}from $\mu$} that are in the opposite direction as the contaminants (in this case, the deviations below $\mu$; Figure~11).  Since the direction of the contaminants might not be known a priori, or since the contaminants might not be fully one-sided, instead being between the cases presented in \textsection3.1 and \textsection3.2, we measure $\sigma$ both below and above $\mu$,\footnote{When computing $\sigma$ below or above $\mu$, if a measurement equals $\mu$, we include it in both the below and above calculations, but with 50\% weight for each (see \textsection6).} and use the smaller of these two measurements when rejecting outliers (Figure~12).  Note, using the smaller of these two measurements should only be done if the uncontaminated measurements are {\color{black}symmetrically} distributed (see \textsection3.3.1).
	
	For the same techniques presented in \textsection3.1, except now computing $\sigma$ both below and above $\mu$ and adopting the smaller of the two, and for the same sample sizes presented in \textsection3.1, we plot the average recovered $\mu_1$ in Figure~13, the uncertainty in the recovered $\mu_1$ in Figure~14, the average recovered $\sigma_1$ in Figure~15, and the uncertainty in the recovered $\sigma_1$ in Figure~16.    
	
	\begin{sidewaysfigure*}
		\centering
		\vspace{3.75in}
		\includegraphics[width=0.9\textwidth]{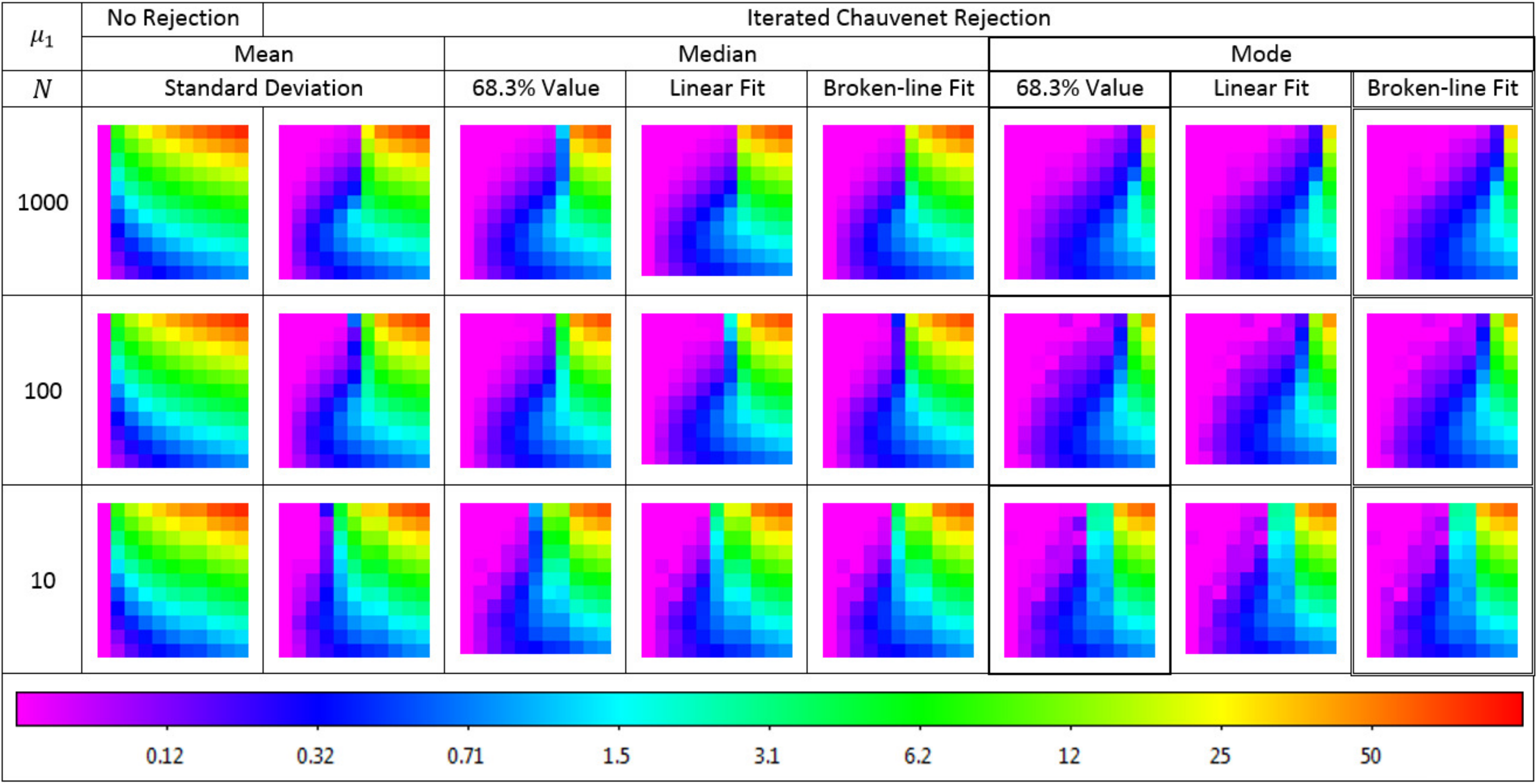}
		\caption{Average recovered $\mu_1$ for increasingly robust measurement techniques and decreasing sample sizes ($N$), for one-sided contaminants.  See Figure~5 for contaminant strength ($\sigma_2$) vs.\@ fraction of sample ($f_2$) axis information.  The effect of the contaminants, without rejection, can be seen in the first column:  Larger contaminant fractions and strengths result in larger recovered values of $\mu_1$.  However, for a fixed $\sigma$-measurement technique, our increasingly robust $\mu$-measurement techniques are increasingly effective at rejecting outliers in large-$f_2$ samples, allowing $\mu_1$ to be measured significantly more accurately.  However, when $\mu_1$ cannot be measured accurately, as is the case with the mean and the median when $f_2$ is large (Figures 10, 11, and 12), our (otherwise) increasingly robust $\sigma$-measurement techniques are \textit{decreasingly} effective at rejecting outliers (see Figure~17).  This can be seen in columns 3 -- 5, which use the 68.3-percentile deviation as measured by technique~1 from \textsection2.2, the 68.3-percentile deviation as measured by technique~2 from \textsection2.2, and the 68.3-percentile deviation as measured by technique~3 from \textsection2.2, respectively.  However, the mode can measure $\mu_1$ significantly more accurately (Figures 10, 11, and 12), even when $f_2$ is large, though with decreasing effectiveness in the low-$N$ limit.  In any case, when $\mu_1$ is measured accurately, all of these techniques are nearly equally effective, because $\sigma_1$ is measured on the nearly uncontaminated side of each sample's distribution.  {\color{black}All things considered (Figures 13 -- 16; \textsection3.2), the best-performing technique for Chauvenet-rejecting one-sided contaminants is highlighted with a bold outline, and the best-performing technique for Chauvenet-rejecting contaminants that are neither one-sided nor two-sided, but that are in-between these cases, is highlighted with a double outline.}  The colors are scaled logarithmically, between 0.02 and 100.}
	\end{sidewaysfigure*}
	
	\begin{sidewaysfigure*}
		\centering
		\vspace{3.75in}
		\includegraphics[width=0.9\textwidth]{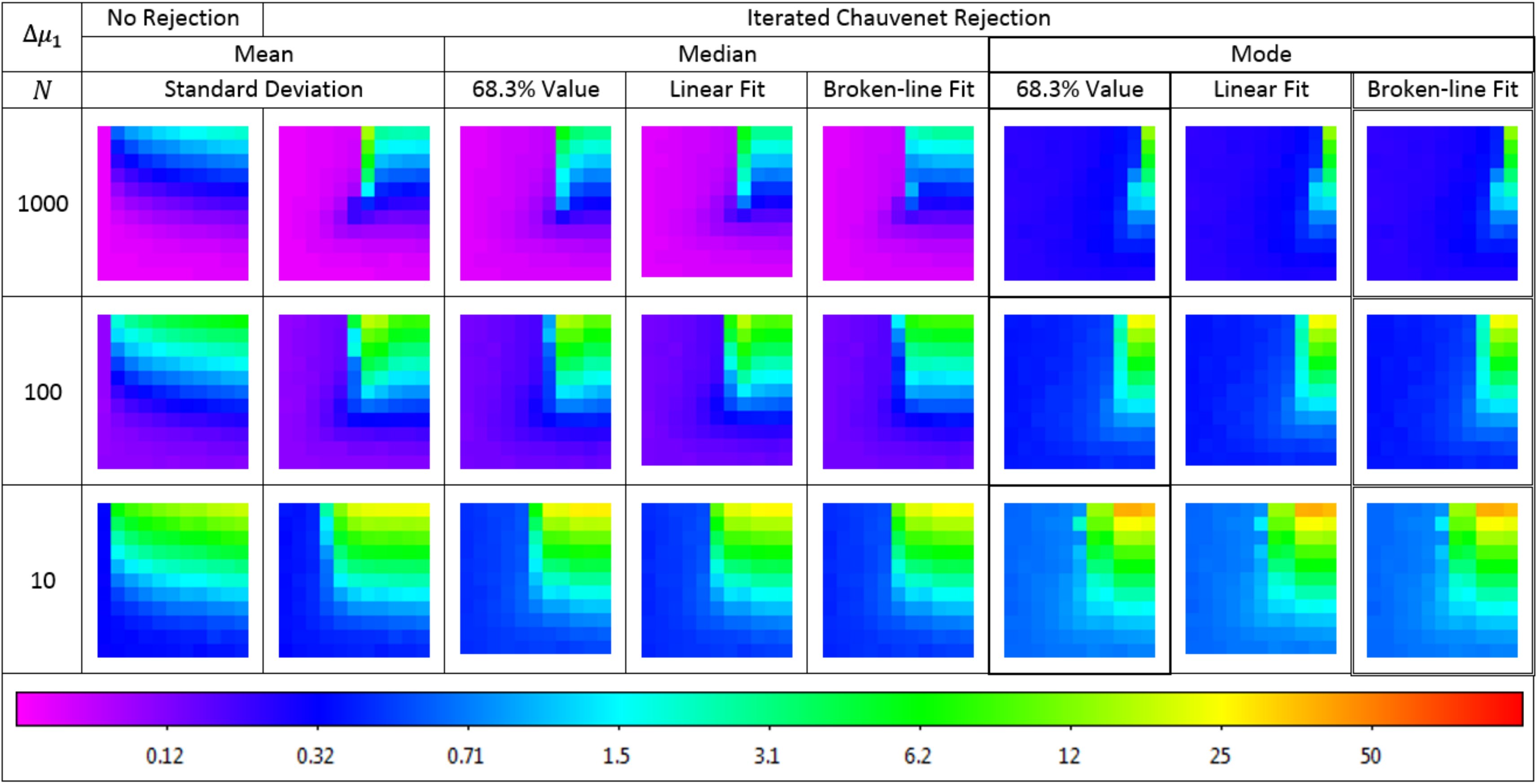}
		\caption{Uncertainty in the recovered $\mu_1$ for increasingly robust measurement techniques and decreasing sample sizes ($N$), for one-sided contaminants.  See Figure~5 for contaminant strength ($\sigma_2$) vs.\@ fraction of sample ($f_2$) axis information.  The effect of the contaminants, without rejection, can be seen in the first column:  Larger contaminant fractions and strengths, as well as smaller sample sizes, result in less precise recovered values of $\mu_1$.  However -- to the degree that $\mu_1$ can be measured accurately (Figure~13) -- all of our Chauvenet rejection techniques are effective at removing outliers (and nearly equally so, since $\sigma_1$ is measured on the nearly uncontaminated side of each sample's distribution), allowing $\mu_1$ to be measured significantly more precisely.  Note that, as in the case of two-sided contaminants (Figure~7), when applied to uncontaminated samples ($f_2=0$), these techniques recover $\mu_1$ with degrading precisions, of $\Delta\mu_1/\sigma_1\approx1.0N^{-1/2}$, $1.0N^{-1/2}$, $1.3N^{-1/2}$, $1.3N^{-1/2}$, $1.3N^{-1/2}$, $7.4N^{-1/2}$, $7.4N^{-1/2}$, and $7.4N^{-1/2}$, respectively.  {\color{black}However, all things considered (Figures 13 -- 16; \textsection3.2), the best-performing technique for Chauvenet-rejecting one-sided contaminants is highlighted with a bold outline, and the best-performing technique for Chauvenet-rejecting contaminants that are neither one-sided nor two-sided, but that are in-between these cases, is highlighted with a double outline.}  The colors are scaled logarithmically, between 0.02 and 100.}
	\end{sidewaysfigure*}
	
	\begin{sidewaysfigure*}
		\centering
		\vspace{3.75in}
		\includegraphics[width=0.9\textwidth]{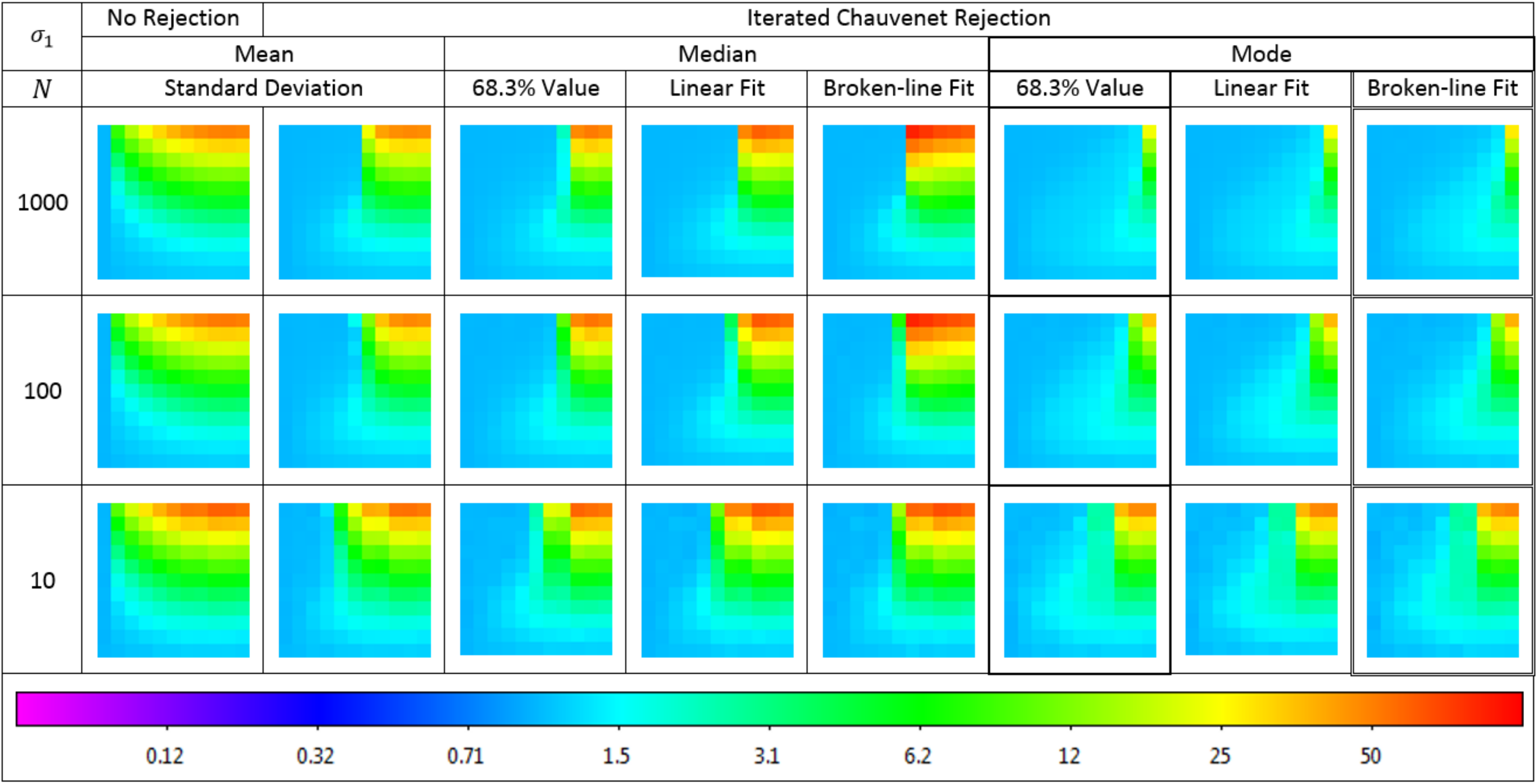}
		\caption{Average recovered $\sigma_1$ for increasingly robust measurement techniques and decreasing sample sizes ($N$), for one-sided contaminants.  See Figure~5 for contaminant strength ($\sigma_2$) vs.\@ fraction of sample ($f_2$) axis information.  The effect of the contaminants, without rejection, can be seen in the first column:  Larger contaminant fractions and strengths generally result in larger recovered values of $\sigma_1$.  However -- to the degree that $\mu_1$ can be measured accurately (Figure~13) -- all of our Chauvenet rejection techniques are effective at removing outliers (and nearly equally so, since $\sigma_1$ is measured on the nearly uncontaminated side of each sample's distribution), allowing $\sigma_1$ to be measured significantly more accurately.  {\color{black}All things considered (Figures 13 -- 16; \textsection3.2), the best-performing technique for Chauvenet-rejecting one-sided contaminants is highlighted with a bold outline, and the best-performing technique for Chauvenet-rejecting contaminants that are neither one-sided nor two-sided, but that are in-between these cases, is highlighted with a double outline.}  The colors are scaled logarithmically, between 0.02 and 100.}
	\end{sidewaysfigure*}
	
	\begin{sidewaysfigure*}
		\centering
		\vspace{3.75in}
		\includegraphics[width=0.9\textwidth]{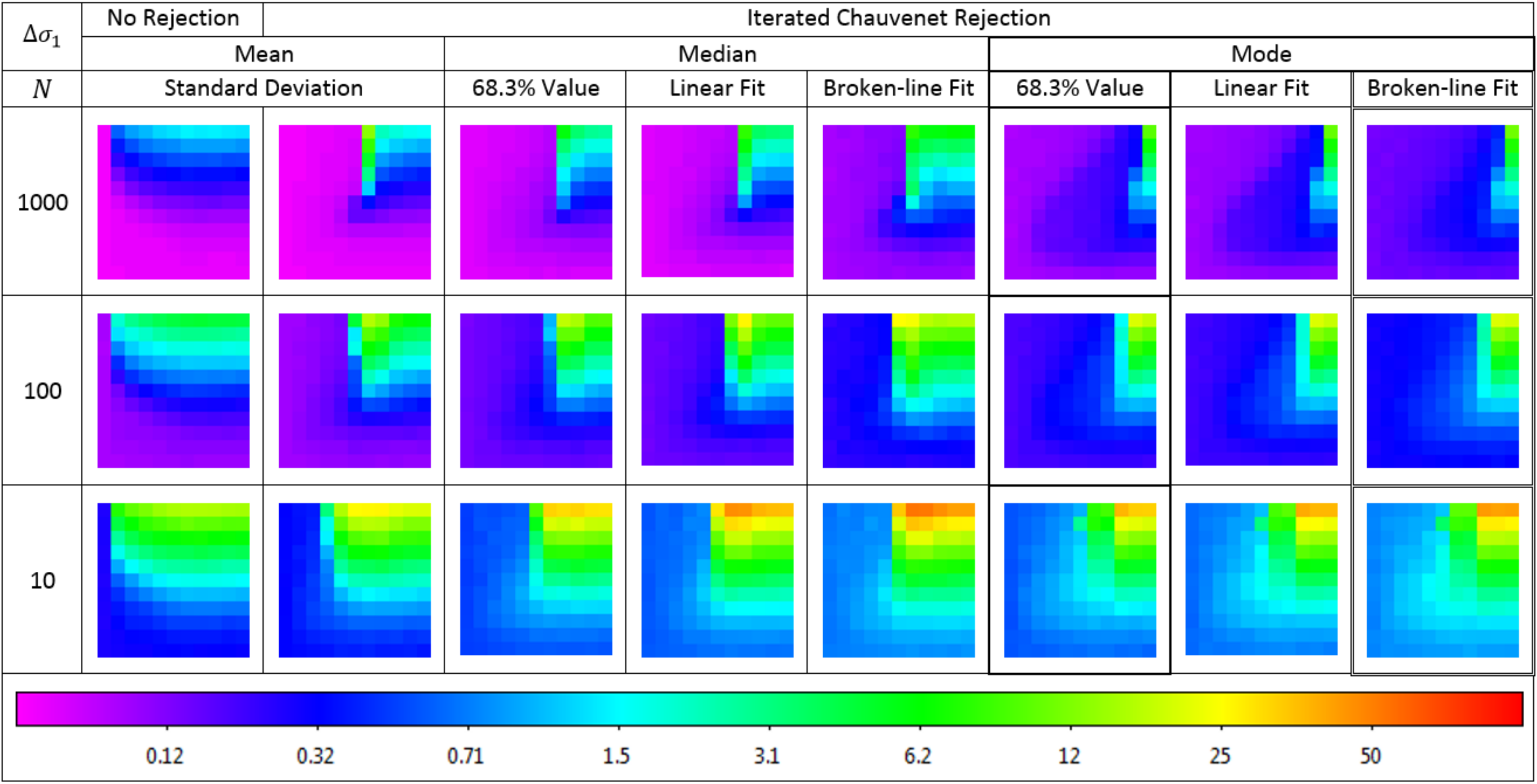}
		\caption{Uncertainty in the recovered $\sigma_1$ for increasingly robust measurement techniques and decreasing sample sizes ($N$), for one-sided contaminants.  See Figure~5 for contaminant strength ($\sigma_2$) vs.\@ fraction of sample ($f_2$) axis information.  The effect of the contaminants, without rejection, can be seen in the first column:  Larger contaminant fractions and strengths, as well as smaller sample sizes, generally result in less precise recovered values of $\sigma_1$.  However -- to the degree that $\mu_1$ can be measured accurately (Figure~13) -- all of our Chauvenet rejection techniques are effective at removing outliers (and nearly equally so, since $\sigma_1$ is measured on the nearly uncontaminated side of each sample's distribution), allowing $\sigma_1$ to be measured significantly more precisely.  Note that, as in the case of two-sided contaminants (Figure~9), when applied to uncontaminated samples ($f_2=0$), these techniques recover $\sigma_1$ with degrading precisions of $\Delta\sigma_1/\sigma_1\approx0.8N^{-1/2}$, $0.9N^{-1/2}$, $1.2N^{-1/2}$, $1.3N^{-1/2}$, $2.7N^{-1/2}$, $2.5N^{-1/2}$, $2.8N^{-1/2}$, and $3.9N^{-1/2}$, respectively.  {\color{black}However, all things considered (Figures 13 -- 16; \textsection3.2), the best-performing technique for Chauvenet-rejecting one-sided contaminants is highlighted with a bold outline, and the best-performing technique for Chauvenet-rejecting contaminants that are neither one-sided nor two-sided, but that are in-between these cases, is highlighted with a double outline.}  The colors are scaled logarithmically, between 0.02 and 100.}
	\end{sidewaysfigure*}
	
	With one-sided contaminants, all four of these are susceptible to contamination, especially when $f_2$ and $\sigma_2$ are large.  However, for a fixed $\sigma$-measurement technique, our increasingly robust $\mu$-measurement techniques are increasingly effective at rejecting outliers in large-$f_2$ samples, allowing $\mu_1$ and $\sigma_1$ to be measured both significantly more accurately and significantly more precisely.  However, when $\mu_1$ cannot be measured accurately, as is the case with the mean and the median when $f_2$ is large (Figures 10, 11, and 12), our (otherwise) increasingly robust $\sigma$-measurement techniques are \textit{decreasingly} effective at rejecting outliers (see Figure~17).  However, the mode can measure $\mu_1$ significantly more accurately (Figures 10, 11, and 12), even when $f_2$ is large, though with decreasing effectiveness in the low-$N$ limit.  In any case, when $\mu_1$ is measured accurately, all of these techniques are nearly equally effective, because $\sigma_1$ is measured on the nearly uncontaminated side of each sample's distribution.  Given this, when {\color{black}Chauvenet-rejecting} one-sided contaminants, we recommend using (1)~the mode, and (2)~the 68.3-percentile deviation as measured by technique~1 from \textsection2.2 (the 68.3\% value, because it is essentially as accurate as the other techniques (in this case), more precise,\footnote{As in the case of two-sided contaminants, when applied to uncontaminated samples, our increasingly robust measurement techniques recover $\mu_1$ and $\sigma_1$ with degrading precisions (Figures 14 and 16), but again, this is a drawback that we largely eliminate in \textsection4.} and computationally faster).  {\color{black}This technique is highlighted in Figures 13 -- 16 with a bold outline.}
	
	\begin{figure*}
		\plottwo{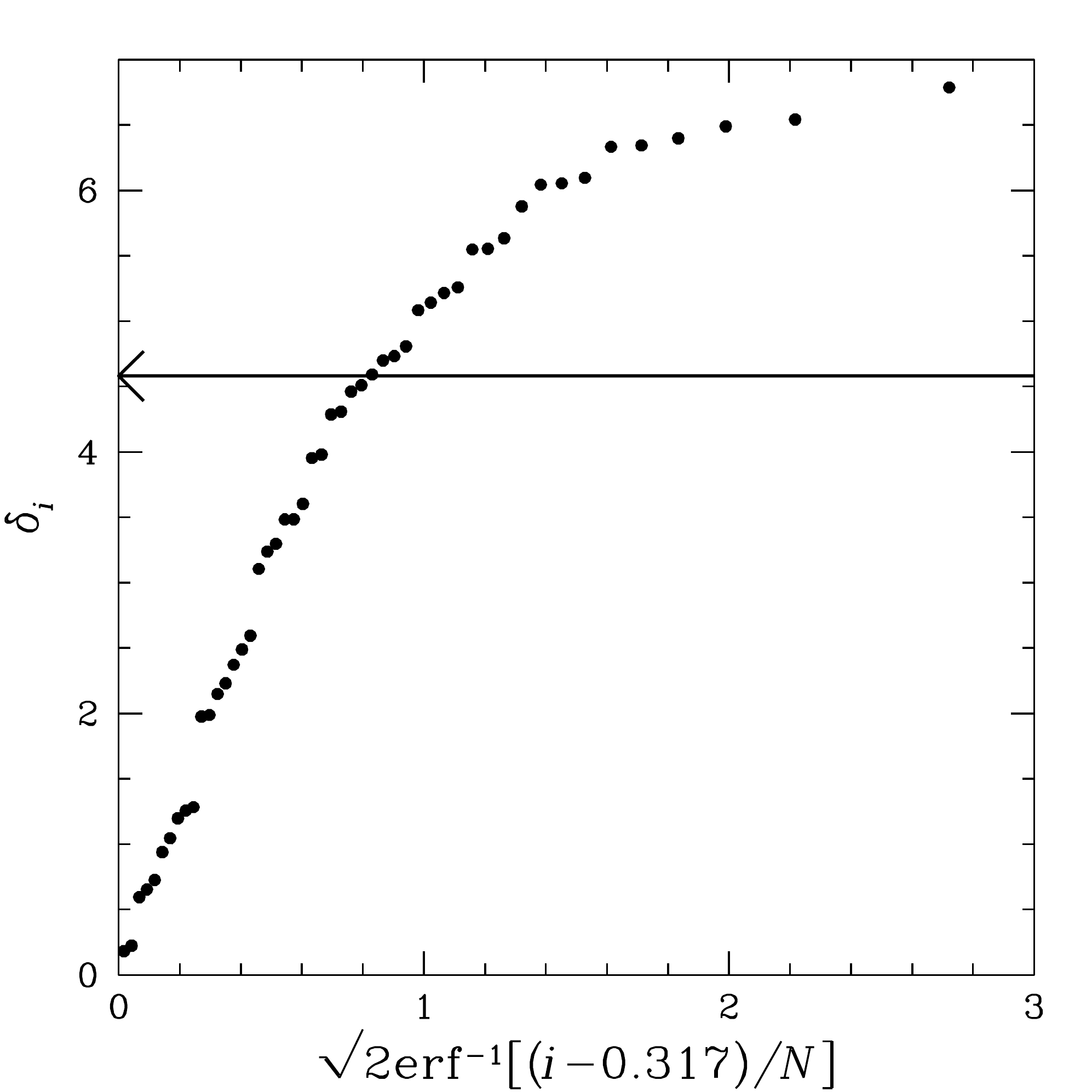}{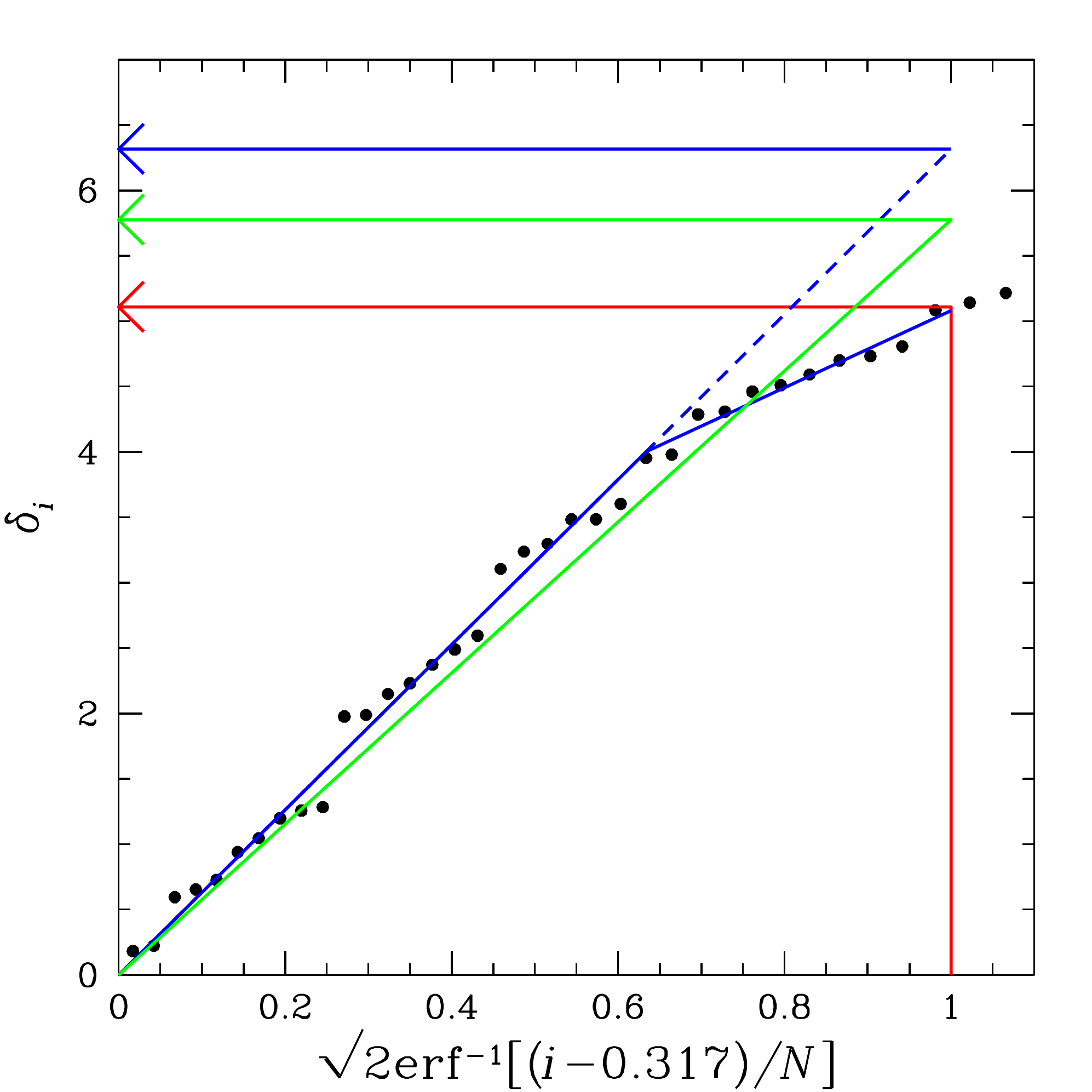}
		\caption{\textbf{Left:}  Sorted deviations from below the median of 100 measurements.  A fraction $f_1=0.15$ of these measurements are drawn from a Gaussian distribution of mean $\mu_1=0$ and standard deviation $\sigma_1=1$, and a fraction $f_2=0.85$, representing contaminated measurements, are drawn from the positive side of a Gaussian distribution of mean $\mu_2=0$ and standard deviation $\sigma_2=10$, and added to uncontaminated measurements, drawn as above.  The standard deviation, measured below the median, is marked (black arrow).  \textbf{Right:}  Zoom-in of the left panel, but with the 68.3-percentile deviation, also measured below the median, using technique~1 from \textsection2.2 (68.3\% value, red), using technique~2 from \textsection2.2 (linear fit, green), and using technique~3 from \textsection2.2 (broken-line fit, blue), instead marked.  In this case, the median significantly overestimates $\mu_1$, measuring 5.81 instead of 0, and consequently the curve breaks downward instead of upward.  When this happens, our normally increasingly robust $\sigma$-measurement techniques are \textit{decreasingly} accurate, measuring $\sigma_1=4.58$, 5.11, 5.78, and 6.32, respectively, instead of 1.  In other words, these techniques are only increasingly robust if $\mu_1$ is measured sufficiently accurately.  This is the case with the mode, even when $f_2$ is large, but is not the case with the mean and the median when $f_2$ is large (Figures 10 and 11), even post-rejection (Figure~12).}
	\end{figure*}
	
	When {\color{black}Chauvenet-rejecting} contaminants that are neither one-sided nor two-sided, but {\color{black}that are in-between these cases, with values that are both positive and negative,} but not in equal proportion or strength, we recommend using the smaller of the below- and above-measured 68.3-percentile deviations, as in the one-sided case, but recommend using (1)~the mode (which is just as effective as the median at eliminating two-sided contaminants (\textsection3.1), but more effective at eliminating one-sided contaminants), and (2)~the 68.3-percentile deviation as measured by technique~3 from \textsection2.2 (the broken line fit, which is more effective than the other techniques at eliminating two-sided contaminants (\textsection3.1) and essentially as effective at eliminating one-sided contaminants).  {\color{black}This technique is highlighted in Figures 13 -- 16 with a double outline.\\    
		
		\subsection{Non-Normally Distributed Uncontaminated Measurements with Contaminants}}
	
	In \textsection3.1 and \textsection3.2, we assumed that the uncontaminated measurements were drawn from a Gaussian distribution.  Although this is often a reasonable assumption, sometimes one might need to admit the possibility of an asymmetric (see \textsection3.3.1) or a peaked or flat-topped (see \textsection3.3.2) distribution for the uncontaminated measurements.  
	
	\subsubsection{Asymmetric Uncontaminated Distributions}
	
	In this case, it is better to use the $\sigma$ (equal to the standard deviation or the 68.3-percentile deviation, as measured by any of the techniques presented in \textsection2.2) measured from the deviations below $\mu$ (equal to the mean, the median, or the mode) to reject outliers below $\mu$, and the $\sigma$ measured from the deviations above $\mu$ to reject outliers above $\mu$, assuming that the distribution is only mildly non-normal, even if this means not always using the smaller of the two $\sigma$ values, as can be done with normally distributed uncontaminated measurements (\textsection3.2).
	
	However, this weakens one's ability to reject outliers, particularly when one-sided contaminants dominate the sample.  Even if the uncontaminated measurements are not asymmetrically distributed, simply admitting the possibility {\color{black}can reduce} one's ability to remove contaminants, so this is a decision that should be made with care.  
	
	To demonstrate this, we {\color{black}repeated} the analysis of \textsection3.2, not changing the uncontaminated measurements, but changing the assumption that we {\color{black}made} about their distribution, instead admitting the possibility of asymmetry.  We {\color{black}then plotted} the average recovered {\color{black}$\mu_1$,} the uncertainty in the recovered {\color{black}$\mu_1$,} the average recovered below-measured {\color{black}$\sigma_{1-}$,} the uncertainty in the recovered {\color{black}$\sigma_{1-}$,} the average recovered above-measured {\color{black}$\sigma_{1+}$,} and the uncertainty in the recovered {\color{black}$\sigma_{1+}$, and compared these to those from \textsection3.1 and \textsection3.2.
		
		As one might expect, (1)~the plots for $\mu_1$, $\Delta\mu_1$, $\sigma_{1-}$, and $\Delta\sigma_{1-}$ resembled the one-sided contaminant results (Figures 13 -- 16, respectively), and (2)~the plots for $\sigma_{1+}$ and $\Delta\sigma_{1+}$ resembled the two-sided contaminant results (Figures 15 and 16, respectively, but for about half as many measurements), where the latter can be less effective in the limit of large $f_2$ and $\sigma_2$ (but still significantly more effective than traditional Chauvenet rejection).}  Since this case approximates both one-sided and two-sided results, when {\color{black}Chauvenet-rejecting} contaminants, we recommend using (1)~the mode and (2)~the 68.3-percentile deviation as measured by technique~3 from \textsection2.2 (the broken-line fit) for the same reasons that we recommend using this combination when rejecting in-between contaminants from normally distributed uncontaminated measurements (\textsection3.2).
	
	It should be noted that if we also change the uncontaminated measurements to be asymmetrically distributed, instead of merely admitting the possibility that they are asymmetrically distributed, the mean, median, and mode then mean different things, in the sense that they mark different parts of the distribution, even in the limit of large $N$ and no contaminants.  Furthermore, deviations, however measured, from each of these $\mu$ measurements likewise then mean different things.  A deeper exploration of these differences, and of their effects on contaminant removal, is beyond the scope of this paper.  However, as long as the asymmetry is mild, the effectiveness of this technique should not differ greatly from what has been presented here.{\color{black}\footnote{\color{black}And what has been presented here is treating each side of the distribution as a pure Gaussian, but of different $\sigma$ (which, technically, is a discontinuous approximation of the true distribution.)}}
	
	It should also be noted that in the simpler case of two-sided contaminants, this technique differs very little from what has been presented in \textsection3.1, except that $\sigma_{1-}$, $\Delta\sigma_{1-}$, $\sigma_{1+}$, and $\Delta\sigma_{1+}$ are each determined with about half as many measurements (the measurements on each quantity's side of $\mu_1$).
	
	{\color{black}Finally, it should be noted that this technique is less prone to runaway over-rejection than the techniques presented in \textsection3.1 and \textsection3.2.  The calibration of these techniques that we introduced in \textsection2.3 is intended to, and largely does, prevent this from happening, but it can still happen in the limit of very-low $N$, if two measurements happen to be unusually close together (in which case all other measurements are rejected).  For uncontaminated, Gaussian-random data, and the techniques presented in \textsection3.1 and \textsection3.2, this happens $\approx$25\% -- 33\%, $\approx$3.6\% -- 9.4\%, and $\approx$0.30\% -- 2.3\% of the time when $N=5$, 10, and 20, respectively.  This is not surprising, given that very-low $N$ distributions can be very non-Gaussian in appearance, in which case this, asymmetric technique may be more appropriate.  In this case, again applied to uncontaminated, Gaussian-random data, runaway over-rejection happens only $\approx$0.014\% of the time when $N=5$, and never when $N\geq10$.}
	
	\subsubsection{Peaked or Flat-Topped Uncontaminated Distributions}
	
	Consider the following generalization of the Gaussian (technically called an exponential power distribution):
	
	\begin{equation}
	p(\delta) = \frac{\sqrt{\kappa}}{\sqrt{2\pi}\sigma}e^{-\frac{1}{2}\left|\frac{\delta}{\sigma}\right|^{2\kappa}},
	\end{equation}
	
	\noindent
	which reduces to a Gaussian when $\kappa=1$, but results in peaked (positive-kurtosis) distributions when $\kappa<1$ and flat-topped (negative-kurtosis) distributions when $\kappa>1$ (see Figure~18).  The standard deviation of this distribution is $\sigma/\sqrt{\kappa}$.
	
	\begin{figure}
		\plotone{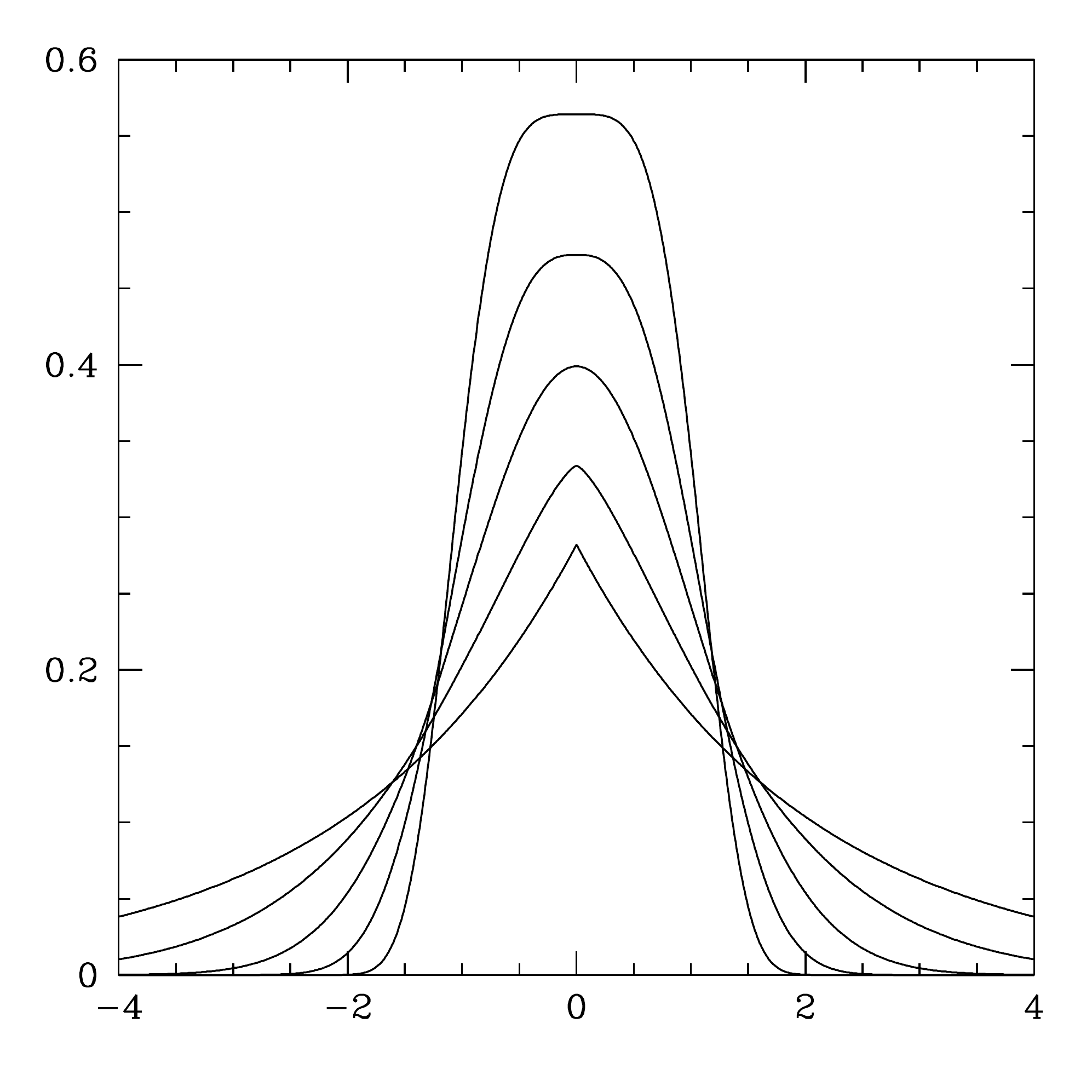}
		\caption{Exponential power distribution (Equation 5), for $\kappa=0.5$ (peaked), 0.7, 1 (Gaussian), 1.4, and 2 (flat-topped).}
	\end{figure}
	
	For this distribution, Chauvenet's criterion (Equation 1) implies that measurements are rejected if their deviations are greater than a certain number of $\sigma/\sqrt{\kappa}$ (standard deviations), instead of $\sigma$, as in the pure Gaussian case.    
	
	Furthermore, Equation 4 becomes:
	
	\begin{equation}
	\delta_i = \frac{\sigma}{\sqrt{\kappa}}\left[\sqrt{2}\mathrm{erf}^{-1}\left(\frac{i-0.317}{N}\right)\right],
	\end{equation}
	
	\noindent
	which is proportional to $\sigma/\sqrt{\kappa}$, instead of $\sigma$.  Consequently, the techniques presented in this paper work identically if the uncontaminated measurements are distributed not normally but peaked or flat-topped -- in this specific way.
	
	Of course, not all peaked and flat-topped distributions are of this specific form.  However, if only mildly peaked or flat-topped, this form is a good, first-order approximation, and consequently we conclude that the techniques presented in this paper are not overly sensitive to our assumption of Gaussianity, for the uncontaminated measurements.
	
	{\color{black}We summarize all of the recommended, or best-option, robust techniques of \textsection3 in Figure~19.
		
		\begin{sidewaysfigure*}
			\centering
			\vspace{3.75in}
			\includegraphics[width=0.9\textwidth]{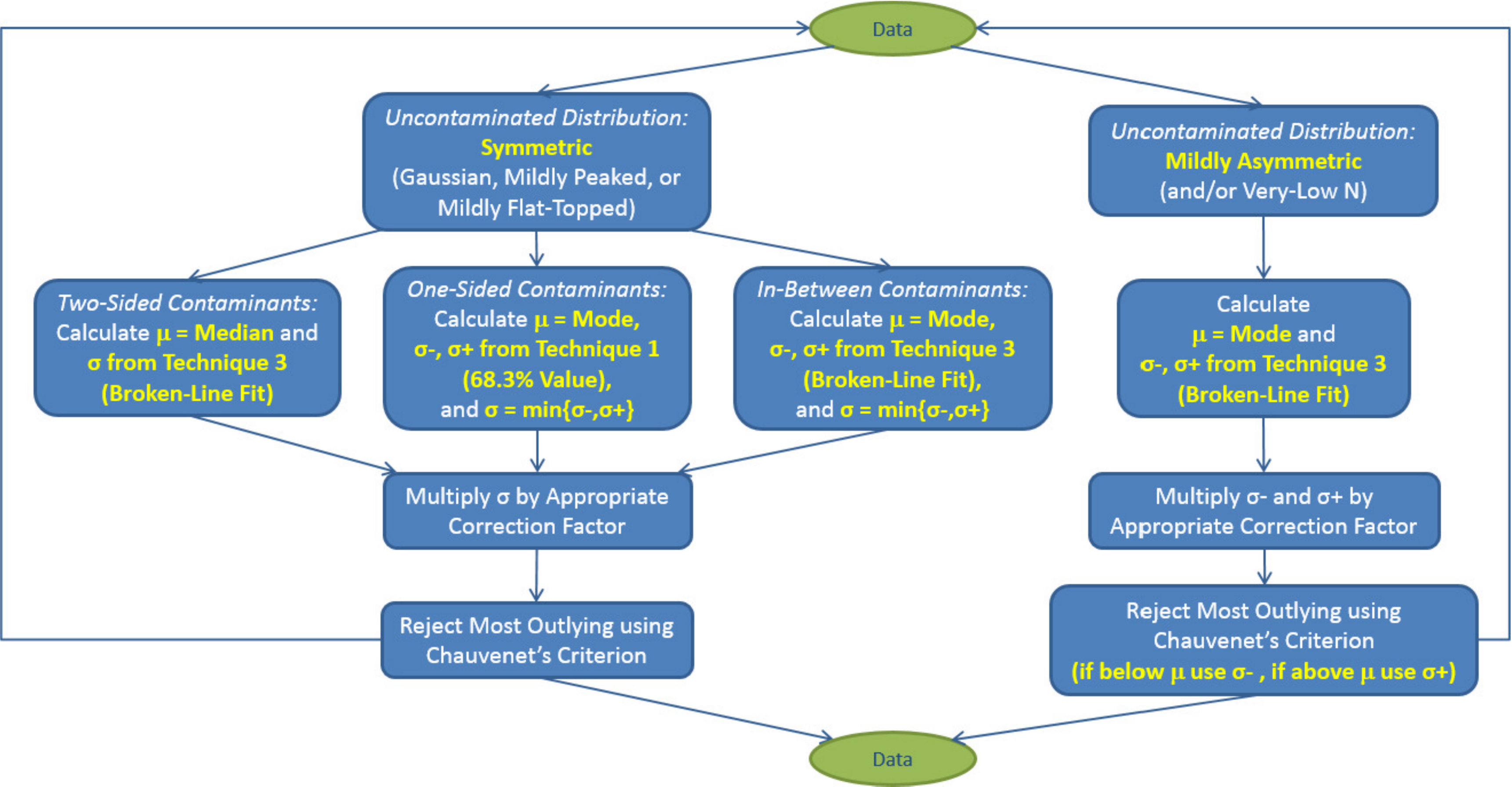}
			\caption{{\color{black}Best-option robust techniques for different uncontaminated distributions and contaminant types.  The mildly asymmetric technique is also robust against runaway over-rejection in very small samples (\textsection3.3.1).  The most discrepant outlier is rejected each iteration, and one iterates until no outliers remain.  $\mu$ and $\sigma$ (or $\sigma_{-}$ and $\sigma_{+}$, depending on {\color{black}whether the uncontaminated distribution is symmetric or asymmetric, and on the} contaminant type) are recalculated after each iteration, and the latter is multiplied by the appropriate correction factor (see {\color{black}Figure~4}) before being used to reject the next outlier.}}
		\end{sidewaysfigure*}
		
		\begin{sidewaysfigure*}
			\centering
			\vspace{3.75in}
			\includegraphics[width=0.9\textwidth]{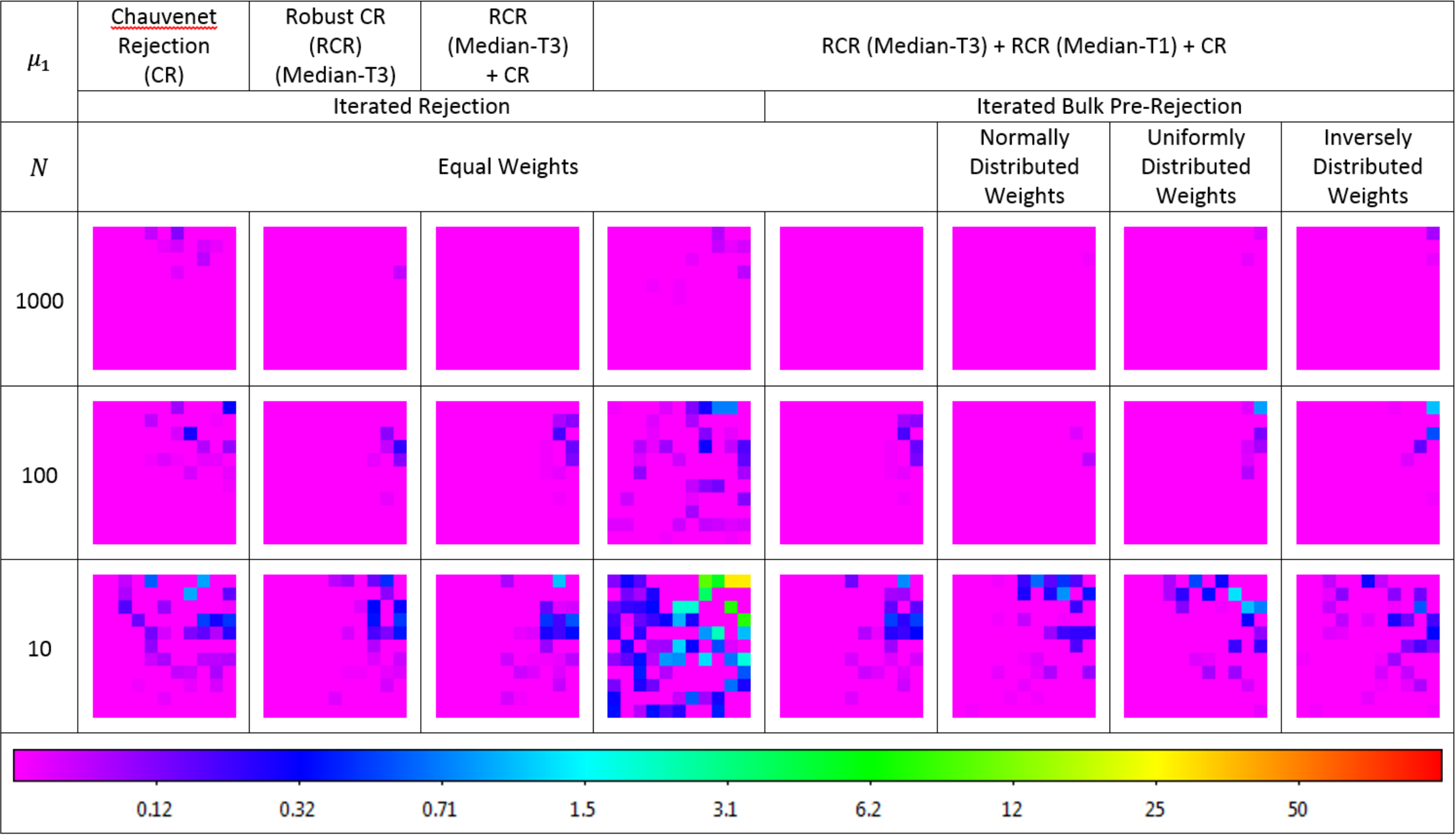}
			\caption{Average recovered $\mu_1$ given two-sided contaminants, for, from left to right:  (1)~{\color{black}traditional} Chauvenet rejection; (2)~{\color{black}our} best-option robust {\color{black}technique} (\textsection3.1, \textsection4); (3)~column~2 followed by column~1 (\textsection4); (4)~column~2 followed by our most{\color{black}-}precise robust {\color{black}technique} followed by column~1, plotted as improvement over column~3, multiplied by 100 (\textsection4); (5)~{\color{black}our} best-option {\color{black}bulk pre-rejection technique} followed by column~4 (\textsection5); (6)~same as column~5, except for weighted data, with weights distributed normally with standard deviation as a fraction of the mean $\sigma_w/\mu_w=0.3$ (\textsection6); (7)~same as column~5, except for weights distributed uniformly from zero, corresponding to $\sigma_w/\mu_w\approx0.58$ (\textsection6); and (8)~same as column~5, except for weights distributed inversely over one dex, corresponding to $\sigma_w/\mu_w\approx0.73$ (\textsection6).  See Figure~5 for $\sigma_2$ vs.\@ $f_2$ axis labels.  The colors are scaled logarithmically, between 0.02 and 100.}
		\end{sidewaysfigure*}
		
		\begin{sidewaysfigure*}
			\centering
			\vspace{3.75in}
			\includegraphics[width=0.9\textwidth]{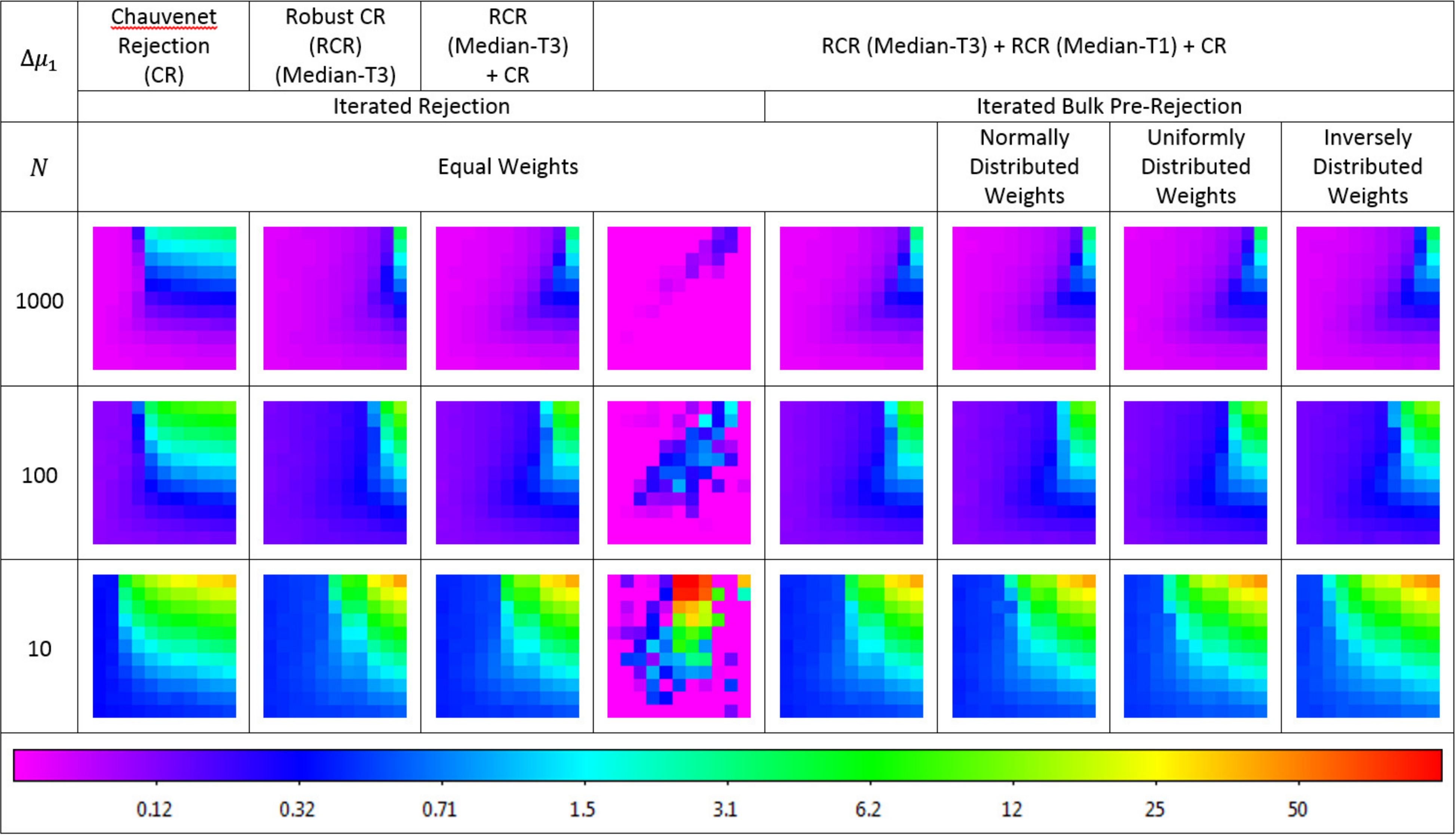}
			\caption{Uncertainty in recovered $\mu_1$ given two-sided contaminants, for, from left to right:  (1)~{\color{black}traditional} Chauvenet rejection; (2)~{\color{black}our} best-option robust {\color{black}technique} (\textsection3.1, \textsection4); (3)~column~2 followed by column~1 (\textsection4); (4)~column~2 followed by our most{\color{black}-}precise robust {\color{black}technique} followed by column~1, plotted as improvement over column~3, multiplied by 100 (\textsection4); (5)~{\color{black}our} best-option {\color{black}bulk pre-rejection technique} followed by column~4 (\textsection5); (6)~same as column~5, except for weighted data, with weights distributed normally with standard deviation as a fraction of the mean $\sigma_w/\mu_w=0.3$ (\textsection6); (7)~same as column~5, except for weights distributed uniformly from zero, corresponding to $\sigma_w/\mu_w\approx0.58$ (\textsection6); and (8)~same as column~5, except for weights distributed inversely over one dex, corresponding to $\sigma_w/\mu_w\approx0.73$ (\textsection6).  See Figure~5 for $\sigma_2$ vs.\@ $f_2$ axis labels.  The colors are scaled logarithmically, between 0.02 and 100.}
		\end{sidewaysfigure*}
		
		\begin{sidewaysfigure*}
			\centering
			\vspace{3.75in}
			\includegraphics[width=0.9\textwidth]{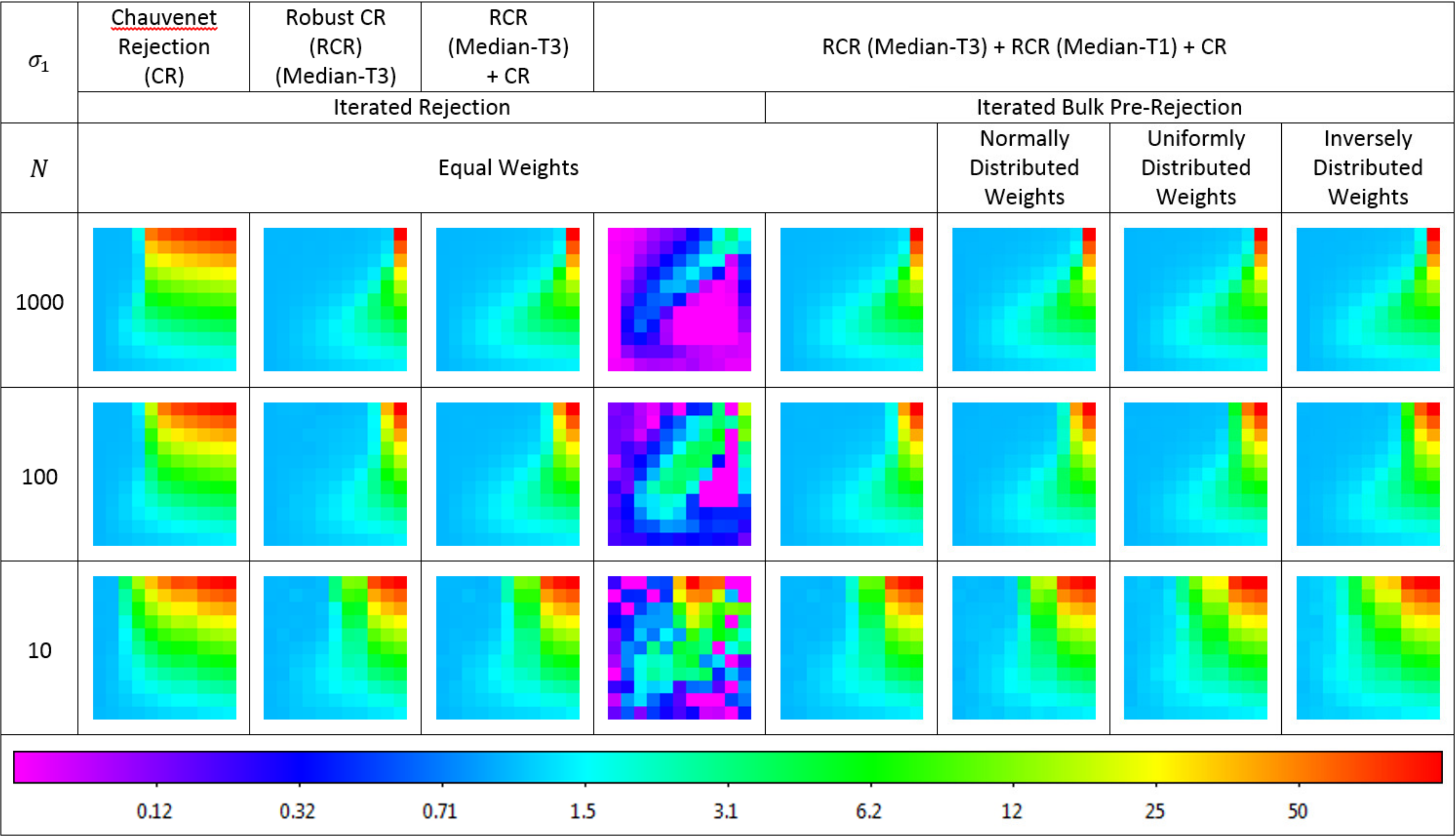}
			\caption{Average recovered $\sigma_1$ given two-sided contaminants, for, from left to right:  (1)~{\color{black}traditional} Chauvenet rejection; (2)~{\color{black}our} best-option robust {\color{black}technique} (\textsection3.1, \textsection4); (3)~column~2 followed by column~1 (\textsection4); (4)~column~2 followed by our most{\color{black}-}precise robust {\color{black}technique} followed by column~1, plotted as improvement over column~3, multiplied by 100 (\textsection4); (5)~{\color{black}our} best-option {\color{black}bulk pre-rejection technique} followed by column~4 (\textsection5); (6)~same as column~5, except for weighted data, with weights distributed normally with standard deviation as a fraction of the mean $\sigma_w/\mu_w=0.3$ (\textsection6); (7)~same as column~5, except for weights distributed uniformly from zero, corresponding to $\sigma_w/\mu_w\approx0.58$ (\textsection6); and (8)~same as column~5, except for weights distributed inversely over one dex, corresponding to $\sigma_w/\mu_w\approx0.73$ (\textsection6).  See Figure~5 for $\sigma_2$ vs.\@ $f_2$ axis labels.  The colors are scaled logarithmically, between 0.02 and 100.}
		\end{sidewaysfigure*}
		
		\begin{sidewaysfigure*}
			\centering
			\vspace{3.75 in}
			\includegraphics[width=0.9\textwidth]{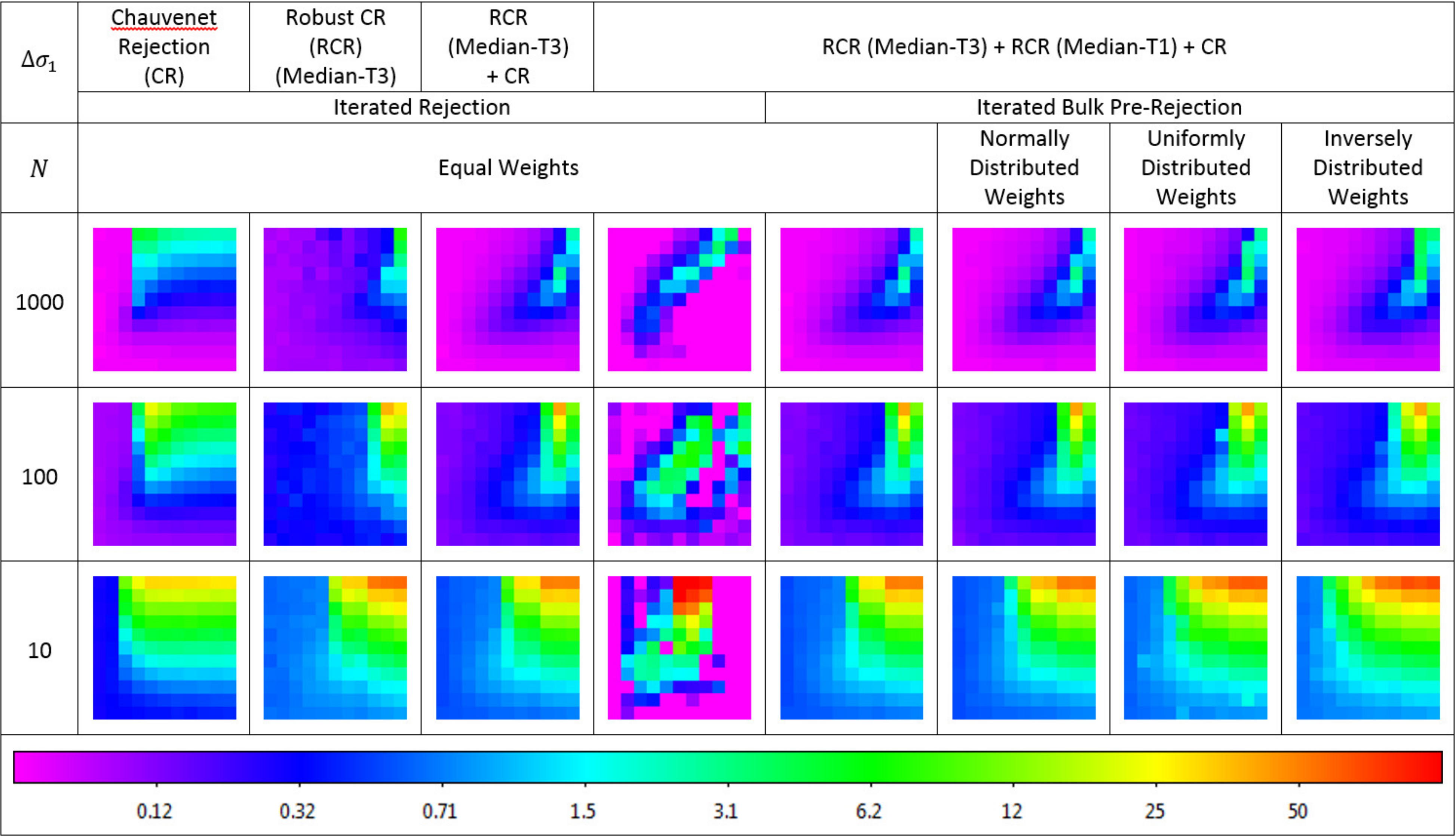}
			\caption{Uncertainty in recovered $\sigma_1$ given two-sided contaminants, for, from left to right:  (1)~{\color{black}traditional} Chauvenet rejection; (2)~{\color{black}our} best-option robust {\color{black}technique} (\textsection3.1, \textsection4); (3)~column~2 followed by column~1 (\textsection4); (4)~column~2 followed by our most{\color{black}-}precise robust {\color{black}technique} followed by column~1, plotted as improvement over column~3, multiplied by 100 (\textsection4); (5)~{\color{black}our} best-option {\color{black}bulk pre-rejection technique} followed by column~4 (\textsection5); (6)~same as column~5, except for weighted data, with weights distributed normally with standard deviation as a fraction of the mean $\sigma_w/\mu_w=0.3$ (\textsection6); (7)~same as column~5, except for weights distributed uniformly from zero, corresponding to $\sigma_w/\mu_w\approx0.58$ (\textsection6); and (8)~same as column~5, except for weights distributed inversely over one dex, corresponding to $\sigma_w/\mu_w\approx0.73$ (\textsection6).  See Figure~5 for $\sigma_2$ vs.\@ $f_2$ axis labels.  The colors are scaled logarithmically, between 0.02 and 100.}
		\end{sidewaysfigure*}
		
		\begin{sidewaysfigure*}
			\centering
			\vspace{3.75in}
			\includegraphics[width=0.9\textwidth]{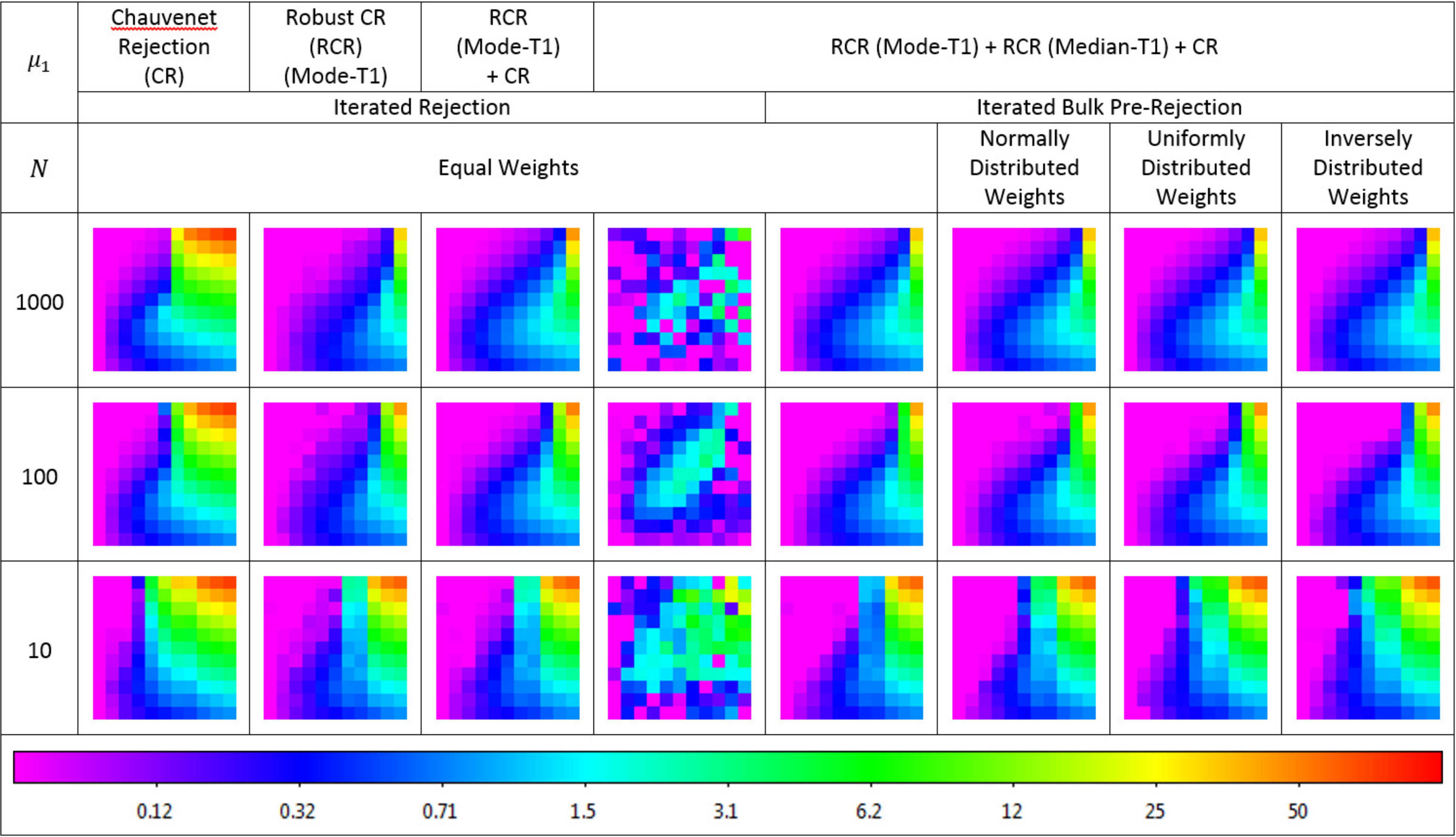}
			\caption{Average recovered $\mu_1$ given one-sided contaminants, for, from left to right:  (1)~{\color{black}traditional} Chauvenet rejection; (2)~{\color{black}our} best-option robust {\color{black}technique} (\textsection3.1, \textsection4); (3)~column~2 followed by column~1 (\textsection4); (4)~column~2 followed by our most{\color{black}-}precise robust {\color{black}technique} followed by column~1, plotted as improvement over column~3, multiplied by 100 (\textsection4); (5)~{\color{black}our} best-option {\color{black}bulk pre-rejection technique} followed by column~4 (\textsection5); (6)~same as column~5, except for weighted data, with weights distributed normally with standard deviation as a fraction of the mean $\sigma_w/\mu_w=0.3$ (\textsection6); (7)~same as column~5, except for weights distributed uniformly from zero, corresponding to $\sigma_w/\mu_w\approx0.58$ (\textsection6); and (8)~same as column~5, except for weights distributed inversely over one dex, corresponding to $\sigma_w/\mu_w\approx0.73$ (\textsection6).  See Figure~5 for $\sigma_2$ vs.\@ $f_2$ axis labels.  The colors are scaled logarithmically, between 0.02 and 100.}
		\end{sidewaysfigure*}
		
		\begin{sidewaysfigure*}
			\centering
			\vspace{3.75in}
			\includegraphics[width=0.9\textwidth]{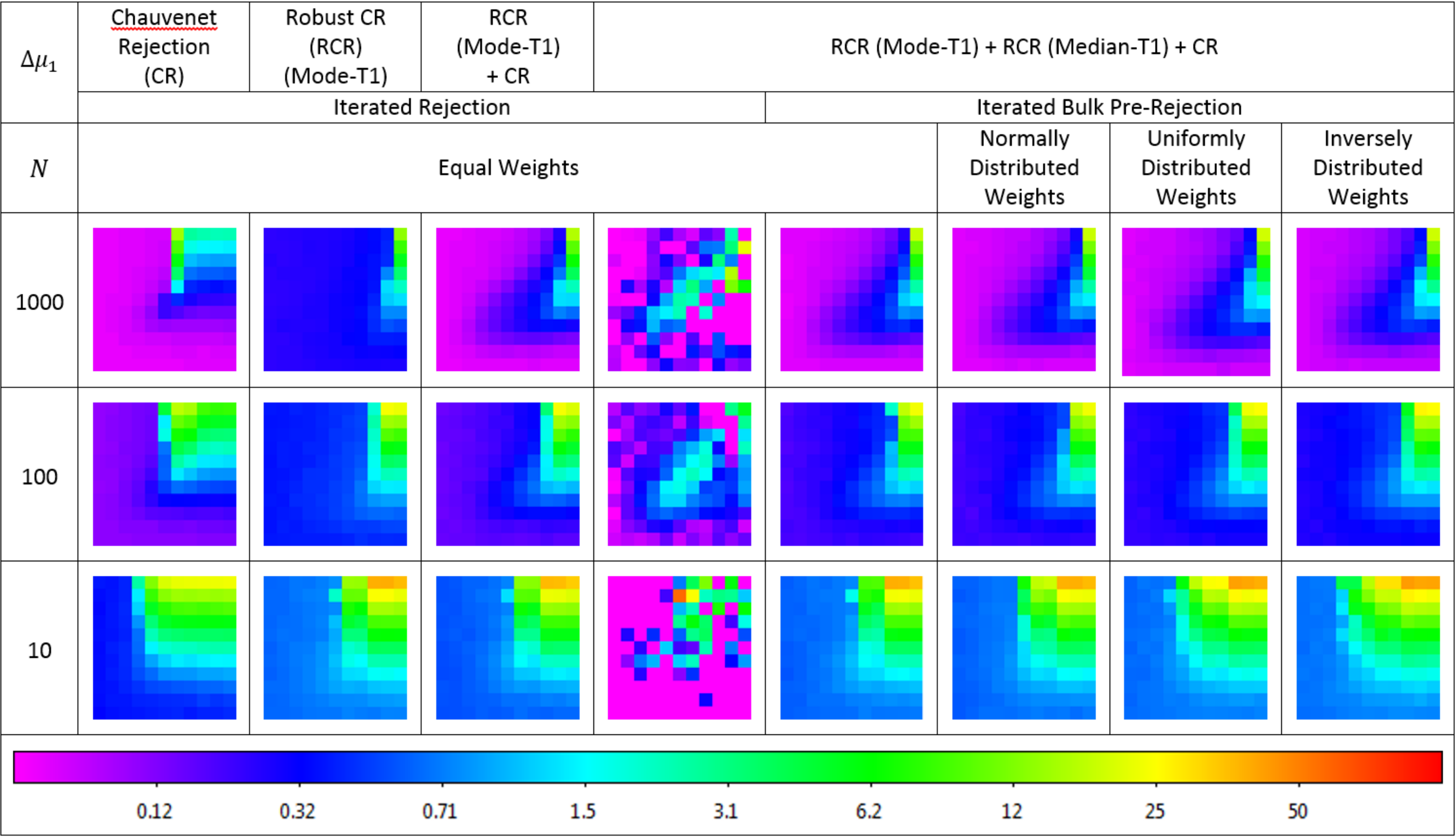}
			\caption{Uncertainty in recovered $\mu_1$ given one-sided contaminants, for, from left to right:  (1)~{\color{black}traditional} Chauvenet rejection; (2)~{\color{black}our} best-option robust {\color{black}technique} (\textsection3.1, \textsection4); (3)~column~2 followed by column~1 (\textsection4); (4)~column~2 followed by our most{\color{black}-}precise robust {\color{black}technique} followed by column~1, plotted as improvement over column~3, multiplied by 100 (\textsection4); (5)~{\color{black}our} best-option {\color{black}bulk pre-rejection technique} followed by column~4 (\textsection5); (6)~same as column~5, except for weighted data, with weights distributed normally with standard deviation as a fraction of the mean $\sigma_w/\mu_w=0.3$ (\textsection6); (7)~same as column~5, except for weights distributed uniformly from zero, corresponding to $\sigma_w/\mu_w\approx0.58$ (\textsection6); and (8)~same as column~5, except for weights distributed inversely over one dex, corresponding to $\sigma_w/\mu_w\approx0.73$ (\textsection6).  See Figure~5 for $\sigma_2$ vs.\@ $f_2$ axis labels.  The colors are scaled logarithmically, between 0.02 and 100.}
		\end{sidewaysfigure*}
		
		\begin{sidewaysfigure*}
			\centering
			\vspace{3.75in}
			\includegraphics[width=0.9\textwidth]{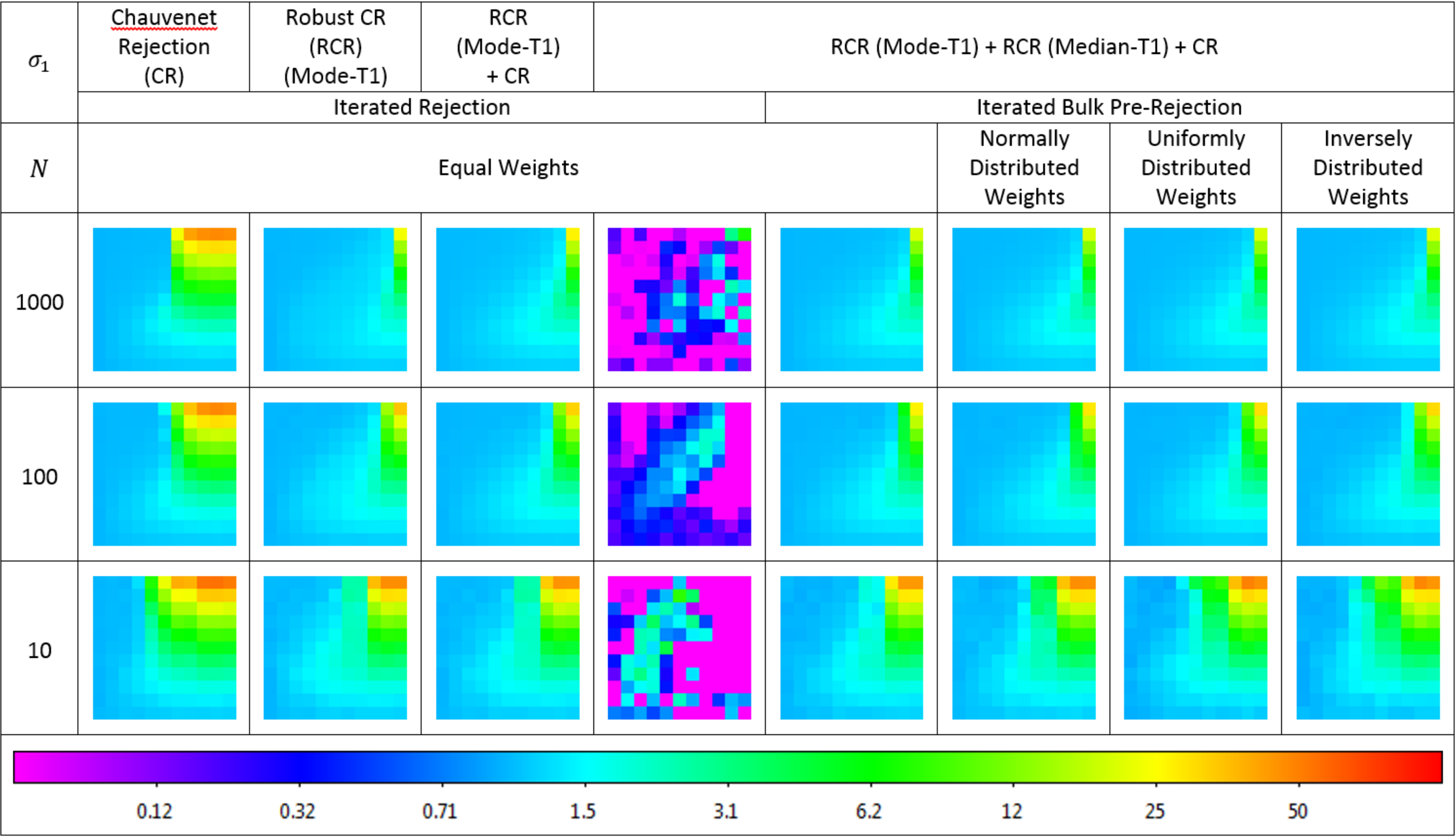}
			\caption{Average recovered $\sigma_1$ given one-sided contaminants, for, from left to right:  (1)~{\color{black}traditional} Chauvenet rejection; (2)~{\color{black}our} best-option robust {\color{black}technique} (\textsection3.1, \textsection4); (3)~column~2 followed by column~1 (\textsection4); (4)~column~2 followed by our most{\color{black}-}precise robust {\color{black}technique} followed by column~1, plotted as improvement over column~3, multiplied by 100 (\textsection4); (5)~{\color{black}our} best-option {\color{black}bulk pre-rejection technique} followed by column~4 (\textsection5); (6)~same as column~5, except for weighted data, with weights distributed normally with standard deviation as a fraction of the mean $\sigma_w/\mu_w=0.3$ (\textsection6); (7)~same as column~5, except for weights distributed uniformly from zero, corresponding to $\sigma_w/\mu_w\approx0.58$ (\textsection6); and (8)~same as column~5, except for weights distributed inversely over one dex, corresponding to $\sigma_w/\mu_w\approx0.73$ (\textsection6).  See Figure~5 for $\sigma_2$ vs.\@ $f_2$ axis labels.  The colors are scaled logarithmically, between 0.02 and 100.}
		\end{sidewaysfigure*}
		
		\begin{sidewaysfigure*}
			\centering
			\vspace{3.75 in}
			\includegraphics[width=0.9\textwidth]{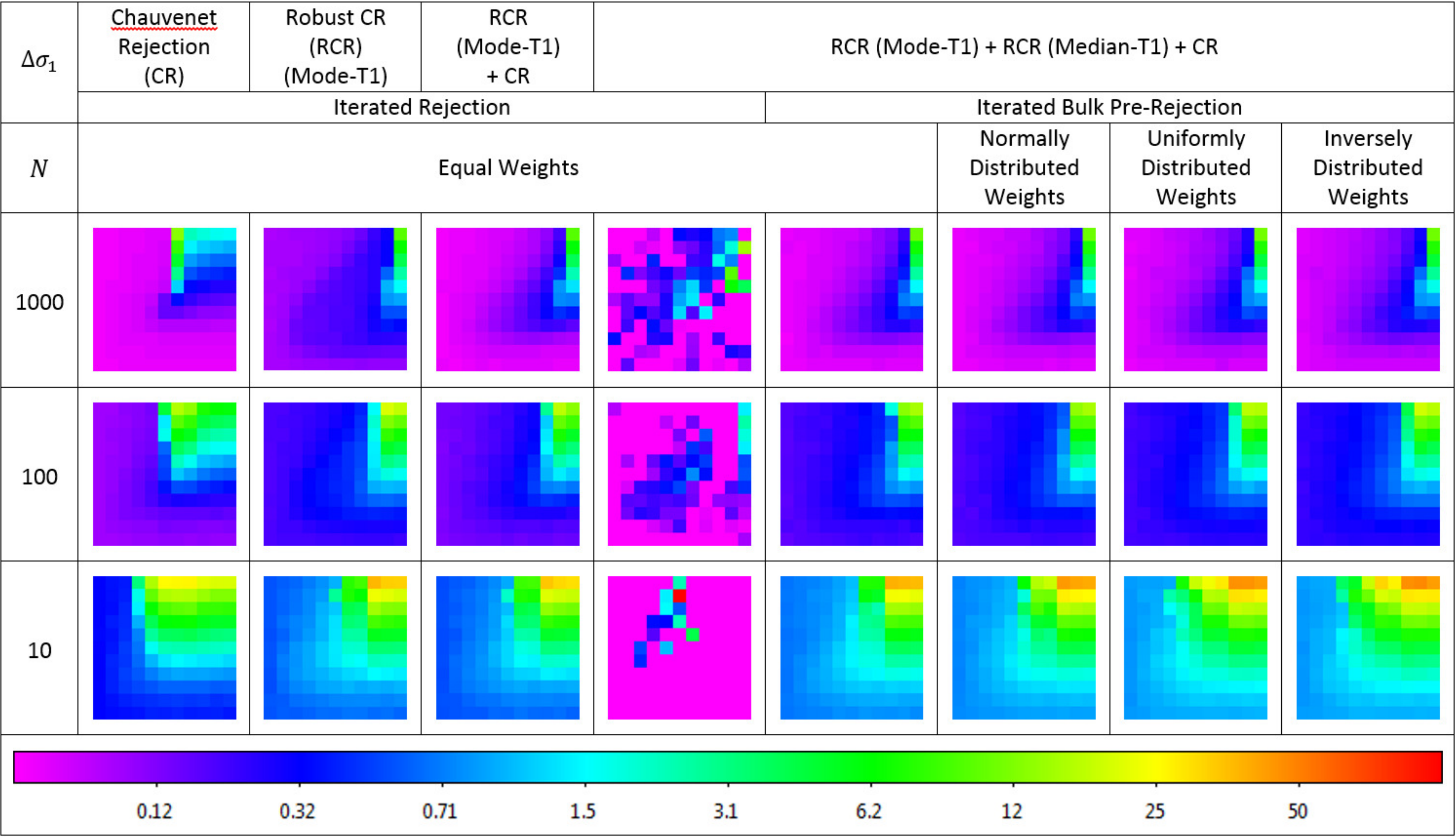}
			\caption{Uncertainty in recovered $\sigma_1$ given one-sided contaminants, for, from left to right:  (1)~{\color{black}traditional} Chauvenet rejection; (2)~{\color{black}our} best-option robust {\color{black}technique} (\textsection3.1, \textsection4); (3)~column~2 followed by column~1 (\textsection4); (4)~column~2 followed by our most{\color{black}-}precise robust {\color{black}technique} followed by column~1, plotted as improvement over column~3, multiplied by 100 (\textsection4); (5)~{\color{black}our} best-option {\color{black}bulk pre-rejection technique} followed by column~4 (\textsection5); (6)~same as column~5, except for weighted data, with weights distributed normally with standard deviation as a fraction of the mean $\sigma_w/\mu_w=0.3$ (\textsection6); (7)~same as column~5, except for weights distributed uniformly from zero, corresponding to $\sigma_w/\mu_w\approx0.58$ (\textsection6); and (8)~same as column~5, except for weights distributed inversely over one dex, corresponding to $\sigma_w/\mu_w\approx0.73$ (\textsection6).  See Figure~5 for $\sigma_2$ vs.\@ $f_2$ axis labels.  The colors are scaled logarithmically, between 0.02 and 100.}
		\end{sidewaysfigure*}
		
		\begin{figure}
			\plotone{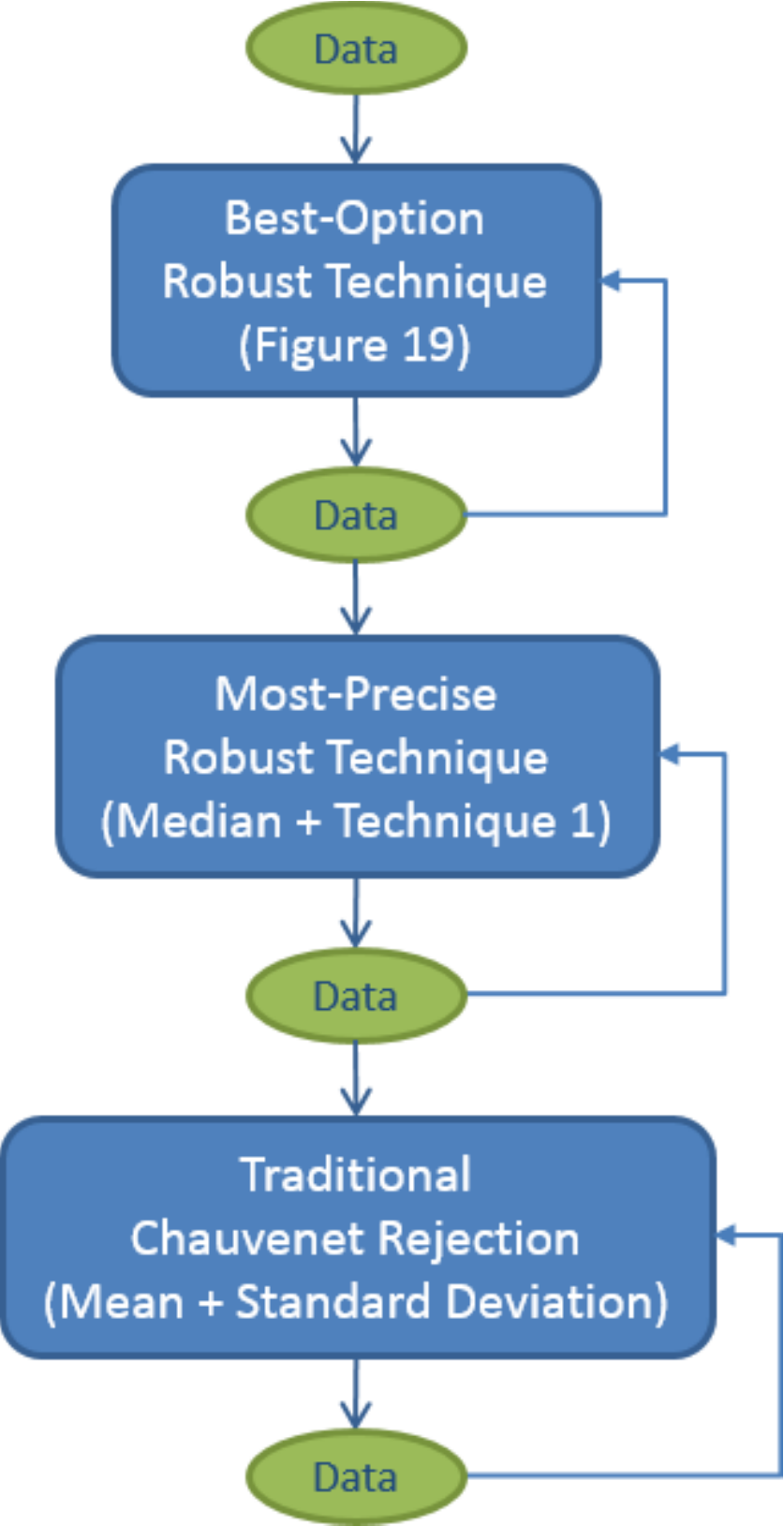}
			\caption{{\color{black}Flowchart of our algorithm, without bulk pre-rejection (see Figure~30).  The most discrepant outlier is rejected each iteration, and one iterates until no outliers remain before moving on to the next step.  $\mu$ and $\sigma$ (or $\sigma_{-}$ and $\sigma_{+}$, depending on {\color{black}whether the uncontaminated distribution is symmetric or asymmetric, and on the} contaminant type; Figure~19) are recalculated after each iteration, and the latter is multiplied by the appropriate correction factor (see Figure~29) before being used to reject the next outlier.  $\mu$ and $\sigma$ (or $\sigma_{-}$ and $\sigma_{+}$) may be calculated in different ways in different steps, but how they are used to reject outliers depends on {\color{black}whether the uncontaminated distribution is symmetric or asymmetric, and on the} contaminant type, and consequently does not change from step to step (Figure~19).}}
		\end{figure}
		
		\section{Robust Chauvenet Rejection:  Accuracy \textit{and} Precision}}
	
	In general, we have found that the mode is just as accurate (in the case of two-sided contaminants) or more accurate (in the case of one-sided contaminants) than the median, yet the mode is up to $\approx$5.8 times less precise than the median, and up to $\approx$7.7 times less precise than the mean.  We have also found that when $\mu$ (equal to the median or the mode) is measured accurately, our increasingly robust 68.3-percentile deviation measurement techniques are either equally accurate (in the case of one-sided contaminants) or increasingly accurate (in the case of two-sided contaminants), yet technique~3 (the broken-line fit) is up to $\approx$2.2 times less precise than technique~2 (the linear fit), up to $\approx$2.4 times less precise than technique~1 (the 68.3\% value), and up to $\approx$3.6 times less precise than the standard deviation.
	
	Consequently, there appears to be a {\color{black}tradeoff} between accuracy and precision.  But can we {\color{black}have both}?  In this section, we {\color{black}demonstrate that we can, by applying (1)~our robust improvements to traditional Chauvenet rejection (\textsection3), and (2)~traditional Chauvenet rejection (\textsection1) in sequence.  Traditional} Chauvenet rejection uses the mean and {\color{black}the} standard deviation, {\color{black}and is consequently the least robust of these techniques, but it} is also the most precise, {\color{black}at least} when not significantly contaminated by outliers.  By applying {\color{black}our robust techniques first, we eliminate the outliers that most significantly affect traditional} Chauvenet rejection, allowing us to {\color{black}then} capitalize on its precision without its inaccuracy.
	
	We demonstrate the success of this approach {\color{black}first using only our best-option robust techniques, for each of the following contaminant types:}\\
	
	\begin{itemize}
		\item The median $+$ technique~3 (the broken-line fit) is {\color{black}our} best option {\color{black}\textit{for two-sided contaminants}, which are contaminants that are both positive and negative, in equal proportion and strength (}\textsection3.1).  We plot the average recovered $\mu_1$, the uncertainty in the recovered $\mu_1$, the average recovered $\sigma_1$, and the uncertainty in the recovered $\sigma_1$ {\color{black}for this technique followed by traditional Chauvenet rejection} in the third column of Figures 20 -- 23, respectively.
		\item The mode $+$ technique~1 (the 68.3\% value) is {\color{black}our} best option {\color{black}\textit{for one-sided contaminants}, which are contaminants that are all positive (the case presented here) or all negative (}\textsection3.2).  We plot the average recovered $\mu_1$, the uncertainty in the recovered $\mu_1$, the average recovered $\sigma_1$, and the uncertainty in the recovered $\sigma_1$ {\color{black}for this technique followed by traditional Chauvenet rejection} in the third column of Figures 24 -- 27, respectively.
		\item The mode $+$ technique~3 (the broken-line fit) is {\color{black}our} best option {\color{black}(1)~\textit{for in-between cases}, in which contaminants are both positive and negative, but not in equal proportion or strength (\textsection3.2), and/or (2)~\textit{if the uncontaminated distribution is taken to be asymmetric} (\textsection3.3.1).  The former case behaves very similarly to Figures 20 -- 23 in the limit of two-sided contaminants, and very similarly to Figures 24 -- 27 in the limit of (positive) one-sided contaminants.}  The latter case behaves very similarly to {\color{black}Figure~20 ($\mu_1$), Figure~21 ($\Delta\mu_1$), Figure~22 ($\sigma_{1-}$ and $\sigma_{1+}$), and Figure~23 ($\Delta\sigma_{1-}$ and $\Delta\sigma_{1+}$)} in the limit of two-sided contaminants, and similarly to Figure~24 ($\mu_1$), Figure~25 ($\Delta\mu_1$), Figure~26 ($\sigma_{1-}$), Figure~27 ($\Delta\sigma_{1-}$), Figure~22 ($\sigma_{1+}$), and Figure~23 ($\Delta\sigma_{1+}$), in the limit of (positive) one-sided contaminants{\color{black}.  (Consequently, we will not plot these cases separately.)}
	\end{itemize}
	
	In all cases, {\color{black}our best-option robust techniques} followed by {\color{black}traditional} Chauvenet rejection results in vastly improved precisions -- comparable to those of {\color{black}traditional} Chauvenet rejection when not significantly contaminated by outliers -- with only small compromises in accuracy.  The small compromises in accuracy, when they occur, are due to {\color{black}our best-option} robust techniques not eliminating enough outliers before {\color{black}traditional} Chauvenet rejection is applied.
	
	We further improve {\color{black}this approach} by sequencing (1)~our best-option robust technique from above, (2)~our most{\color{black}-}precise robust technique -- the median $+$ technique~1 (the 68.3\% value) -- to eliminate more outliers before {\color{black}applying} (3)~{\color{black}traditional} Chauvenet rejection {\color{black}(see Figure~28 for a flowchart)}.  In nearly all cases, this either leaves the accuracies and the precisions the same, or improves them, by as much as {\color{black}$\approx$}30\%.  These are worthwhile gains, particularly given the computational efficiency of the additional step, but they are also difficult to see given the logarithmic scaling that we use in Figures 20 -- 27.  Consequently, we instead plot the improvement over column~3, multiplied by 100, in column~4.
	
	Both of these sequencing techniques, as well as a bulk-rejection variant of the latter technique that we present in \textsection5, require the calculation of new correction factors, which we do as in \textsection2.3 and plot in Figure~29.
	
	\begin{figure*}
		\plotone{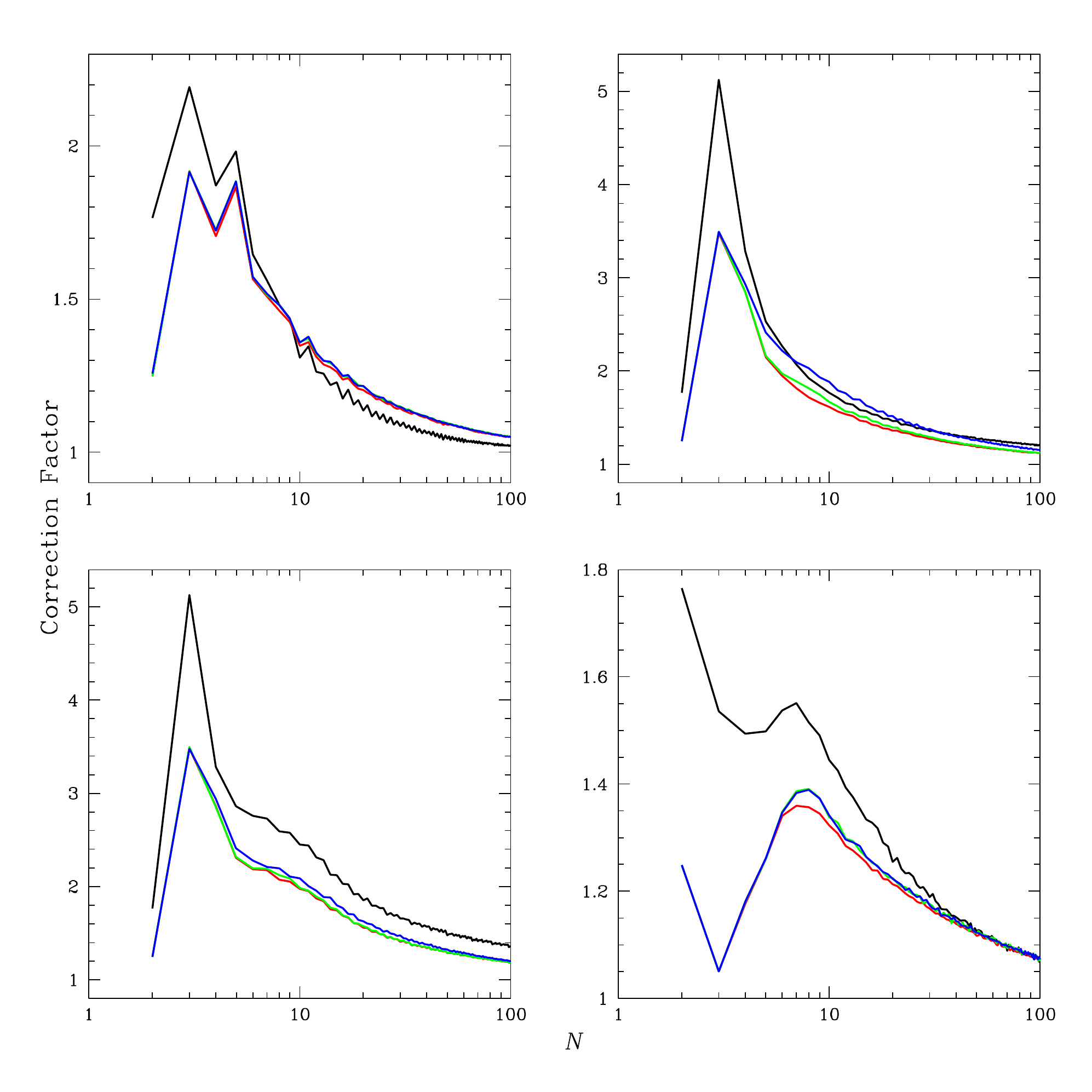}
		\caption{Correction factors by which standard and 68.3-percentile deviations, measured from uncontaminated data, need to be multiplied to yield the correct result, on average, and to avoid overaggressive rejection {\color{black}(although this can still happen in sufficiently small samples; see §3.3.1)}, (1)~for the case of our best-option robust techniques (see below; black curves, from Figure~4); (2)~for the case of (1)~followed by {\color{black}traditional} Chauvenet rejection (red curves); (3)~for the case of (1)~followed by our most{\color{black}-}precise robust technique -- the median $+$ technique~1 (the 68.3\% value) -- followed by {\color{black}traditional} Chauvenet rejection (green curves); and (4)~for the case of bulk rejection (see \textsection5) followed by (3)~(blue curves).  \textbf{Upper left:}  For our best-option robust technique for two-sided contaminants -- the median $+$ technique~3 (the broken-line fit) -- in which we compute a single $\sigma$ using the deviations both below and above $\mu$ (\textsection3.1).  \textbf{Upper right:}  For our best-option robust technique for one-sided contaminants -- the mode $+$ technique~1 (the 68.3\% value) -- in which we compute separate $\sigma$ below and above $\mu$ ($\sigma_-$ and $\sigma_+$, respectively) and use the smaller of the two when rejecting outliers (\textsection3.2).  \textbf{Lower left:}  For our best-option robust technique for in-between cases -- the mode $+$ technique~3 (the broken-line fit) -- in which we also use the smaller of $\sigma_-$ and $\sigma_+$ when rejecting outliers (\textsection3.2).  \textbf{Lower right:}  For our best-option robust technique if the uncontaminated distribution is taken to be asymmetric -- the mode $+$ technique~3 (the broken-line fit) -- in which we use $\sigma_-$ to reject outliers below $\mu$ and $\sigma_+$ to reject outliers above $\mu$ (\textsection3.3.1).  We use look-up tables for $N\leq100$ and power-law approximations for $N>100$ (see Appendix B).}
	\end{figure*}
	
	\section{Bulk Rejection}
	
	So far, we have rejected only one outlier -- the most discrepant outlier -- at a time, recomputing {\color{black}$\mu$} and {\color{black}$\sigma$ (or $\sigma_{-}$ and $\sigma_{+}$, depending on {\color{black}whether the uncontaminated distribution is symmetric or asymmetric, and on the} contaminant type; Figure~19)} after each rejection.  This can be time-consuming, computationally, particularly with large samples, so now we evaluate the effectiveness of bulk rejection.  In this case, we reject all measurements that meet Chauvenet's criterion each iteration {\color{black}(however, see Footnote~3)}, recomputing {\color{black}$\mu$} and {\color{black}$\sigma$} once per iteration instead of once per rejection.  

However, bulk rejection works only if $\sigma_1$ is never significantly underestimated.  If this happens, even if only for a single iteration, significant over-rejection can occur.  Furthermore, each of the techniques that we have presented can fail in this way, under the right (or wrong) conditions:

\begin{itemize}
	\item{With one-sided contaminants, when $\mu_1$ {\color{black}\textit{cannot}} be measured accurately (Figure~13), the standard deviation underestimates the 68.3-percentile deviation as measured by technique~1 (the 68.3\% value), which underestimates the 68.3-percentile deviation as measured by technique~2 (the linear fit), which underestimates the 68.3-percentile deviation as measured by technique~3 (the broken-line fit; Figure~17).  In this case, the latter technique overestimates $\sigma_1$.  However, the former three techniques can either overestimate $\sigma_1$ or underestimate it, sometimes significantly.}
	\item{With one-sided or two-sided contaminants, when $\mu_1$ {\color{black}\textit{can}} be measured accurately, technique~3 (the broken-line fit) is as accurate (\textsection3.2) or more accurate (\textsection3.1) than the other techniques, but it is also the least precise (\textsection4), meaning that it is as likely to underestimate $\sigma_1$ as overestimate it, and, again, sometimes significantly.}  
\end{itemize}

Note also that one can transition between these two cases:  $\mu_1$ often begins inaccurately measured but ends accurately measured, after iterations of rejections (Figures 11 and 12).

A solution that works in all cases is to measure $\sigma_1$ using both techniques 2 (the linear fit) and 3 (the broken-line fit), and adopt the larger of the two for bulk rejection.  When $\mu_1$ cannot be measured accurately, the deviation curve breaks downward, and the broken-line fit is the most conservative option (Figure~17).  When $\mu_1$ can be measured accurately, the deviation curve breaks upward, and the linear fit is a sufficiently conservative option (Figures 2 and 3).  (Technique~1, the 68.3\% value, is in this case a more conservative option, but can be overly conservative, bulk{\color{black}-}rejecting too few points per iteration.)  

We use the same $\mu$-measurement technique as we use for individual rejection.  Finally, once bulk rejection is done, we follow up with individual rejection, as described in the second to last paragraph of \textsection4 {\color{black}(see Figure~30 for a flowchart)}.  Individual rejection (1)~is significantly faster now that most of the outliers have already been bulk pre-rejected, and (2)~ensures accuracy with precision (\textsection4).  We plot the results in column~5 of Figures 20 -- 27, and, desirably, they do not differ significantly from those of column~4.  Speed-up times are presented in Table~1.\\

\section{Weighted Data}

We now consider the case of weighted data.  In this case, the mean is given by:

\begin{equation}
\mu = \frac{\sum\limits_{i=1}^{N}{w_ix_i}}{\sum\limits_{i=1}^{N}{w_i}},
\end{equation}

\noindent where $x_i$ are the data and $w_i$ are the weights.  When the mean is measured from the sample, the standard deviation is given by:

\begin{equation}
\sigma = \sqrt{\frac{\sum\limits_{i=1}^{N}{w_i(x_i-\mu)^2}}{\sum\limits_{i=1}^{N}{w_i-\Delta\frac{\sum\limits_{i=1}^{N}{w_i^2}}{\sum\limits_{i=1}^{N}{w_i}}}}},
\end{equation}

\noindent where $\Delta=1$ when summing over data both below and above the mean, and we take $\Delta=0.5$ when summing over data either only below or only above the mean.

\begin{figure}
	\plotone{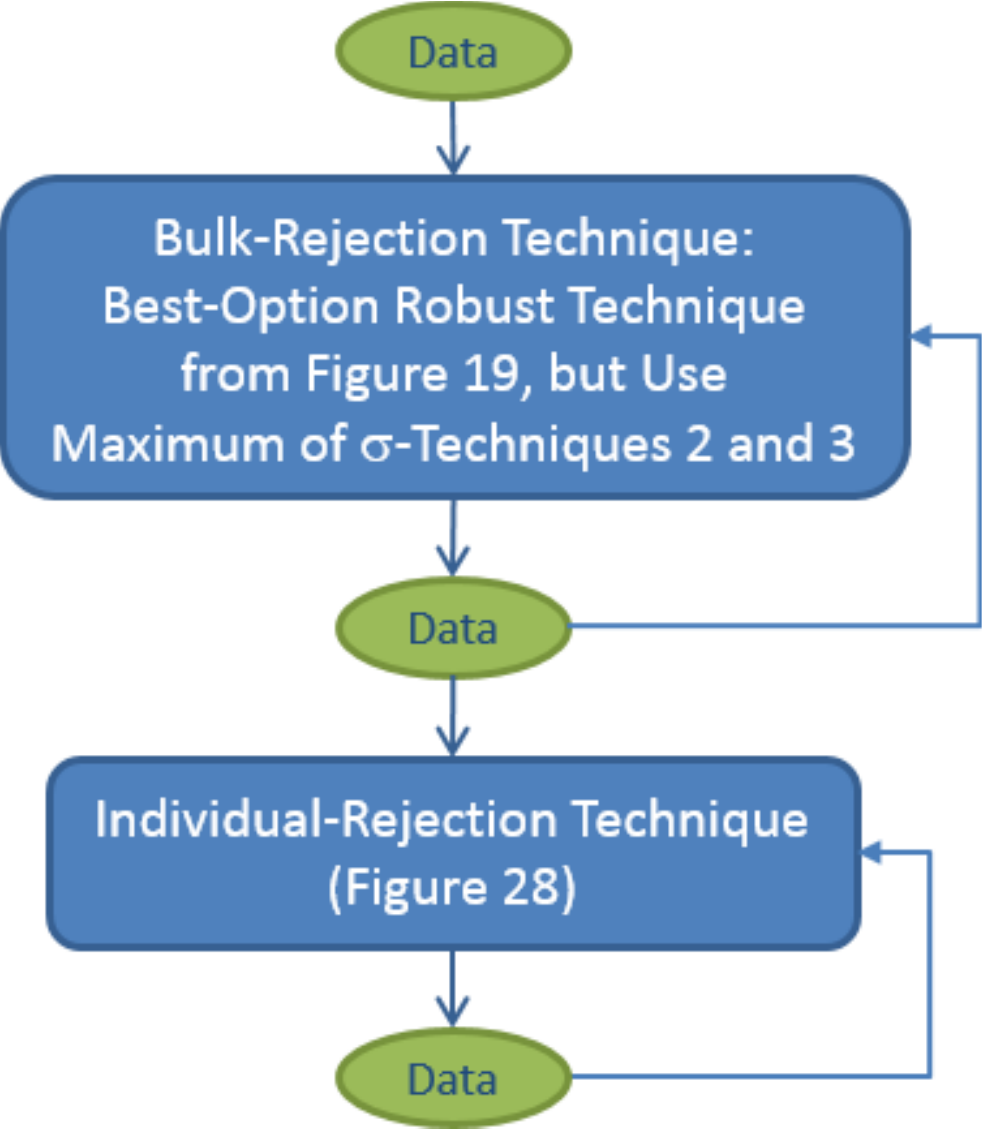}
	\caption{{\color{black}Flowchart of our algorithm, with bulk pre-rejection.  The first step is bulk rejection, in which all outliers are rejected each iteration, and one iterates until no more outliers are identified.  $\mu$ and $\sigma$ (or $\sigma_{-}$ and $\sigma_{+}$, depending on {\color{black}whether the uncontaminated distribution is symmetric or asymmetric, and on the} contaminant type; Figure~19) are recalculated after each iteration, and the latter is multiplied by the appropriate correction factor (Figure~29) before being used to reject more outliers.  The second step is our individual-rejection algorithm (Figure~28), which ensures accuracy with precision (\textsection4)}}
\end{figure}

To determine the weighted median, sort the data and the weights by $x_i$.  First, consider the following, crude definition:  Let $j$ be the smallest integer such that:

\begin{equation}
\sum\limits_{i=1}^{j}{w_i} \geq 0.5\sum\limits_{i=1}^{N}{w_i}.
\end{equation}

\noindent The weighted median could then be given by $\mu=x_j$, but this definition would be very sensitive to edge effects.  Instead, we define the weighted median as follows.  Let:

\begin{equation}
s_j=\sum\limits_{i=1}^{j}{(0.5w_{i-1}+0.5w_i)},
\end{equation}

\noindent where $w_0=0$, and let $j$ be the smallest integer such that:

\begin{equation}
s_j \geq 0.5\sum\limits_{i=1}^{N}{w_i}.
\end{equation}

\noindent The weighted median is then given by interpolation:

\begin{equation}
\mu = x_{j-1}+(x_j - x_{j-1})\frac{0.5\sum\limits_{i=1}^{N}{w_i} - s_{j-1}}{s_j - s_{j-1}},
\end{equation}

\noindent where $s_0=0$.  

To determine the weighted mode, we again follow {\color{black}an iterative half-sample} approach (\textsection2.1).  For every $j$ such that:

\begin{equation}
s_j \leq 0.5\sum\limits_{i=1}^{N}{w_i},
\end{equation}

\noindent let $k$ be the largest integer such that:

\begin{equation}
s_k \leq s_j + 0.5\sum\limits_{i=1}^{N}{w_i},
\end{equation}

\noindent and for every $k$ such that:

\begin{equation}
s_k \geq 0.5\sum\limits_{i=1}^{N}{w_i},
\end{equation}

\noindent let $j$ be the smallest integer such that:

\begin{equation}
s_j \geq s_k - 0.5\sum\limits_{i=1}^{N}{w_i}.
\end{equation}

\noindent Of these ($j$,$k$) combinations, select the one for which $|x_k-x_j|$ is smallest.  If multiple combinations meet this criterion, let $j$ be the smallest of their $j$ values and $k$ be the largest of their $k$ values.  Restricting oneself to only the $k-j+1$ values between and including $j$ and $k$, repeat this procedure, iterating to completion.  Take the weighted median of the final $k-j+1$ values.

\begin{deluxetable*}{ccccccccc}
	\tablewidth{0pt}
	\tablecaption{Time in Milliseconds to Measure $\mu_1$ and $\sigma_1$\tablenotemark{a}}
	\tablehead{
		\colhead{Contaminant Type:} & \multicolumn{2}{c}{2-Sided} & \multicolumn{2}{c}{1-Sided} & \multicolumn{4}{c}{In-Between} \\
		& & & & & \multicolumn{2}{c}{(2-Sided Limit)} & \multicolumn{2}{c}{(1-Sided Limit)}\B \\ 
		\colhead{Post-Bulk Rejection Technique:\tablenotemark{b}} & \multicolumn{2}{c}{RCR (Median-T3)} & \multicolumn{2}{c}{RCR (Mode-T1)} & \multicolumn{4}{c}{RCR (Mode-T3)}\T\B \\
		\colhead{Corresponding Figures:} & \multicolumn{2}{c}{19 -- 22} & \multicolumn{2}{c}{23 -- 26} & \multicolumn{2}{c}{---} & \multicolumn{2}{c}{---}\T\B \\
		\colhead{Bulk Pre-Rejection:} & \colhead{No} & \colhead{Yes} & \colhead{No} & \colhead{Yes} & \colhead{No} & \colhead{Yes} & \colhead{No} & \colhead{Yes}\T\B \\
		\colhead{Corresponding Column:} & \colhead{4} & \colhead{5} & \colhead{4} & \colhead{5}\T & --- & --- & --- & ---
	}
	\startdata
	$N = 1000$ & 73 & 29 & 160 & 5.6 & 160 & 59 & 190 & 8.7\B \\
	$N = 100$ & 0.86 & 0.50 & 1.6 & 0.40 & 1.9 & 0.98 & 2.0 & 2.2\T\B \\
	$N = 10$ & 0.027 & 0.030 & 0.042 & 0.037 & 0.047 & 0.042 & 0.048 & 0.078\T
	\enddata
	\tablenotetext{a}{Averaged over the $11 \times 11 \times 100 = 121,000$ samples in each $\sigma_2$ vs.\@ $f_2$ figure in columns 4 vs.\@ 5 of Figures {\color{black}20 -- 23 (2-sided case), Figures 24 -- 27 (1-sided case), and in corresponding (but similar looking, and hence unplotted; \textsection4) figures for the in-between case, in both the 2-sided and 1-sided limits}, using a single, AMD Opteron 6168 processor.  Measuring the mode is $\approx$1.6$N^{0.05}$ times slower than measuring the median, and technique~3 (the broken-line fit) is $\approx$1.2 times slower than technique~1 (the 68.3\% value), but bulk pre-rejection is $\approx${\color{black}$(N/7.8)^{0.21}$} (2-sided) to $\approx${\color{black}$(N/12)^{0.73}$} (1-sided) times faster than no bulk pre-rejection, where $N$ is the sample size.  Time to completion is proportional to $N^\alpha$, where $\alpha \approx2$ (no bulk pre-rejection) or $1<\alpha<2$ (bulk pre-rejection), plus an overhead constant, which dominates when $N\lesssim5 - 500$.  In the case of weighted data (see \textsection6), completion times are roughly $1+0.7N^{-0.4}$ times longer.}
	\tablenotetext{b}{+ RCR (Median-T1) + CR (\textsection5)}
\end{deluxetable*}

\begin{figure*}
	\plotone{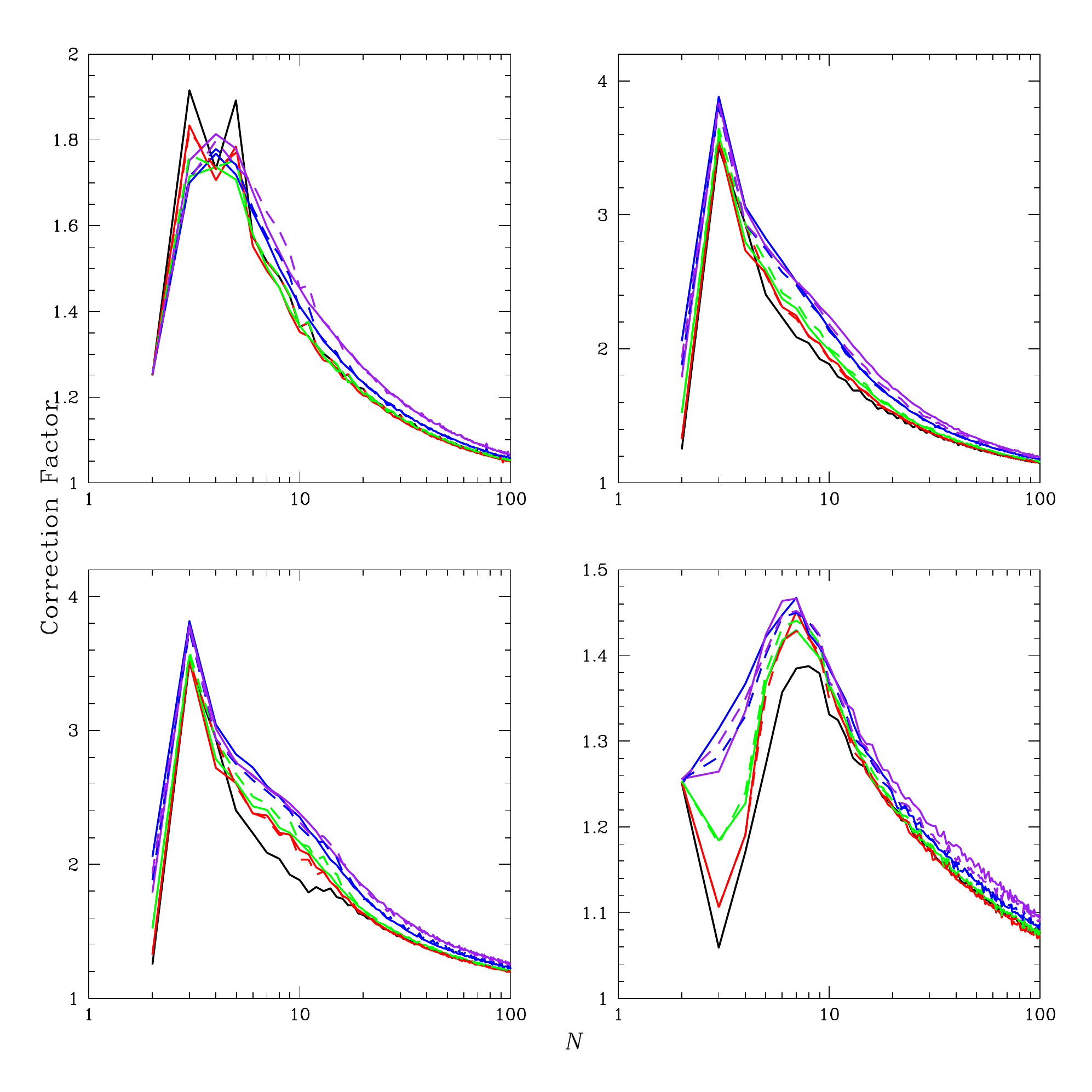}
	\caption{Same as the blue curves from Figure~29, but for five representative weight distributions:  (1)~all weights equal (solid black curves -- same as the blue curves from Figure~29); (2)~weights distributed normally with standard deviation as a fraction of the mean $\sigma_w/\mu_w=0.1$ (solid red curves); (3)~weights distributed normally with $\sigma_w/\mu_w=0.3$ (solid green curves); (4)~weights distributed uniformly from zero (i.e., low-weight points as common as high-weight points; solid blue curves), corresponding to $\sigma_w/\mu_w\approx0.58$; and (5)~weights distributed inversely over one dex (i.e., low-weight points more common than high-weight points, with the sum of the weights of the low-weight points as impactful as the sum of the weights of the high-weight points; solid purple curves), corresponding to $\sigma_w/\mu_w\approx0.73$.  From these, we have produced empirical approximations, as functions of (1)~$N$ and (2)~$\sigma_w/\mu_w$ of the $x_i=\sqrt{2}\mathrm{erf}^{-1}(s_i/\sum_{i=1}^{N}w_i)<1$ points, which can be used with any sample of similarly distributed weights (dashed curves; see Appendix B).}
\end{figure*}

To determine the weighted 68.3-percentile deviation, measured either from the weighted median or the weighted mode, sort the deviations $\delta_i=|x_i-\mu|$ and the weights by $\delta_i$.  Analogously to the weighted median above, first consider the following, crude definition:  Let $j$ be the smallest integer such that:

\begin{equation}
\sum\limits_{i=1}^{j}{w_i} \geq 0.683\sum\limits_{i=1}^{N}{w_i}.
\end{equation}

\noindent The weighted 68.3-percentile deviation could then be given by $\sigma=\delta_j$, but, again, this definition would be very sensitive to edge effects.  Instead, we define the weighted 68.3-percentile deviation, for technique~1 (the 68.3\% value), as follows.  Let:\footnote{We center these not halfway through each bin, as we do for the weighted median and weighted mode, but 68.3\% of the way through each bin.  The need for this can be seen in the case of $\mu$ being known a priori, in the limit of one measurement having significantly more weight than the rest, or in the limit of $N\rightarrow1$.} 

\begin{equation}
s_j=\sum\limits_{i=1}^{j}{(0.317w_{i-1}+0.683w_i)},
\end{equation}

\noindent where $w_0=0$, and let $j$ be the smallest integer such that:

\begin{equation}
s_j \geq 0.683\sum\limits_{i=1}^{N}{w_i}.
\end{equation}

\noindent The weighted 68.3-percentile deviation, for technique~1, is then given by interpolation:

\begin{equation}
\sigma = \delta_{j-1}+(\delta_j - \delta_{j-1})\frac{0.683\sum\limits_{i=1}^{N}{w_i} - s_{j-1}}{s_j - s_{j-1}},
\end{equation}

\noindent where $s_0=0$.  For techniques 2 (the linear fit) and 3 (the broken-line fit), the 68.3-percentile deviation is given by plotting $\delta_i$ vs.\@ $\sqrt{2}\mathrm{erf}^{-1}(s_i/\sum_{i=1}^{N}{w_i})$ and fitting as before (\textsection2.2), except to weighted data (e.g., Appendix A).

Note that as defined here, all of these measurement techniques reduce to their unweighted counterparts (\textsection2.1 and \textsection2.2) when all of the weights, $w_i$, are equal.

Note also that the correction factors (\textsection2.3) that one uses depend on the weights of the data.  To this end, for each of the four scenarios that we consider in \textsection4, corresponding to the four panels of Figure~29, we have computed correction factors for the case of bulk rejection (\textsection5) followed by individual rejection as described in the second to last paragraph of \textsection4, for five representative weight distributions:  (1)~all weights equal (see Figure~31, solid black curves -- same as Figure~29, blue curves); (2)~weights distributed normally with standard deviation as a fraction of the mean $\sigma_w/\mu_w=0.1$ (Figure~31, solid red curves); (3)~weights distributed normally with $\sigma_w/\mu_w=0.3$ (Figure~31, solid green curves); (4)~weights distributed uniformly from zero (i.e., low-weight points as common as high-weight points; Figure~31, solid blue curves), corresponding to $\sigma_w/\mu_w\approx0.58$; and (5)~weights distributed inversely over one dex (i.e., low-weight points more common than high-weight points, with the sum of the weights of the low-weight points as impactful as the sum of the weights of the high-weight points; Figure~31, solid purple curves), corresponding to $\sigma_w/\mu_w\approx0.73$.  

{\color{black}The differences between these are small, but monotonically increasing with $\sigma_w/\mu_w$, at each $N$.   Furthermore, we have tried other-shaped weight distributions, but with similar $\sigma_w/\mu_w$, to similar results:  The small differences that we do see appear to be more about the effective width of these distributions -- which can be easily measured from any sample of weighted measurements -- than about the specific shape of these distributions.
	
	Consequently, using these five representative weight distributions}, we have produced empirical approximations, as functions of (1)~$N$ and (2)~$\sigma_w/\mu_w$ of the $x_i=\sqrt{2}\mathrm{erf}^{-1}(s_i/\sum_{i=1}^{N}w_i)<1$ points, which can be used with any sample of similarly distributed weights (Figure~31, dashed curves; see Appendix B).  We demonstrate these for the latter three weight distributions listed above in columns 6, 7, and 8, respectively, of Figures 20 -- 27, and, desirably, they do not differ significantly from those of column~5, in which $\sigma_w/\mu_w=0$, although there is some decrease in effectiveness in the low-$N$, high-$\sigma_w/\mu_w$ limit.

{\color{black}It is this combination of (1)~sequencing robust improvements to traditional Chauvenet rejection with traditional Chauvenet rejection, to achieve both accuracy and precision (\textsection4), (2)~bulk pre-rejection, to significantly decrease computing times in large samples (\textsection5), and (3)~the ability to handle weighted data (\textsection6) that we typically refer to as robust Chauvenet rejection (RCR).}

\section{Example:  Aperture Photometry}

The Skynet Robotic Telescope Network is a global network of fully automated, or robotic, volunteer telescopes, scheduled through a common web interface.\footnote{https://skynet.unc.edu}   Currently, our optical telescopes range in size from 14 to 40 inches, and span four continents.  Recently, {\color{black}we added} Skynet's first radio telescope, Green Bank Observatory's 20-meter {\color{black}diameter dish, in West Virginia (Martin et al.\@ 2018).    
	
	We have been} incorporating {\color{black}RCR} into Skynet's image-processing library, beginning with our single-dish mapping algorithm (Martin et al.\@ 2018).  Here, we use {\color{black}RCR} extensively:  (1)~to eliminate contaminants during gain calibration; (2)~to measure the noise level of the data along {\color{black}each scan}, and as a function of time, to aid in background subtraction along the scans; (3)~to combine {\color{black}locally fitted}, background-level models into global models, for background subtraction along {\color{black}each scan}; (4)~to eliminate contaminants if signal and telescope-position clocks must be synchronized post facto from the background-subtracted data; (5)~to measure the noise level of the background-subtracted data across {\color{black}each scan}, and as a function of time, to aid in radio-frequency interference (RFI) cleaning; and (6)~to combine {\color{black}locally fitted} models of the background-subtracted, RFI-cleaned signal into a global model, describing the entire observation.  After this, we locally model and fit a ``surface'' to the background-subtracted, time-delay corrected, RFI-cleaned data, filling in the gaps between the signal measurements to produce the final image (e.g., {\color{black}see} Figure~32).  Furthermore, each pixel in the final image is weighted, equal to the proximity-weighted number of data points that contributed to its determination (e.g., Figure~32{\color{black}, lower right}).    

\begin{figure*}
	\plotone{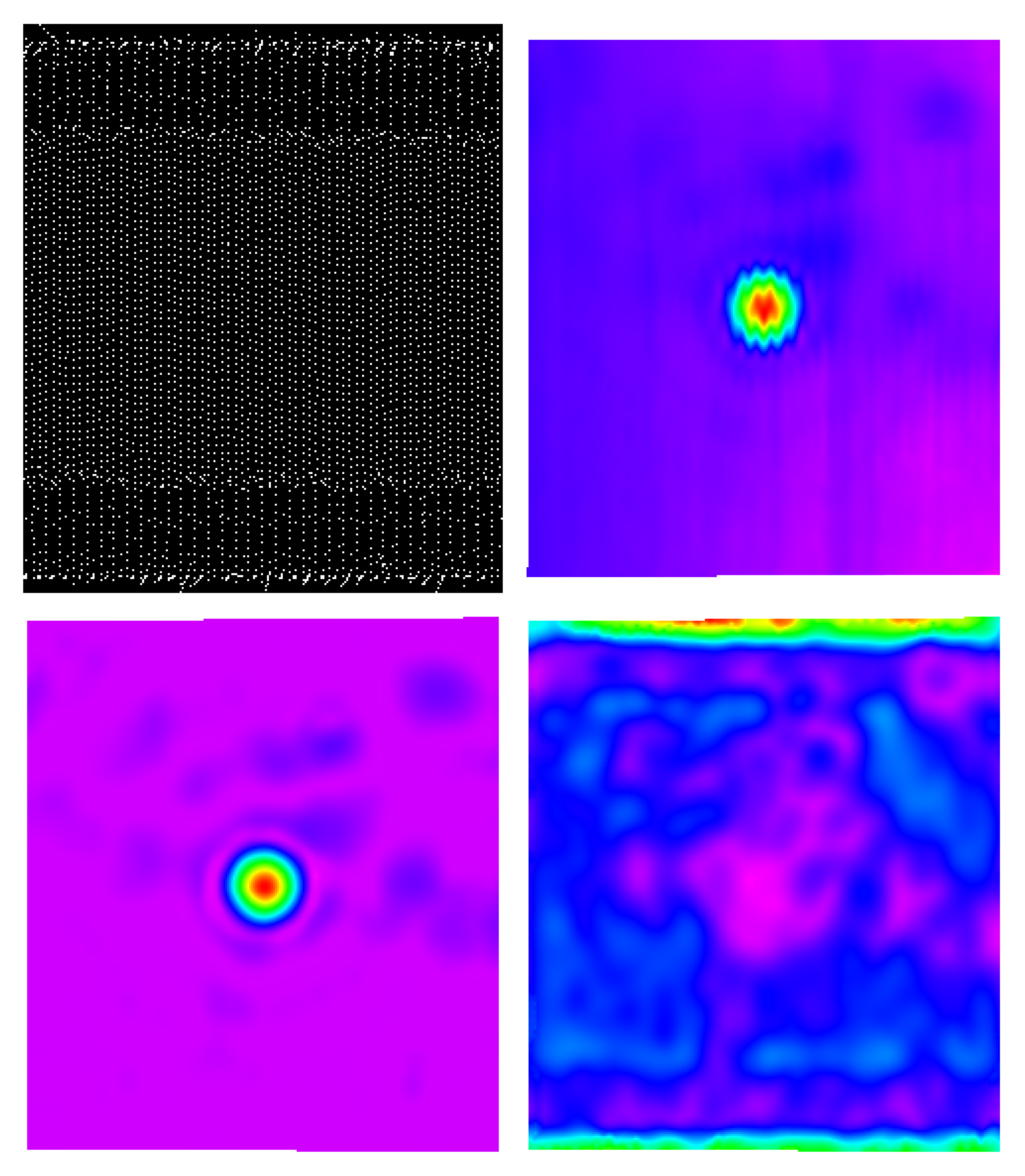}
	\caption{\textbf{Upper left:}  Signal{\color{black}-}measurement positions from an on-the-fly raster mapping of Cas A, made with Green Bank Observatory's 20-meter diameter telescope, in L band {\color{black}(gaps at the top and bottom are due to the telescope jumping ahead to get back on schedule, after losing time reversing direction at the ends of scans)}.  \textbf{{\color{black}Upper right}:}  Raw image, which has been surface modeled (to fill in the gaps between the signal measurements, without additionally blurring the image), but has not been background subtracted, time-delay corrected, or RFI cleaned.  \textbf{Lower {\color{black}left}:}  Final image, which has been background subtracted, time-delay corrected, RFI cleaned, and then surface {\color{black}modeled.  \textbf{Lower right:}  Proximity-weighted number of data points that contributed to the surface model at each pixel.  Weights are lower in the vicinity of signal, due to the RFI-cleaning algorithm (Martin et al.\@ 2018).  In the latter three panels, square-root scaling is used to enhance the visibility of fainter structures.}}
\end{figure*}

Here, we demonstrate another application of {\color{black}RCR}:  aperture photometry, in this case of the primary source, Cas A, in the lower-{\color{black}left} panel of Figure~32.  We have centered the aperture on the source, and have selected its radius to match that of the minimum between the source and its first Airy ring (see Figure~33).  We sum all of the values in the aperture, but from each we must also subtract off the average background-level value, which we measure from the surrounding annulus.  

\begin{figure*}
	\plottwo{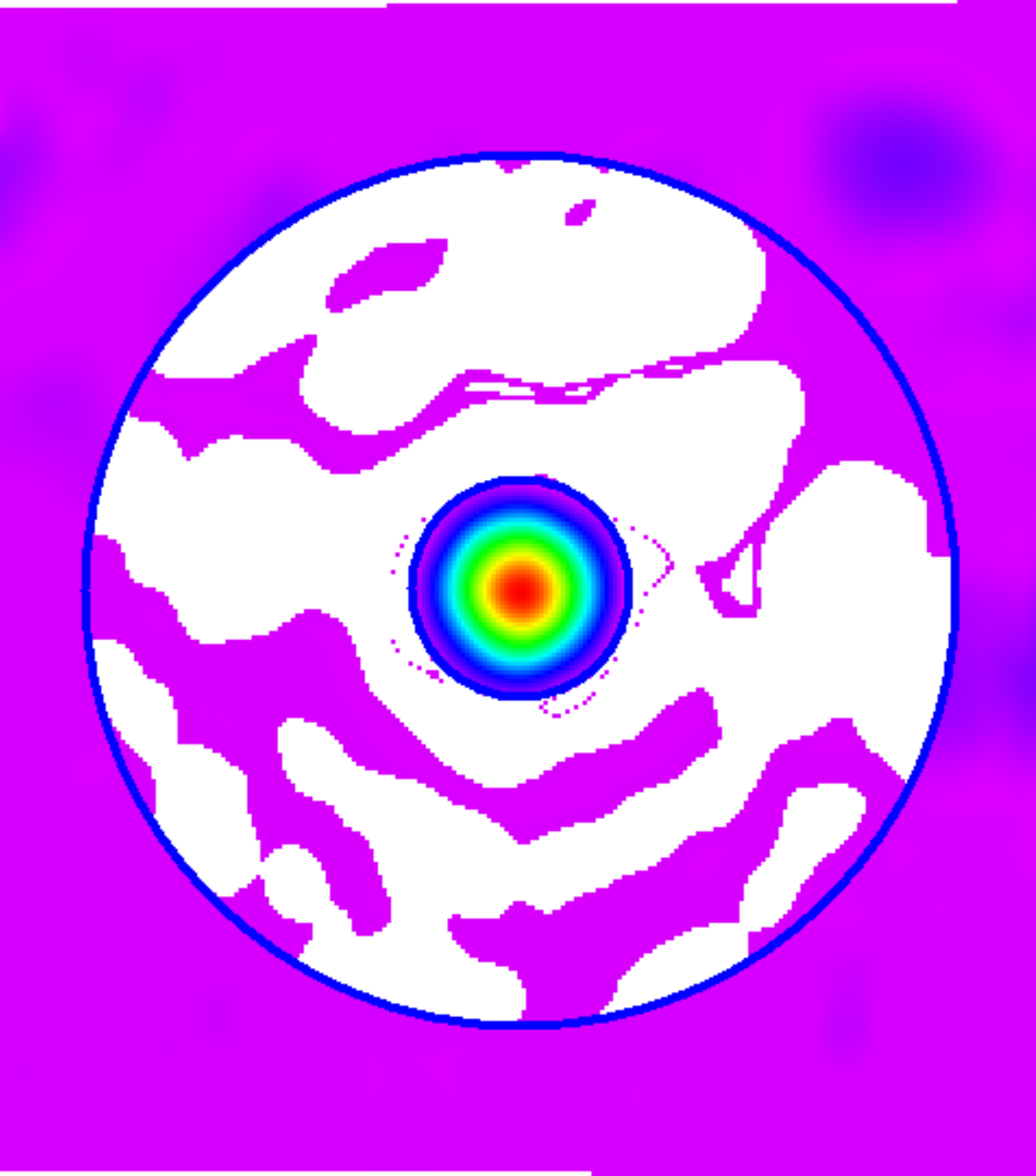}{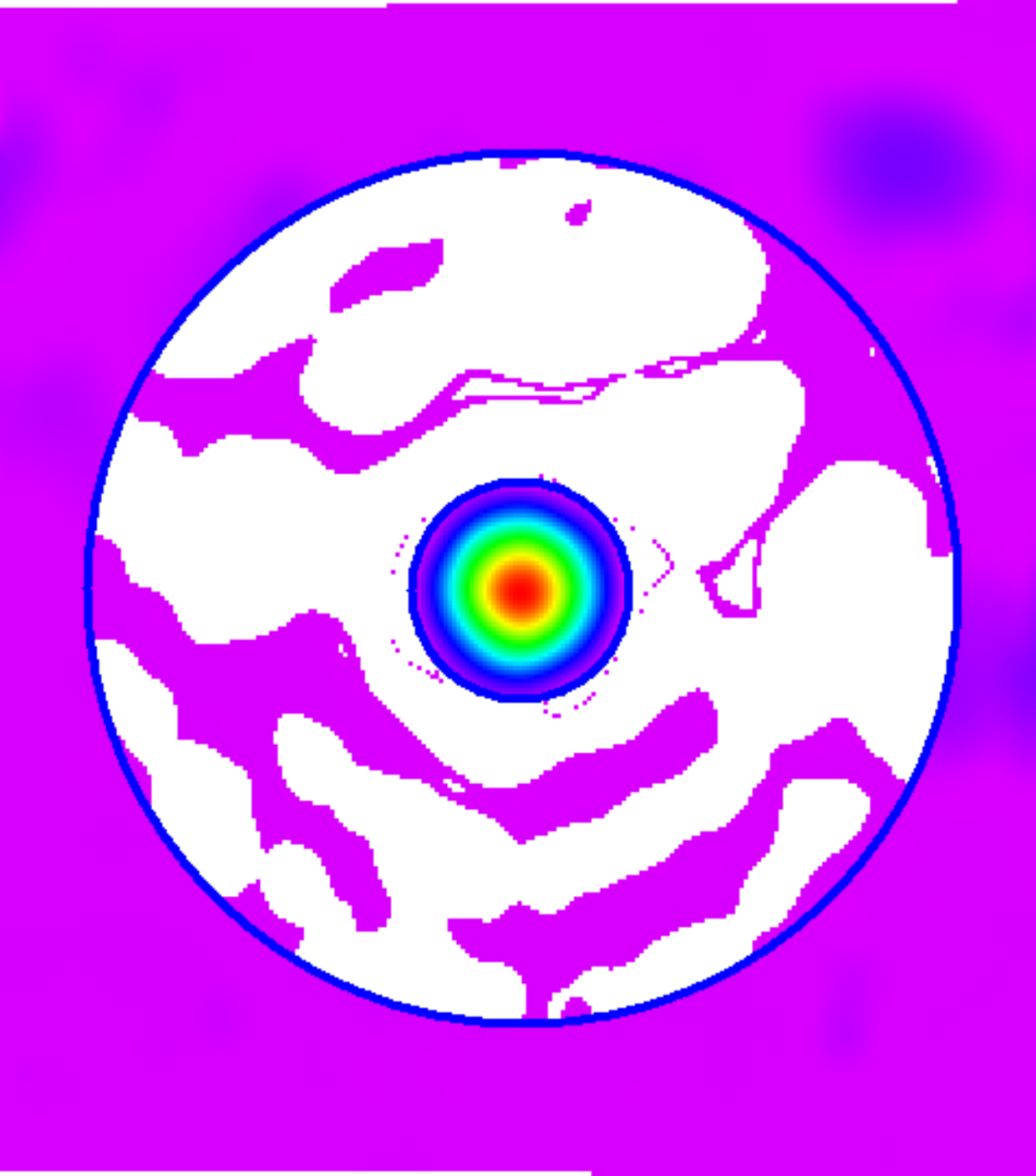}
	\caption{Same as the lower-{\color{black}left} panel of Figure~32, except that contaminated pixels (contaminated by other sources, Airy rings, {\color{black}diffraction spikes,} etc.) have been robust{\color{black}-}Chauvenet rejected within an annulus in which we are measuring the background level, (1)~assuming that the contaminants are one sided (left), and (2)~assuming that the contaminants are an in-between case, with some {\color{black}negative} contaminants as well (right).}
\end{figure*}

The annulus we have selected to extend from the radius of the aperture to {\color{black}10 beamwidths} (Figure~33).  However, it is heavily contaminated, by the source's Airy rings {\color{black}and diffraction spikes,} and by other sources.  This is a good case to demonstrate {\color{black}RCR}, because (1)~a large fraction, $f_2$, of the pixels in the annulus are contaminated, and (2)~they are strongly contaminated, $\sigma_2$, compared to the background{\color{black}-}noise level, $\sigma_1$.  It is also a good case to demonstrate bulk pre-rejection (\textsection5), because there {\color{black}is} a large number of pixels in the annulus, and to demonstrate {\color{black}RCR's ability to handle} weighted data (\textsection6, Figure~32{\color{black}, lower right}). 

These are one-sided contaminants, so we follow bulk pre-rejection with {\color{black}``}RCR (Mode -- Technique~1) + RCR (Median -- Technique~1) + CR{\color{black}''} (\textsection4, Figures 24 -- 27).  The rejected pixels have been excised from the left panel of Figure~33.  

If one suspected an in-between case, with some {\color{black}negative} contaminants as well, we would instead follow bulk pre-rejection with {\color{black}``}RCR (Mode -- Technique~3) + RCR (Median -- Technique~1) + CR{\color{black}'' (\textsection4)}.  The rejected pixels for this case have been excised from the right panel of Figure~33.  

For these two cases, the post-rejection background level is measured to be $-0.00002 \pm 0.00047$ and $-0.00003 \pm 0.00045$, respectively, which is a significant improvement over the pre-rejection value, $0.023 \pm 0.040$ (gain{\color{black}-}calibration units).  

It is also a significant improvement over what {\color{black}traditional} Chauvenet rejection yields:  $0.022 \pm 0.036$, which is nearly identical to the pre-rejection value.  I.e., {\color{black}traditional} Chauvenet rejection fails to eliminate most of the outliers, resulting in biased, and additionally uncertain, photometry.  In this case, {\color{black}traditional} Chauvenet rejection is equivalent to sigma clipping with a 4.35$\sigma$ threshold, given the number of pixels in the annulus (Equation 1).  This demonstrates that something as fundamental to astronomy as aperture photometry can be improved upon, in the limit of contaminated, or crowded, fields.

Lastly, we point out that RCR has already been successfully employed by Trotter et al.\@ (2017), who made many measurements of Cas~A, and other bright radio sources, with Skynet's 20-meter telescope, and calibrated these with measurements of Cyg~A, observed as closely in time as possible, but not always on the same day.  RCR was used to reject measurements that were outlying, because of variations in the receiver's gain between the primary and calibration observations.  In some cases, in particular when the timescale between these observations was longer, up to 35\% of these samples were contaminated, necessitating the use of RCR instead of {\color{black}traditional} Chauvenet rejection/sigma clipping.  (Trotter {\color{black}et al.} additionally used RCR to eliminate occasional pointing errors when modeling systematic focus differences between these sources, from drift-scan data taken with a different, transit radio telescope.) 

\section{Model Fitting}

So far, we have only {\color{black}considered} cases where uncontaminated measurements are distributed, either {\color{black}normally} (\textsection3.1, \textsection3.2) or {\color{black}non-normally} (\textsection3.3), about a single, parameterized value, $y$.  In particular, we have introduced increasingly robust ways of measuring $y$, or to put it differently, of fitting $y$ to measurements, namely:  the mean, the median, and the mode (\textsection2.1, \textsection6).  We have also introduced techniques:  (1)~to more robustly identify outlying deviations from $y$, for rejection (\textsection2.2 -- \textsection3, \textsection6); (2)~to more precisely measure $y$, without sacrificing robustness (\textsection4); and (3)~to more rapidly measure $y$ (\textsection5).  

In this section, we show that RCR can also be applied when measurements are distributed not about a single, parameterized value, but about a parameterized model, $y\left(\left\{x\right\}|\left\{\theta\right\}\right)$, where $\left\{x\right\}$ are the model's independent variables, and $\left\{\theta\right\}$ are the model's parameters.  But first, we must introduce new, increasingly robust ways of fitting $y\left(\left\{x\right\}|\left\{\theta\right\}\right)$ to measurements, now given by $\left\{\left\{{x_i}_{-\sigma_{x-,i}}^{+\sigma_{x+,i}}\right\}, {y_i}_{-\sigma_{y-,i}}^{+\sigma_{y+,i}}\right\}$. Specifically, these will be generalizations of the mean, the median, and the mode, that reduce to these in the limit of a single-parameter fit, but that result in best-fit, or baseline, models from which deviations can be calculated otherwise.  Consequently, these will be able to replace the mean, the median, and the mode in the RCR algorithm, with no other modification to the algorithm being necessary.
		
		\subsection{Generalized Measures of Central Tendency}
	
	Usually, models are fitted to measurements by maximizing a likelihood function.\footnote{Or, by maximizing the product of a likelihood function and a prior probability distribution, if the latter is available.}   For example, if:
	
	\begin{equation}
	\sigma_{x-,i} \approx \sigma_{x+,i} \approx 0,
	\end{equation}
	
	\begin{equation}
	\sigma_{y-,i} \approx \sigma_{y+,i} \approx \sigma_{y,i},
	\end{equation}
	
	\noindent and  
	
	\begin{equation}
	\chi^2 = \sum_i \left[\frac{y_i - y\left(\left\{x\right\}|\left\{\theta\right\}\right)}{\sigma_{y,i}}\right]^2 \approx N - M,
	\end{equation}
	
	\noindent where $N$ is the number of independent measurements, and $M$ is the number of non-degenerate model parameters, this function is simple:  ${\cal L}\propto e^{-\chi^2/2}$, in which case maximizing $\cal{L}$ is equivalent to minimizing $\chi^2$.  If these conditions are not met, $\cal{L}$, and its maximization, can be significantly more involved (e.g., Reichart 2001; Trotter 2011).  Regardless, such, maximum-likelihood, approaches are generalizations of the mean, and consequently are not {\color{black}robust.  
		
		To} see this, again consider the simple case of the single-parameter model:  $y\left(\left\{x\right\}|\left\{\theta\right\}\right)=y$.  Minimizing Equation 23 with respect to $y$ (i.e., solving $\partial\chi^2/\partial y=0$ for $y$) yields a best-fit parameter value, and a best-fit model, of $y=\left(\sum_i y_i/\sigma_{y,i}^2\right)/\left(\sum_i 1/\sigma_{y,i}^2\right)=\left(\sum_i w_iy_i\right)/\left(\sum_i w_i\right)$.  This is just the weighted mean of the measurements (Equation 7), which is not robust.
	
	One could imagine iterating between (1)~maximizing $\cal{L}$ to establish a best-fit model, and (2)~applying robust {\color{black}outlier} rejection to the deviations from this {\color{black}model, but} given that (1)~is not robust, this would be little better than iterating with {\color{black}traditional} Chauvenet rejection, which relies on the weighted mean.  Instead, we retain {\color{black}the RCR algorithm}, but replace the weighted mean, the weighted median, and the weighted mode with generalized versions, maintaining the robustness, and precision, of each.  We generalize the weighted mean as above, with maximum-likelihood model fitting.  We generalize the weighted median and the weighted mode as follows.
	
	First, consider the case of an $M$-parameter model where for any combination of $M$ measurements, a unique set of parameter values, $\left\{\theta\right\}_j$, can be {\color{black}determined}.{\color{black}\footnote{{\color{black}In the event of redundant independent-variable information, fewer than $M$ parameter values can be determined, and we address this case in \textsection8.3.2.  In the event of a periodic model, multiple $M$-parameter solutions can be determined (some equivalent to each other, some not), and we address this case in \textsection8.3.3.}}}  Furthermore, imagine doing this for all $1\leq j\leq N!/\left[M!(N-M)!\right]$ combinations of $M$ measurements,{\color{black}\footnote{{\color{black}Or for as large of a randomly drawn (but without repetitions) subset of these as is computationally reasonable.  We switch over to random draws, where each measurement is drawn in proportion to its weight, when $N!/\left[M!(N-M)!\right]>20,000$.  For $M=2$, this corresponds to $N>200$.  For $M=3$, this corresponds to $N>50$.}}} and weighting each calculated parameter value {\color{black}by how accurately it could be determined (see \textsection8.2).  \textbf{Our generalizations are then given by:  (1)~the weighted median of $\left\{\left\{\theta\right\}_j\right\}$, and (2)~the weighted mode of $\left\{\left\{\theta\right\}_j\right\}$.}}
	
	Although more sophisticated implementations can be imagined, here we define these quantities simply, and such that they reduce to the weighted median and the weighted mode, respectively, in the limit of the single-parameter model, just as the maximum-likelihood technique above reduces to the weighted mean in this limit{\color{black}:}\\
	
	\begin{itemize}
		\item {\color{black}For} the weighted median of $\left\{\left\{\theta\right\}_j\right\}$, we calculate the weighted median for each model parameter separately.  
		
		\item {\color{black}For} the weighted mode of $\left\{\left\{\theta\right\}_j\right\}$, we determine the half-sample for each model parameter separately, but then include only the intersection of these half-samples in the next iteration.\footnote{In this case, iteration ends either:  (1)~as before, if the next intersection would be unchanged (\textsection2.1, \textsection6), or (2)~if the next intersection would be null. }   
	\end{itemize}
	
	\noindent{\color{black}We} demonstrate these techniques, and the maximum-likelihood technique, for a simple, linear, but contaminated, model in Figure~34.  
	
	\begin{figure*}
		\centering
		\includegraphics[height=0.9\textheight]{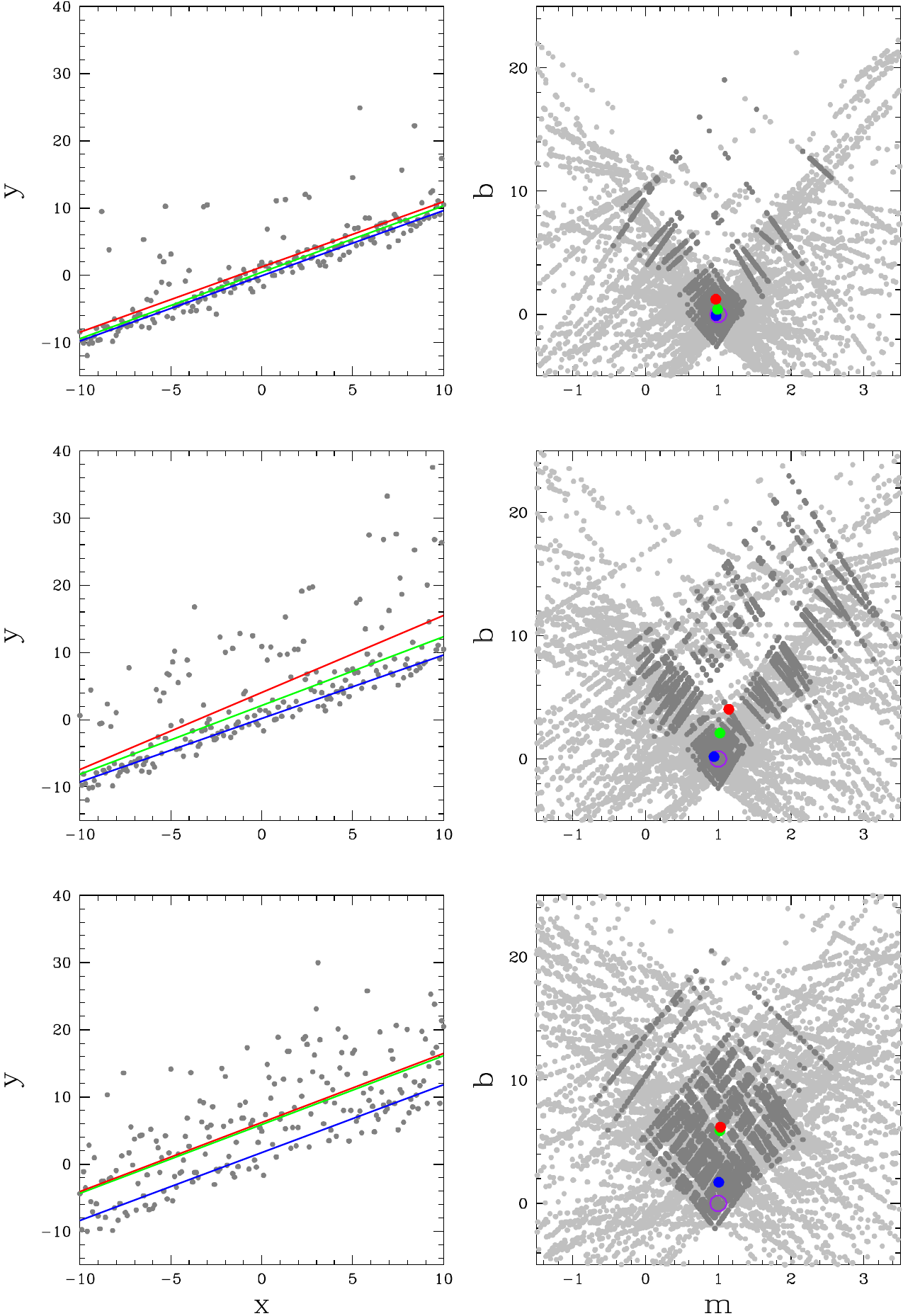}
		\caption{\textbf{Left column:}  201 measurements, with fraction $f_1=1-f_2$ drawn from a Gaussian distribution of mean $y(x)=x$ and standard deviation 1, and fraction $f_2=0.15$ (top row), 0.5 (middle row), and 0.85 (bottom row), representing contaminated measurements, drawn from the positive side of a Gaussian distribution of mean zero and standard deviation 10, and added to uncontaminated measurements, drawn as above.  \textbf{Right column:}  {\color{black}Model solutions}, $\left\{\theta\right\}_j$, calculated from each pair of measurements in the panel to the left, using $y(x)=b+m\left(x-\overline{x}\right)${\color{black}, with $\overline{x}=\sum_iw_ix_i/\sum_iw_i$ (see \textsection8.3.5),} and {\color{black}for} model parameters $b$ and $m$.  Each calculated parameter value is weighted {\color{black}(see \textsection8.2.1 or \textsection8.2.2),} and darker points correspond to models where the product of these weights is in the top 50\%.  {\color{black}The purple circle} corresponds to the original, underlying model, and in both columns, blue corresponds to the weighted mode of $\left\{\left\{\theta\right\}_j\right\}$, green corresponds to the weighted median of $\left\{\left\{\theta\right\}_j\right\}$, and red corresponds to maximum-likelihood model fitting.  The weighted mode of $\left\{\left\{\theta\right\}_j\right\}$ performs the best, especially in the limit of large $f_2$.  Maximum-likelihood model fitting performs the worst.  See Figure~35 for post-rejection versions.}
	\end{figure*}
	
	In Figure~35, we apply {\color{black}RCR} as before (\textsection4, \textsection5), {\color{black}except that we no longer use the weighted mode, the weighted median, and the weighted mean to establish baseline \textit{values} from which the deviations of the measurements can be determined.  Rather, we use our generalizations of these, to establish baseline \textit{functions of $\left\{x\right\}$}, of corresponding robustness and precision, from which these deviations can, as before, be determined}.\footnote{{\color{black}More model parameters means more degrees of freedom, and consequently artificially smaller deviations for the same number of measurements.  To correct for this, we multiply our correction factors (Figures 29 and 31) by (from Equation~8):
			\begin{equation}
			\sqrt{{\left(\sum_{i}{w_i}-\Delta\frac{\sum_{i}{w_i^2}}{\sum_{i}{w_i}}\right)}/{\left(\sum_{i}{w_i}-M\Delta\frac{\sum_{i}{w_i^2}}{\sum_{i}{w_i}}\right)}}
			\end{equation}
			where the sums are over the non-rejected measurements. (For $N$ unweighted measurements, this corresponds to dividing by $\sqrt{N-M}$, instead of by $\sqrt{N-1}$, when calculating a (two-sided) standard deviation, but this correction applies to our 68.3-percentile deviation calculations as well.)  This also prevents over-rejection rates (\textsection3.3.1) from increasing with $M$.}}  Even in the face of heavy contamination, this approach can be very effective at recovering the original, underlying {\color{black}correlation.
		
		\begin{figure*}
			\centering
			\includegraphics[height=0.9\textheight]{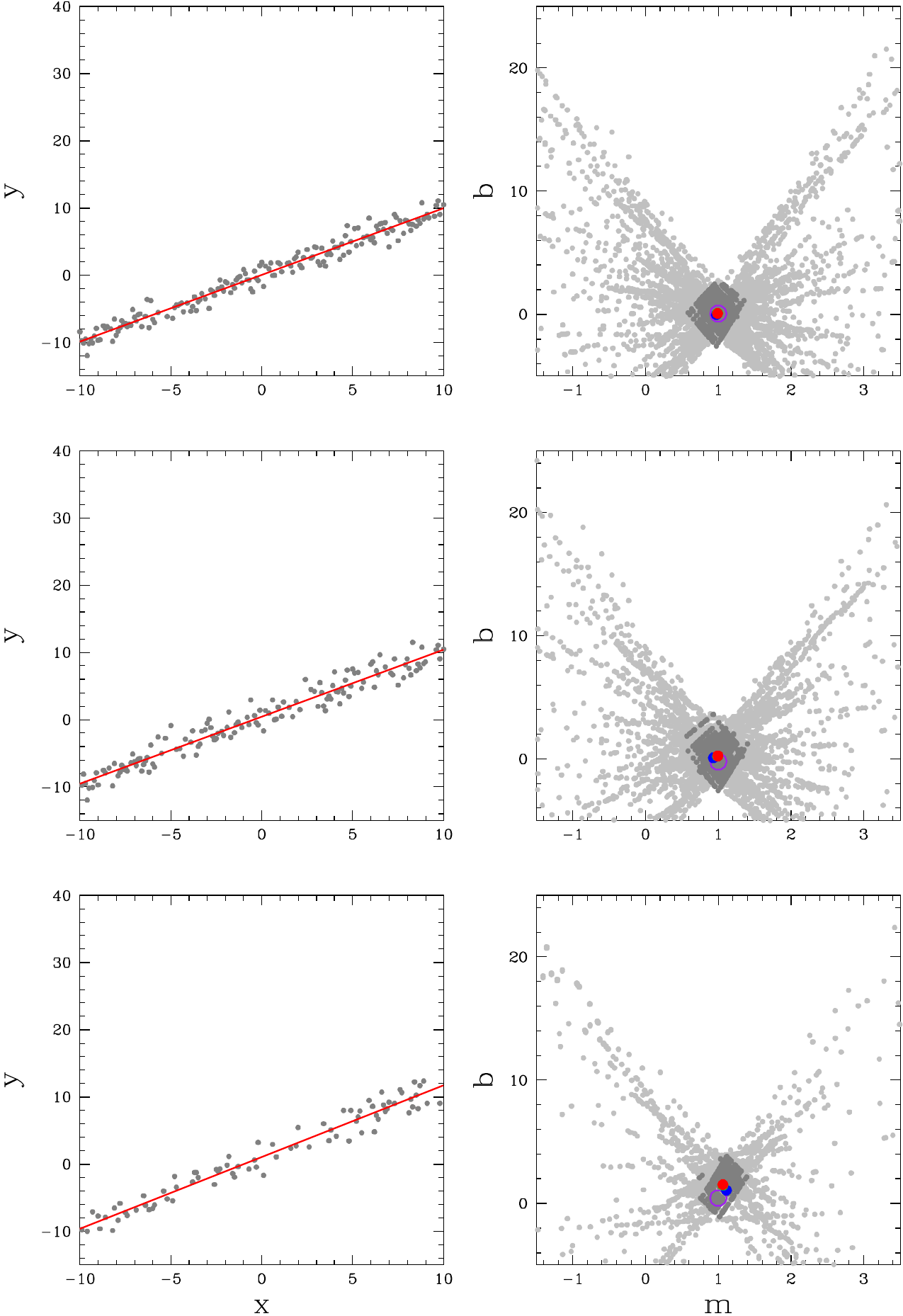}
			\caption{Figure~34, after {\color{black}RCR}.  Here, we have performed bulk rejection {\color{black}as in \textsection5, but} using our generalization of the mode {\color{black}instead of the mode}, followed by {\color{black}individual rejection as in \textsection4, using (1)}~our most{\color{black}-}general robust technique for symmetrically distributed uncontaminated measurements -- {\color{black}now consisting of} our generalization of the mode $+$ technique~3 (the broken-line fit) -- followed by {\color{black}(2)}~our most{\color{black}-}precise robust technique -- {\color{black}now consisting of} our generalization of the median $+$ technique~1 (the 68.3\% value) -- followed by {\color{black}(3)~traditional} Chauvenet rejection{\color{black}, but} using our generalization of the mean {\color{black}instead of the mean (e.g., Figures 30, 28, and 19)}.  {\color{black}RCR} proves effective, even in the face of heavy contamination.}
		\end{figure*}
		
		\subsection{Implementation}
		
		In this section, we describe how $M$ parameter values, $\left\{\theta\right\}_j$, can be calculated from $M$ measurements, both in the simplest $M>1$ case of a linear model (see \textsection8.2.1), and in general (see \textsection8.2.2).$^{12}$  
		
		We also describe how uncertainties, and hence weights, can be calculated for each of these $M$ parameter values.  This depends on the locations and weights of the $M$ measurements, but it also depends on how one models their scatter, about the best-fit model to all of the non-rejected measurements.  In \textsection8.2.1 and \textsection8.2.2, we present the simplest, and most common, model for this scatter, in which its RMS, at least for same-weight, uncontaminated measurements, is taken to be the same, or constant, at all locations, as it is in Figures 34 and 35.  In \textsection8.2.3, we consider non-constant RMS scatter, and present its most common case.
		
		\subsubsection{Simplest $M>1$ Case:  Linear Model with Constant RMS Scatter}
		
		Consider a linear model given by $y(x)=b+m\left(x-\overline{x}\right)$.  For any $M=2$ of the $N$ measurements, $(x_1,y_1)$ and $(x_2,y_2)$, one can calculate $M=2$ parameter values, given by:
		
		\begin{equation}
		m = \frac{y_2-y_1}{x_2-x_1}    
		\end{equation}
		
		\noindent and
		
		\begin{equation}
		b = \frac{x_2-\overline{x}}{x_2-x_1}y_1 - \frac{x_1-\overline{x}}{x_2-x_1}y_2.   
		\end{equation}
		
		\noindent The uncertainties in these values depend not only on the statistical uncertainties in $y_1$ and $y_2$ -- which may or may not be known -- but also on any systematic scatter in the measurements, at $x_1$ and $x_2$.  
		
		Let $\sigma_y(x)$ be a to-be-specified model for the RMS scatter (statistical and/or systematic) of \textit{average-weight}, uncontaminated measurements, about the best-fit model to all of the non-rejected measurements.  
		
		In the limit that $\sigma_y(x)$ is purely statistical, the RMS scatter of \textit{any-weight}, uncontaminated measurements is then given by $(\overline{w}/w)^{1/2}\sigma_y(x)$, where $w$ is measured weight, and $\overline{w}$ is the average value of $w$ for the uncontaminated measurements.
		
		In the limit that $\sigma_y(x)$ is purely systematic, statistical error bars, and hence measured weights, do not matter, and consequently, an unweighted fit should be performed instead.\footnote{\color{black}Note, this is as much the case in \textsection6 as it is here.}   Note, the same expression may be used for the RMS scatter, but in this case, all measured weights should be reset to a common value, such as $w=\overline{w}=1$.\footnote{\color{black}If in-between these two limiting cases, with statistical uncertainty greater than systematic scatter for some measurements, and less than it for the rest, one should also perform an unweighted fit.  In this case, most measurements with statistical uncertainty $\gg$ systematic scatter will be rejected as outlying (e.g., as outliers were rejected in Figure~35), but since these measurements are, by definition, of low measured weight, they were not going to significantly impact the fit anyway.  However, if statistical uncertainties are known, one could then calculate new weights, given by $\left\{1+\left[\sigma_i/\sigma_{sys}(x_i)\right]^2\right\}^{-1}$, where $\sigma_{sys}(x)$ is the RMS scatter of the non-rejected measurements about the unweighted fit, and then perform a weighted fit.} 
		
		Given this expression for the RMS scatter, Equations 25 and 26, and standard propagation of uncertainties, the uncertainties in $m$ and $b$ are then given by:
		
		\begin{equation}
		\sigma_m = \sqrt{\frac{\frac{\overline{w}\sigma_y^2(x_2)}{w_2}+\frac{\overline{w}\sigma_y^2(x_1)}{w_1}}{(x_2-x_1)^2}}    
		\end{equation}
		
		\noindent and
		
		\begin{equation}
		\sigma_b = \sqrt{\frac{\frac{\overline{w}\sigma_y^2(x_1)}{w_1}(x_2-\overline{x})^2+\frac{\overline{w}\sigma_y^2(x_2)}{w_2}(x_1-\overline{x})^2}{(x_2-x_1)^2}}. 
		\end{equation}
		
		\noindent Consequently, we weight $m$ by $w_m\propto\sigma_m^{-2}$ and $b$ by $w_b\propto\sigma_b^{-2}$.  Since $\overline{w}$ factors out, and is constant, it can be ignored.  
		
		In the simplest, and most common, case, $\sigma_y(x)=\sigma_y$ is also constant, as it is in Figures 34 and 35.  In this case, it also factors out and can be ignored, yielding weights for $m$ and $b$ that depend only on the locations and weights of the $M=2$ measurements from which they were calculated:\footnote{\color{black}With non-linear models, $\overline{w}$ and $\sigma_y$ (if constant) also factor out and can be ignored.  However, these weights, on the calculated parameter values, can also depend on the model parameters themselves (see \textsection8.2.2).} 
		
		\begin{equation}
		w_m \propto \frac{(x_2-x_1)^2}{w_1^{-1}+w_2^{-1}}
		\end{equation}
		
		\noindent and
		
		\begin{equation}
		w_b \propto \frac{(x_2-x_1)^2}{\frac{(x_1-\overline{x})^2}{w_2}+\frac{(x_2-\overline{x})^2}{w_1}}. 
		\end{equation}
		
		\noindent And again, (1)~if all $N$ of the measurements have the same weight, and/or (2)~if $\sigma_y$ is dominated by systematic scatter, these equations simplify even further, with $w_1=w_2=1$.
		
		Note, if a parameter's value is known to be more or less probable a priori -- i.e., if there is a prior probability distribution for that parameter -- the $N!/\left[M!(N-M)!\right]$ weights that we calculate for that parameter (given by, e.g., Equation 29 or 30) should be multiplied by the prior probabilities of the $N!/\left[M!(N-M)!\right]$ values that we calculate for that parameter (given by, e.g., Equation 25 or 26), respectively, to up- or down-weight them accordingly, before calculating their generalized median or mode (\textsection8.1).
		
		\subsubsection{General Case}
		
		Although many models can be solved for their $M$ parameters analytically, given $M$ measurements (e.g., as the linear model in \textsection8.2.1 is solved for $m$ and $b$, given two measurements), many models cannot be solved analytically.  And even if a model can be solved analytically, this is not always easy to do, nor can all solvable models be anticipated in advance.  Consequently, in general, we do this numerically, using the Gauss-Newton algorithm,\footnote{\color{black}With one modification:  Each time an iteration results in a poorer fit, (1)~we do not apply the increment vector, and (2)~we shrink it by 50\% in future iterations.    This helps to ensure local convergence, in the case of periodic models (see \textsection8.3.3).}  which requires only that the user supply (1)~the model, (2)~its first partial derivative with respect to each model parameter, to construct its Jacobian, and (3)~an initial guess, which usually has no bearing on the end result (however, see \textsection8.3.3).
		
		Furthermore, the uncertainty, $\sigma_{\theta_i}$, in each calculated parameter value, $\theta_i$, is straightforward to calculate, from the same matrix that lies at the heart of the Gauss-Newton algorithm, which, when $N=M$, is simply the inverse Jacobian, ${\cal J}^{-1}$.
		
		Let $\vec{\sigma}_y = (\sigma_{y_1}, ..., \sigma_{y_{N=M}})$ be an array of hypothetical errors in each of the $M$ measurements, each drawn from a Gaussian of mean zero and standard deviation $(\overline{w}/w_i)^{1/2}\sigma_y(\{x\}_i)$ (\textsection8.2.1).  This corresponds to an array of errors in the calculated parameter values, given by ${\cal J}^{-1}\vec{\sigma}_y$.  Next, imagine repeating these draws, and recalculating ${\cal J}^{-1}\vec{\sigma}_y$, an infinite number of times.  Each $\sigma_{\theta_i}$ is then given by the RMS of these arrays' $i$th values.  Mathematically, this is straightforward to calculate, and is equivalent to setting each $\sigma_{y_i} = (\overline{w}/w_i)^{1/2}\sigma_y(\{x\}_i)$ and calculating $(\sigma_{\theta_1}, ..., \sigma_{\theta_M}) = {\cal J}^{-1}\vec{\sigma}_y$, except that terms in this matrix-vector multiplication are instead summed in quadrature (it is not difficult to show that in the case of the linear model of \textsection8.2.1, this yields Equations 27 and 28.)  
		
		And as in \textsection8.2.1, the weight of each calculated parameter value is then given by $w_{\theta_i}\propto\sigma_{\theta_i}^{-2}$.  
		
		And as in Equations 29 and 30, $\overline{w}$ again factors out, and since constant, can be ignored.  Likewise, if $\sigma_y(\{x\})$ can again be modeled as constant, it too factors out and can be ignored.  (If $\sigma_y(\{x\})$ is not constant, it may be a function of $\{x\}$, as well as of the model parameters; we offer a common example in \textsection8.2.3.)
		
		However, unlike in Equations 29 and 30, and regardless of how $\sigma_y(\{x\})$ is modeled, each $w_{\theta_i}$ may now depend on the model parameters (through ${\cal J}^{-1}$).  Note however, when calculating these weights, we do not use the calculated parameter values, $\left\{\theta\right\}_j$, from the corresponding $M$-measurement combination.  Rather, we use those of the most recent baseline model, determined from taking the generalized mode, the generalized median, or the generalized mean of $\left\{\left\{\theta\right\}_j\right\}$ in the most recent iteration of the RCR algorithm (e.g., Figures 30 and 28).  This should be a significantly more accurate representation of the underlying model than any individual $\left\{\theta\right\}_j$.  
		
		We also use these, significantly more accurate, parameter values as the starting point for the Gauss-Newton algorithm in the next iteration of the RCR algorithm.  Only the starting point for the very first iteration need be supplied by the user.\footnote{\color{black}As stated above, for most applications, the Gauss-Newton algorithm yields the same result, $\left\{\theta\right\}_j$, regardless of the initial guess.  However, the generalized mode, median, or mean of $\left\{\left\{\theta\right\}_j\right\}$ also depends on each $\left\{\theta\right\}_j$'s corresponding weight, which does depend on the initial guess.  Consequently, before beginning the RCR algorithm, and bulk rejecting outliers, we iteratively measure the generalized mode of $\left\{\left\{\theta\right\}_j\right\}$, without rejecting measurements, and with each iteration implying new weights for $\left\{\left\{\theta\right\}_j\right\}$, until we converge, from the user's initial guess, to a starting point for the RCR algorithm that is maximally consistent with the measurements.}
		
		\subsubsection{Non-Constant RMS Scatter:  Logarithmic Case}
		
		In general, $\sigma_y(\{x\})$ may not be constant, in which case a model must be provided for it by the user, just as a model must be provided for $y(\{x\})$ by the user.  With no additional work, we can support models for $\sigma_y(\{x\})$ that are proportional to any function of (1)~the independent variables, $\{x\}$, as well as (2)~the model parameters.  This is because we already support dependencies on both of these in the inverse Jacobian (\textsection8.2.2).  (As with $\sigma_y$ in \textsection8.2.1 and \textsection8.2.2, the constant of proportionality factors out and can be ignored.)  
		
		How one models $\sigma_y(\{x\})$ depends on the problem at hand.  As stated above, $\sigma_y(\{x\})$ can usually be modeled as constant and ignored.  However, another common case arises when the user has a model $y(\{x\})$ that can be linearized.  For example, exponential and power-law models can be linearized by taking a logarithm of both sides:  e.g., $y(x)=be^{m(x-\overline{x})}$ becomes $\ln{y(x)}=\ln{b}+m(x-\overline{x})$, and $y(x)=b\left(x/e^{\overline{\ln{x}}}\right)^m$ becomes $\ln{y(x)}=\ln{b}+m(\ln{x}-\overline{\ln{x}})$.  
		
		This of course is fine, and even preferable, if the RMS scatter about $\ln{y(\{x\})}$ can be modeled as constant:  i.e., if $\sigma_{\ln{y}}(\{x\})=\sigma_{\ln{y}}$.  However, often $\sigma_y(\{x\})=\sigma_y$ is constant, in which case $\sigma_{\ln{y}}(\{x\})$ is then not constant, and consequently must be modeled.
		
		In this case, $\sigma_{+\ln{y}}(\{x\}) \approx \ln\left[y(\{x\}) + \sigma_y\right] - \ln{y(\{x\})} \rightarrow \sigma_y/y(\{x\})$ and $\sigma_{-\ln{y}}(\{x\}) \approx \ln{y(\{x\})} - \ln\left[y(\{x\}) - \sigma_y\right] \rightarrow \sigma_y/y(\{x\})$ when $\sigma_y \ll y(\{x\})$, and $\sigma_{+\ln{y}}\rightarrow\ln{\left[\sigma_y/y(\{x\})\right]}$ and $\sigma_{+\ln{y}}\rightarrow\infty$ when $\sigma_y \gg y(\{x\})$  Since the $\sigma_y \ll y(\{x\})$ measurements are the most informative, one can approximate $\sigma_{\ln{y}} \approx\sigma_y/y(\{x\})$, which, conservatively, underestimates the weights for the less-informative, $\sigma_y \gtrsim y(\{x\})$ measurements.  
		
		In other words, logarithmic compression of constant RMS scatter results in smaller RMS scatter, and hence higher weights, for high-$\ln{y(\{x\})}$ measurements, and larger RMS scatter, and hence lower weights, for low-$\ln{y(\{x\})}$ measurements.
		
		In the case of the linearized exponential model, Equations 29 and 30 then become:
		
		\begin{equation}
		w_m \propto \frac{(x_2-x_1)^2}{w_1^{-1}y^{-2}(x_1)+w_2^{-1}y^{-2}(x_2)}
		\end{equation}
		
		\noindent and
		
		\begin{equation}
		w_b \propto \frac{(x_2-x_1)^2}{\frac{(x_1-\overline{x})^2}{w_2y^2(x_2)}+\frac{(x_2-\overline{x})^2}{w_1y^2(x_1)}}, 
		\end{equation}
		
		\noindent and in the case of the linearized power-law model, they instead become:
		
		\begin{equation}
		w_m \propto \frac{(\ln{x_2}-\ln{x_1})^2}{w_1^{-1}y^{-2}(x_1)+w_2^{-1}y^{-2}(x_2)}
		\end{equation}
		
		\noindent and
		
		\begin{equation}
		w_b \propto \frac{(\ln{x_2}-\ln{x_1})^2}{\frac{(\ln{x_1}-\overline{\ln{x}})^2}{w_2y^2(x_2)}+\frac{(\ln{x_2}-\overline{\ln{x}})^2}{w_1y^2(x_1)}}. 
		\end{equation}
		
		\noindent Note, these equations depend not only on the independent variable, $x$, but now also on the model parameters, $m$ and $b$, through $y(x)$, and consequently are evaluated as prescribed in the second-to-last paragraph of \textsection8.2.2.
		
		However, although the linearization of these, and other, models allows their parameters to be determined analytically, as in Equations 25 and 26, instead of numerically as in \textsection8.2.2, this really does not gain the user anything, given the speeds of modern computers.  Instead, when possible, we recommend either leaving one's model in, or transforming one's model to, whatever form yields constant, or near-constant, RMS scatter about its best fit to the non-rejected measurements, and then simply applying the all-purpose (linear and non-linear) machinery of \textsection8.2.2.\footnote{\color{black}That said, both approaches usually yield near identical results.  Modeling exponential or power-law data with parameters $b$ and $m$, instead of linearized data with $\ln{b}$ and $m$, yields (1)~a different inverse Jacobian (\textsection8.2.2), and (2)~a different model for the RMS scatter, $\sigma_y(x)$ vs.\@ $\sigma_{\ln{y}}(x)$.  But together these yield the same expressions for $w_b=w_{\ln{b}}$ and $w_m$ (up to factors of proportionality that do not matter).  Consequently, the only difference is how concentrated the calculated parameter values, $\left\{\left\{\theta\right\}_j\right\}$, are, which does not affect the weighted median of $\left\{\left\{\theta\right\}_j\right\}$, but can affect the weighted mode of $\left\{\left\{\theta\right\}_j\right\}$:  Using $b$ instead of $\ln{b}$ favors lower values, but usually only marginally.  This is known as choice of basis, which we return to \textsection8.3.6.}
		
		\subsection{Considerations, Limitations, and Examples}
		
		In this section, we present a few additional considerations and limitations, and examples.  In \textsection8.3.1 -- \textsection8.3.3, we consider special cases that while they do not change how we calculate $M$-parameter solutions, $\left\{\theta\right\}_j$, from $M$ measurements (\textsection8.2.2), they can affect how we calculate the generalized mode, and sometimes also the generalized median, from a full set of $N!/\left[M!(N-M)!\right]$ $M$-parameter solutions, $\left\{\left\{\theta\right\}_j\right\}$ (\textsection8.1).  In \textsection8.3.4, we show that RCR becomes less robust as $M$ increases, and this appears to be a fundamental limitation of our approach.  And finally, in \textsection8.3.5 and \textsection§8.3.6, we discuss the importance of good modeling practices, both in general, but also specifically to RCR.
		
		\subsubsection{Measurements that Cannot Be Described by the Model}}
	
	Due to statistical and/or systematic {\color{black}scatter}, some combinations of $M$ measurements, even $M$ uncontaminated measurements, might not {\color{black}map} to {\color{black}any} combination of {\color{black}values for a model's $M$ parameters.
		
		For} example, consider an exponential {\color{black}model that} asymptotes from positive values to zero {\color{black}as $x\rightarrow\infty$ (e.g., $y(x)=be^{m(x-\overline{x})}$, with} $b>0$ and $m<0$), but with measurements, $y_i$, that are occasionally negative due to statistical and/or systematic {\color{black}scatter}.  Since {\color{black}all combinations of values for $b$ and $m$ yield only-}positive {\color{black}or only-}negative values {\color{black}for} $y(x)$, {\color{black}if presented with an oppositely signed pair of measurements, our Gauss-Newton algorithm (\textsection8.2.2) will instead run away to one of the following, limiting solutions, depending on the values of $x_i$, $y_i$, and $\overline{x}$:  $m=-\infty$ or $\infty$, and $b=-\infty$, $0$, $\infty$, or the value of the positive measurement (if $x_i$ happens to equal $\overline{x}$).  Note, such solutions are easily flagged, since the fitted model does not (cannot) pass through all M of the measurements (e.g., resulting in a non-zero $\chi^2$ value).
		
		Although extreme, and not fully representative of the measurements that produced them, we do not exclude such solutions when calculating} the weighted median of $\left\{\left\{\theta\right\}_j\right\}${\color{black}:  To do so could bias the result (in this particular example,} toward higher values of $b$ and {\color{black}shallower} values of $m${\color{black})}.  At the same time, {\color{black}we \textit{do} exclude such solutions when calculating} the weighted mode of $\left\{\left\{\theta\right\}_j\right\}$, lest {\color{black}any of these parameter values} be returned artificially {\color{black}(e.g., in this case, they could result in a meaningless, but statistically significant, overdensity of $b=0$ values).  
		
		We} demonstrate {\color{black}RCR} applied to such an exponential model in Figures 36 and 37{\color{black}, and despite a fair number of $y_i<0$ measurements at high-$x$ values, it converges to an acceptable solution in all but the most contaminated case.\\
		
		\begin{figure*}
			\centering
			\includegraphics[height=0.9\textheight]{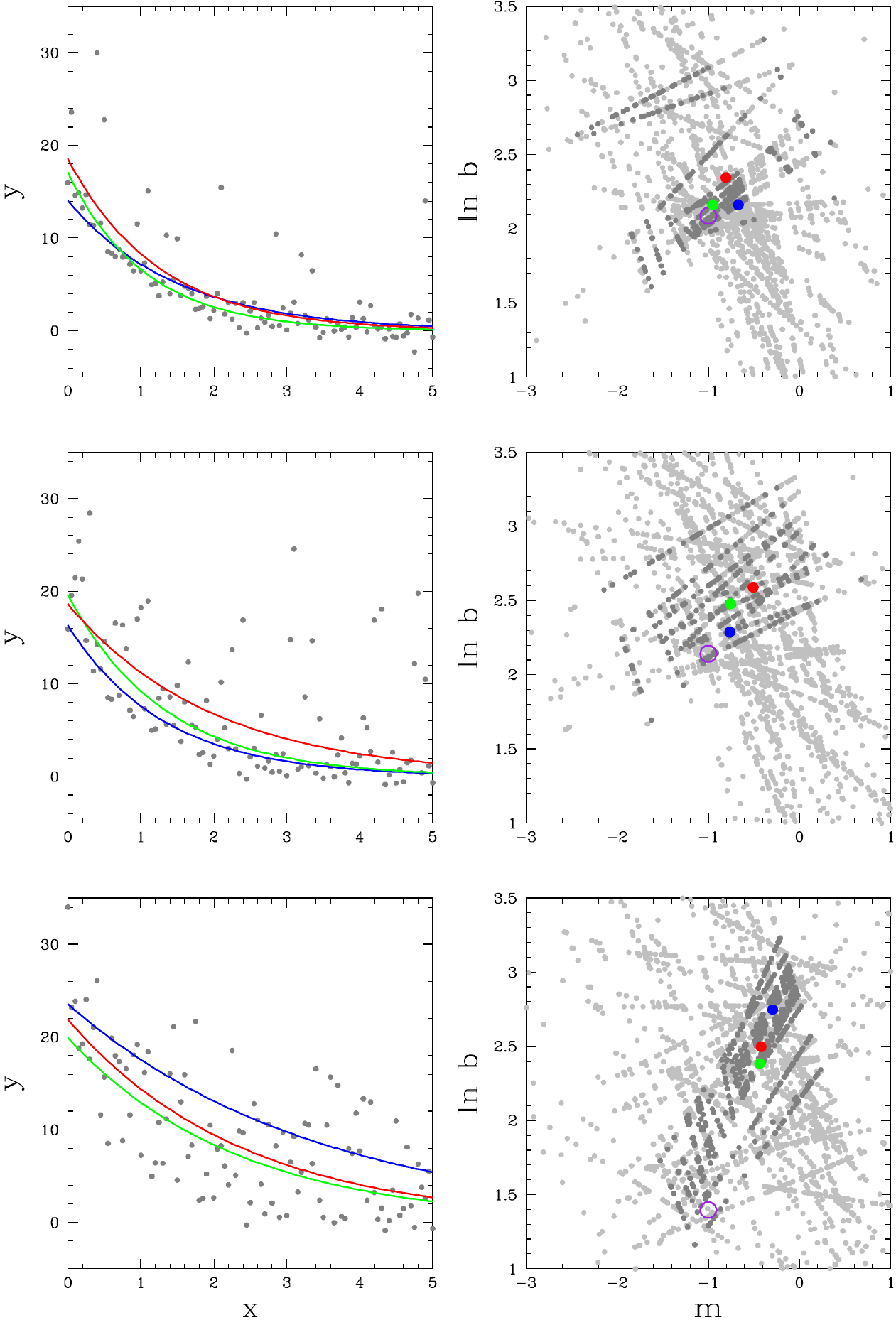}
			\caption{\textbf{Left column:}  101 measurements, with fraction $f_1=1-f_2$ drawn from a Gaussian distribution of mean $y(x)=10e^{-\left(x-0.5\right)}$ and standard deviation 1, and fraction $f_2=0.15$ (top row), 0.5 (middle row), and 0.85 (bottom row), representing contaminated measurements, drawn from the positive side of a Gaussian distribution of mean zero and standard deviation 10, and added to uncontaminated measurements, drawn as above.  \textbf{Right column:}  {\color{black}Model solutions}, $\left\{\theta\right\}_j$, calculated from each pair of measurements in the panel to the left, using $y(x)=be^{m\left(x-\overline{x}\right)}$, {\color{black} with $\overline{x}=\sum_iw_ix_iy^2(x_i)/\sum_iw_iy^2(x_i)$ (see \textsection8.3.5),} and {\color{black}for} model parameters $\ln{b}$ and $m$ (see {\color{black}\textsection8.3.6}).  Each calculated parameter value is weighted {\color{black}(\textsection8.2.2),} and darker points correspond to models where the product of these weights is in the top 50\%.  {\color{black}The purple circle} corresponds to the original, underlying model, and in both columns, blue corresponds to the weighted mode of $\left\{\left\{\theta\right\}_j\right\}$, green corresponds to the weighted median of $\left\{\left\{\theta\right\}_j\right\}$, and red corresponds to maximum-likelihood model fitting.  The contaminants have a greater, relative, effect on the high-$x$/low-$y$ measurements than on the low-$x$/high-$y$ measurements, biasing the calculated models toward shallower slopes and higher normalizations (i.e., toward the upper right, in the panels on the right).  The weighted mode of $\left\{\left\{\theta\right\}_j\right\}$ most successfully overcomes this bias as $f_2\rightarrow0.5$, but all three techniques fail as $f_2\rightarrow0.85$.  See Figure~37 for post-rejection versions.}
		\end{figure*}
		
		\begin{figure*}
			\centering
			\includegraphics[height=0.9\textheight]{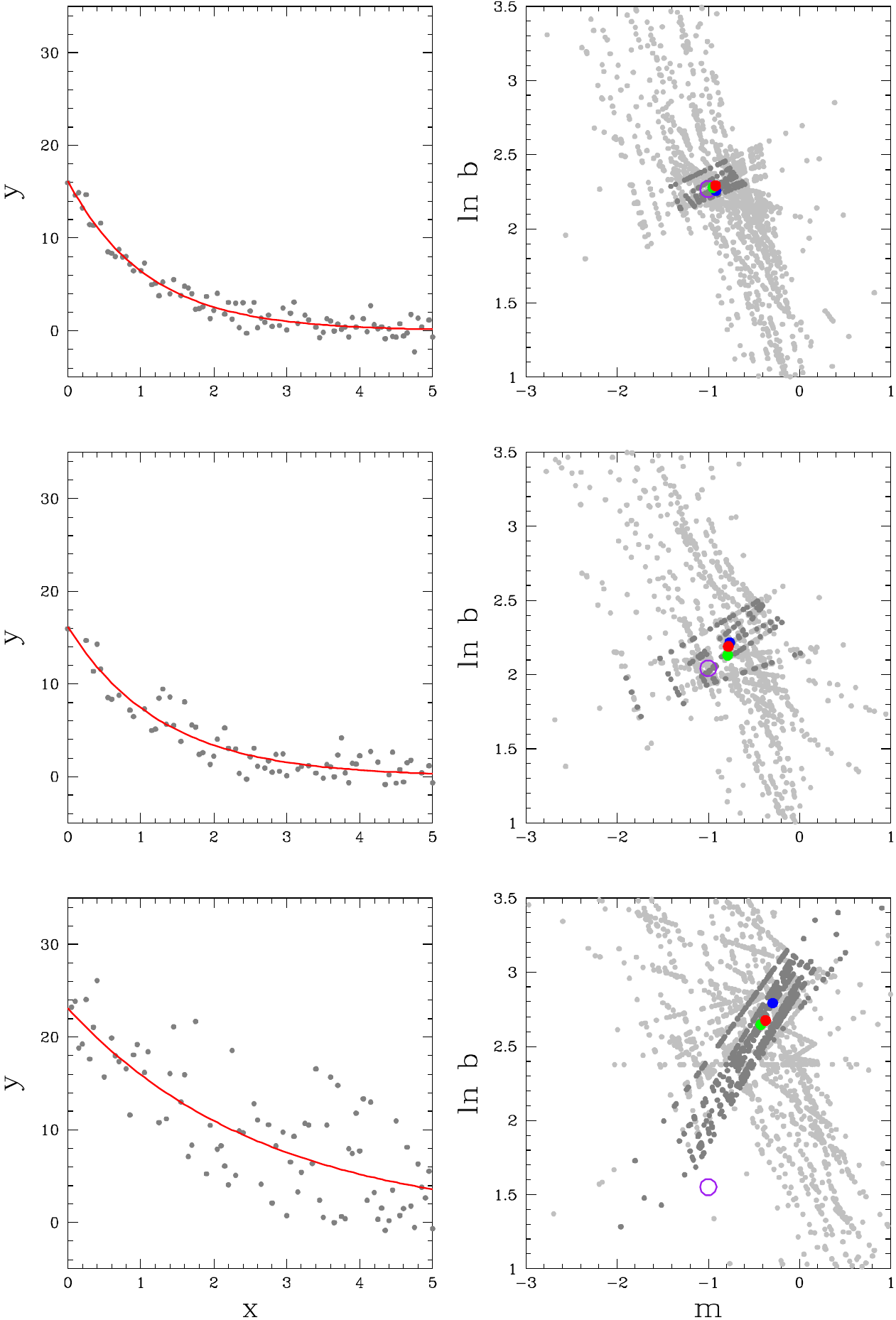}
			\caption{Figure~36, after {\color{black}RCR}.  Here, we have again performed bulk rejection {\color{black}as in \textsection5, but} using our generalization of the mode {\color{black}instead of the mode}, followed by {\color{black}individual rejection as in \textsection4, using (1)}~our most{\color{black}-}general robust technique for symmetrically distributed uncontaminated measurements -- {\color{black}now consisting of} our generalization of the mode $+$ technique~3 (the broken-line fit) -- followed by {\color{black}(2)}~our most{\color{black}-}precise robust technique -- {\color{black}now consisting of} our generalization of the median $+$ technique~1 (the 68.3\% value) -- followed by {\color{black}(3)~traditional} Chauvenet rejection{\color{black}, but} using our generalization of the mean {\color{black}instead of the mean (e.g., Figures 30, 28, and 19)}.  {\color{black}RCR} proves effective in the face of fairly heavy contamination, but is unable to overcome bias introduced by the contaminants (Figure~{\color{black}35)} as $f_2\rightarrow0.85$.}
		\end{figure*}
		
		\subsubsection{Combinations of $M$ Measurements with Redundant Independent-Variable Information}}
	
	Combinations of $M$ measurements with redundant independent-variable information cannot be used to determine all $M$ of {\color{black}a} model's parameters.  Furthermore, if, in this case, any of the model's parameters can be determined, they will be {\color{black}overdetermined.  
		
		For} example, {\color{black}consider a planar model, constrained by three measurements.  If these measurements happen to be co-linear, all three of the model's parameters cannot be determined}.  However, if this line happens to run parallel to one of the model's axes, at least one, and possibly two, of the model's parameters (i.e., the plane's slope along this axis, and the plane's normalization, if defined along this line) can be determined.  But they will be overdetermined, given three measurements for only one or two {\color{black}parameters.  
		
		In} the interest of simplicity, we discard these (usually rare) combinations completely, noting that  uncontaminated measurements selected in this way are unlikely to be preferentially under- or over-estimates, and consequently their exclusion is unlikely to bias calculation of the weighed median of $\left\{\left\{\theta\right\}_j\right\}$, let alone of the weighted mode of $\left\{\left\{\theta\right\}_j\right\}$.  However, more sophisticated implementations can also be imagined.
	
	{\color{black}Note, such cases are also easily flagged, in that the Jacobian in \textsection8.2.2 is not invertible (i.e., its determinant is zero).\\
		
		\subsubsection{Combinations of $M$ Measurements that Can Be Described by Multiple Model Solutions}
		
		Periodic models require a bit more care, in that each combination of $M$ measurements can be described by a countably infinite number of model solutions, including not only solutions that are equivalent to each other, but also shorter-period, overtone solutions that are not.  Both can bias calculation of the weighted median of $\left\{\left\{\theta\right\}_j\right\}$, and of the weighted mode of $\left\{\left\{\theta\right\}_j\right\}$.  
		
		For example, consider the simple, periodic model $y(x)=b\sin{m(x-x_0)}$.  The same measurements can result in model solutions that are equivalent to each other (1)~by reflection about both the $x$ and $y$ axes, (2)~by translation along the $x$ axis, by multiples of $2\pi/m$, and/or (3)~by translation along the $x$ axis by odd multiples of $\pi/m$, in combination with a reflection about the $x$ axis.  Consequently, once the Gauss-Newton algorithm (\textsection8.2.2) finds one of these solutions, we give the user the option to map it to a designated simplest form.  For example, with this model:  (1) If $m<0$, map $m\rightarrow-m$ and $b\rightarrow-b$; (2) then if $m|x_0|\geq2\pi$, map $x_0\rightarrow x_0-\frac{2\pi x_0}{m|x_0|}floor\left(\frac{m|x_0|}{2\pi}\right)$; and (3)~then if $m|x_0|\geq\pi$, map $x_0\rightarrow x_0-\frac{\pi x_0}{m|x_0|}$ and $b\rightarrow-b$.  
		
		In the case of shorter-period/higher-$m$, overtone solutions, which solution the Gauss-Newton algorithm finds depends on the initial guess that it is given.  This is analogous to centroiding algorithms in astrometry.  If a user clicks anywhere in a star's vicinity, such algorithms arrive at the same solution for the star's center.  But if the user clicks too far away, another star's center will be found instead.  We have modified the Gauss-Newton algorithm to help ensure local convergence (Footnote 19), but ultimately it is up to the user to make a reasonable (in this case, low-$m$) initial guess.
		
		We demonstrate RCR applied to this model in Figures 38 and 39, using the same contamination fractions as in Figures 34 -- 37.  The combination of re-mapping equivalent solutions, and of making a reasonable initial guess, results in good outcomes through fairly high contamination fractions (however, see \textsection8.3.4).
		
		\begin{sidewaysfigure*}
			\centering
			\vspace{3.75in}
			\includegraphics[width=0.9\textwidth]{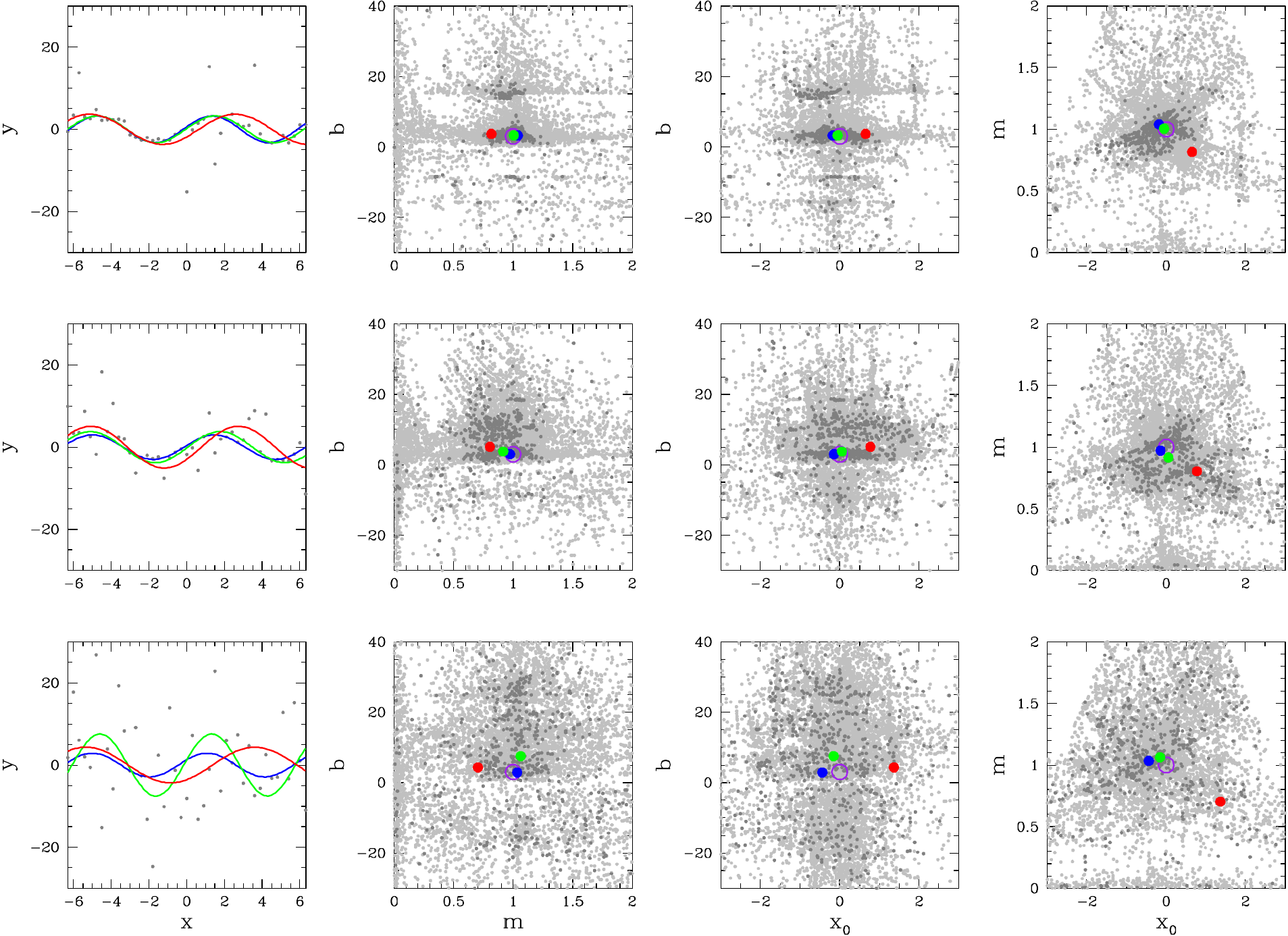}
			\caption{\textbf{Left column:}  43 measurements, with fraction $f_1=1-f_2$ drawn from a Gaussian distribution of mean $y(x)=3\sin x$ and standard deviation 1, and fraction $f_2=0.15$ (top row), 0.5 (middle row), and 0.85 (bottom row), representing contaminated measurements, drawn from a Gaussian distribution of mean zero and standard deviation 10, and added to uncontaminated measurements, drawn as above.  \textbf{Right columns:}  {\color{black}Model solutions}, $\left\{\theta\right\}_j$, calculated from each triplet of measurements in the panel to the left, using $y(x)=b\sin m\left(x-x_0\right)${\color{black}, for} model parameters $b$, $m$, and {\color{black}$x_0$.}  Each calculated parameter value is weighted {\color{black}(\textsection8.2.2),} and darker points correspond to models where the product of these weights is in the top 50\%.  {\color{black}The purple circle} corresponds to the original, underlying model, and in all columns, blue corresponds to the weighted mode of $\left\{\left\{\theta\right\}_j\right\}$, green corresponds to the weighted median of $\left\{\left\{\theta\right\}_j\right\}$, and red corresponds to maximum-likelihood model fitting.  The weighted mode of $\left\{\left\{\theta\right\}_j\right\}$ performs the best, especially in the limit of large $f_2$.  Maximum-likelihood model fitting performs the worst.  See Figure~39 for post-rejection versions.}
		\end{sidewaysfigure*}
		
		\begin{sidewaysfigure*}
			\centering
			\vspace{3.75in}
			\includegraphics[width=0.9\textwidth]{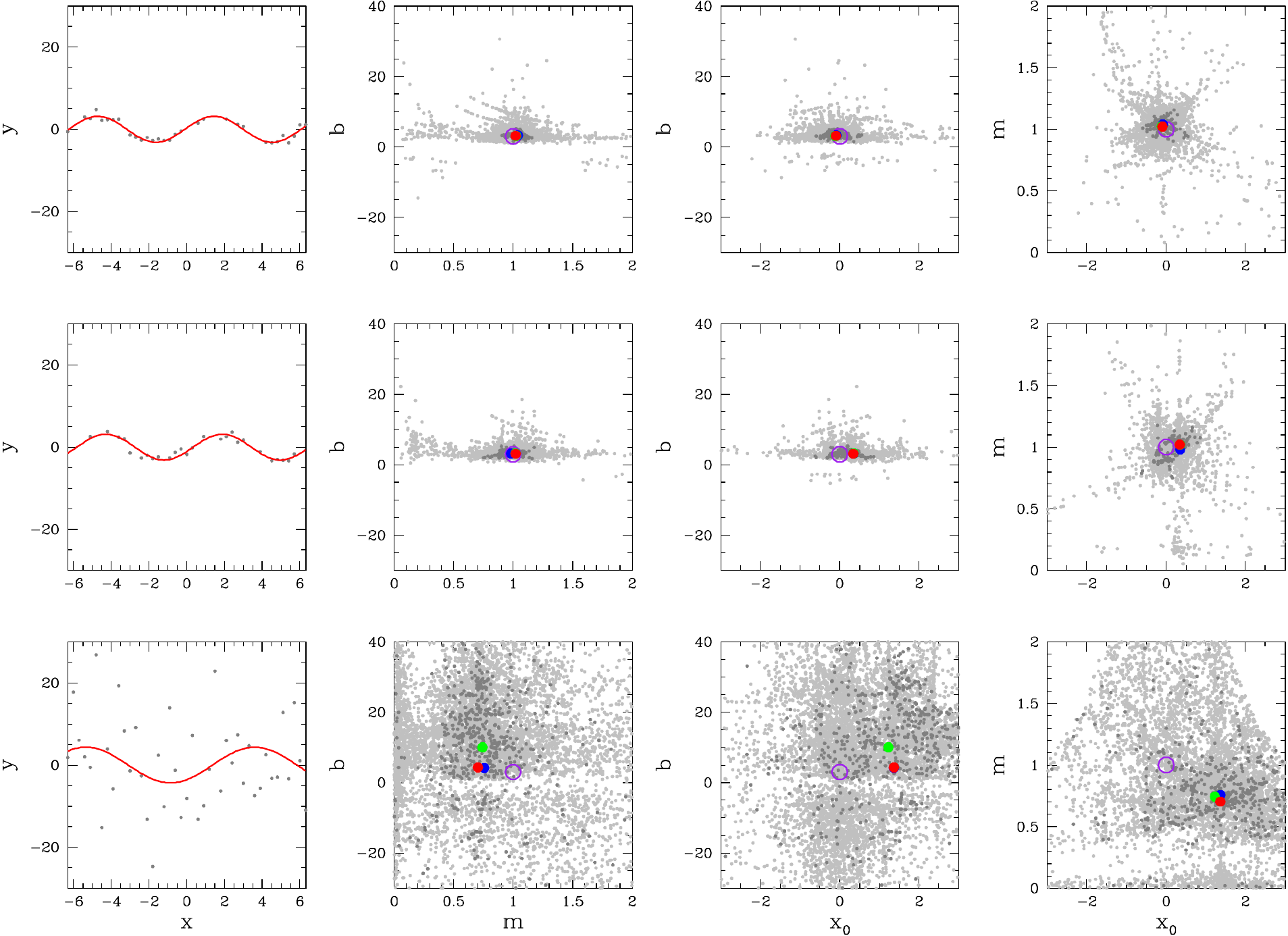}
			\caption{Figure~38, after {\color{black}RCR}.  Here, we have again performed bulk rejection {\color{black}as in \textsection5, but} using our generalization of the mode {\color{black}instead of the mode}, followed by {\color{black}individual rejection as in \textsection4, using (1)}~our most{\color{black}-}general robust technique for symmetrically distributed uncontaminated measurements -- {\color{black}now consisting of} our generalization of the mode $+$ technique~3 (the broken-line fit) -- followed by {\color{black}(2)}~our most{\color{black}-}precise robust technique -- {\color{black}now consisting of} our generalization of the median $+$ technique~1 (the 68.3\% value) -- followed by {\color{black}(3)~traditional} Chauvenet rejection{\color{black}, but} using our generalization of the mean {\color{black}instead of the mean (e.g., Figures 30, 28, and 19)}.  {\color{black}RCR} proves effective in the face of fairly heavy contamination, but is unable to overcome the greater fraction of contaminated {\color{black}model solutions (see \textsection8.3.4)} as $f_2\rightarrow0.85$:  $1-(1-0.85)^3=0.996625$ {\color{black}for $M=3$} vs.\@ $1-(1-0.85)^2=0.9775$ {\color{black}for $M=2$}.}
		\end{sidewaysfigure*}
		
		\subsubsection{RCR Less Robust as M Increases}}
	
	If a fraction, $1-f$, of {\color{black}$N$} measurements is uncontaminated, a smaller fraction, $\left(1-f\right)^M$, of the {\color{black}corresponding $N!/\left[M!(N-M)!\right]$} model solutions, $\left\{\left\{\theta\right\}_j\right\}$, is uncontaminated.  So, the higher the dimension of the model, and hence of the model's parameter space, the more difficult it becomes for our generalization of the mode, in particular, to latch on to a desirable solution.  Or to put it another way, the higher $M$, the lower $f$ beyond which {\color{black}RCR} fails.  {\color{black}This appears to be a fundamental limitation of our approach, and one that can be only partially mitigated by a (significantly) larger number of measurements.\footnote{\color{black}Other approaches can be envisioned, in which combinations of more than $M$ measurements are used to calculate model solutions, with RCR employed at this stage as well, to reduce the fraction of these that are contaminated.  However, this is beyond the scope of this paper.} 
		
		This can be seen by the greater degree of scatter in the $M=3$ parameter-space plots in Figure~38, compared to that of the $M=2$ parameter-space plots in Figures 34 and 36, and by the fact that this greater degree of scatter could not be successfully resolved in the $f=0.85$ row in Figure~39, despite the contaminants not biasing the calculated parameter values in a systematic direction, as they did in Figures 36 and 37.  (See Figures 41 and 42 for another $M=3$ example, with similar results.)\\
		
		\subsubsection{Avoid Introducing Unnecessary Correlations between Calculated Parameter Values through Good Model Design}}
	
	Naturally, our generalization of the mode, in particular, is most effective if the uncontaminated subset of $\left\{\left\{\theta\right\}_j\right\}$ is maximally concentrated.  {\color{black}However, this can depend on how wisely, or poorly, one constructs their model.  
		
		For} example, consider {\color{black}a linear model, given by $y(x)=b+m\left(x-\overline{x}\right)$, with constant RMS scatter, $\sigma_y(x)=\sigma_y$.  In this case, $\overline{x}$ is usually given by $\overline{x}=\sum_iw_ix_i/\sum_iw_i$, which} results in a largely uncorrelated, near-maximally concentrated distribution of, at least the highest-weight, $b$ vs.\@ $m$ values {\color{black}(e.g., Figures 34 and 35).\footnote{\color{black}We calculate $\overline{x}$ using only non-rejected measurements, and consequently, we update $\overline{x}$ after each iteration of the RCR algorithm.}  
		
		However, a significantly different choice for $\overline{x}$ would introduce a correlation between the calculated values of $b$ and $m$, resulting in a dispersed, and hence} not near-maximally concentrated, distribution {\color{black}(we demonstrate this for a different, but similar, case in the bottom row of Figure~40; see below).  This can make our generalization of the mode, in particular, and hence RCR, less precise}, and in this case, {\color{black}unnecessarily.
		
		Note, this is not always the best expression for $\overline{x}$.  For example, consider either an exponential model, given by $y(x)=be^{m(x-\overline{x})}$, or a power-law model, given by $y(x)=b\left(x/e^{\overline{\ln{x}}}\right)^m$, with constant RMS scatter, $\sigma_y(x)=\sigma_y$.  If $m<0$, high-$x$ measurements may be scatter-dominated and not contribute significantly to the fit, and consequently should not contribute significantly to $\overline{x}$ (and vice versa if $m>0$, with low-$x$ measurements).  However, if linearized (\textsection8.2.3), resulting in $\ln{y(x_i)}$ vs.\@ $x_i$ data for the exponential model and $\ln{y(x_i)}$ vs.\@ $\ln{x_i}$ data for the power-law model, all measurements would contribute to the fit, but with additional weights given by $\sigma_{\ln{y}}^{-2}(x_i)\propto y^2(x_i)$ (\textsection8.2.3).  Hence, we take $\overline{x}=\sum_iw_ix_iy^2(x_i)/\sum_iw_iy^2(x_i)$ for the exponential model, and $\overline{\ln{x}}=\sum_iw_i(\ln{x_i})y^2(x_i)/\sum_iw_iy^2(x_i)$ for the power-law model (whether the model has been linearized or not).\footnote{\color{black}Here, $\overline{x}$ and $\overline{\ln{x}}$ additionally depend on model parameters, through $y(x)$.  But as we do when calculating parameter weights in \textsection8.2.3, we use the parameter values of the most recent baseline model, from the most recent iteration of the RCR algorithm.  This should be significantly more accurate than using, say, individual $y_i$ measurements for $y(x_i)$.}  We demonstrate the effectiveness of this in Figure~40.
		
		\begin{figure*}
			\plotone{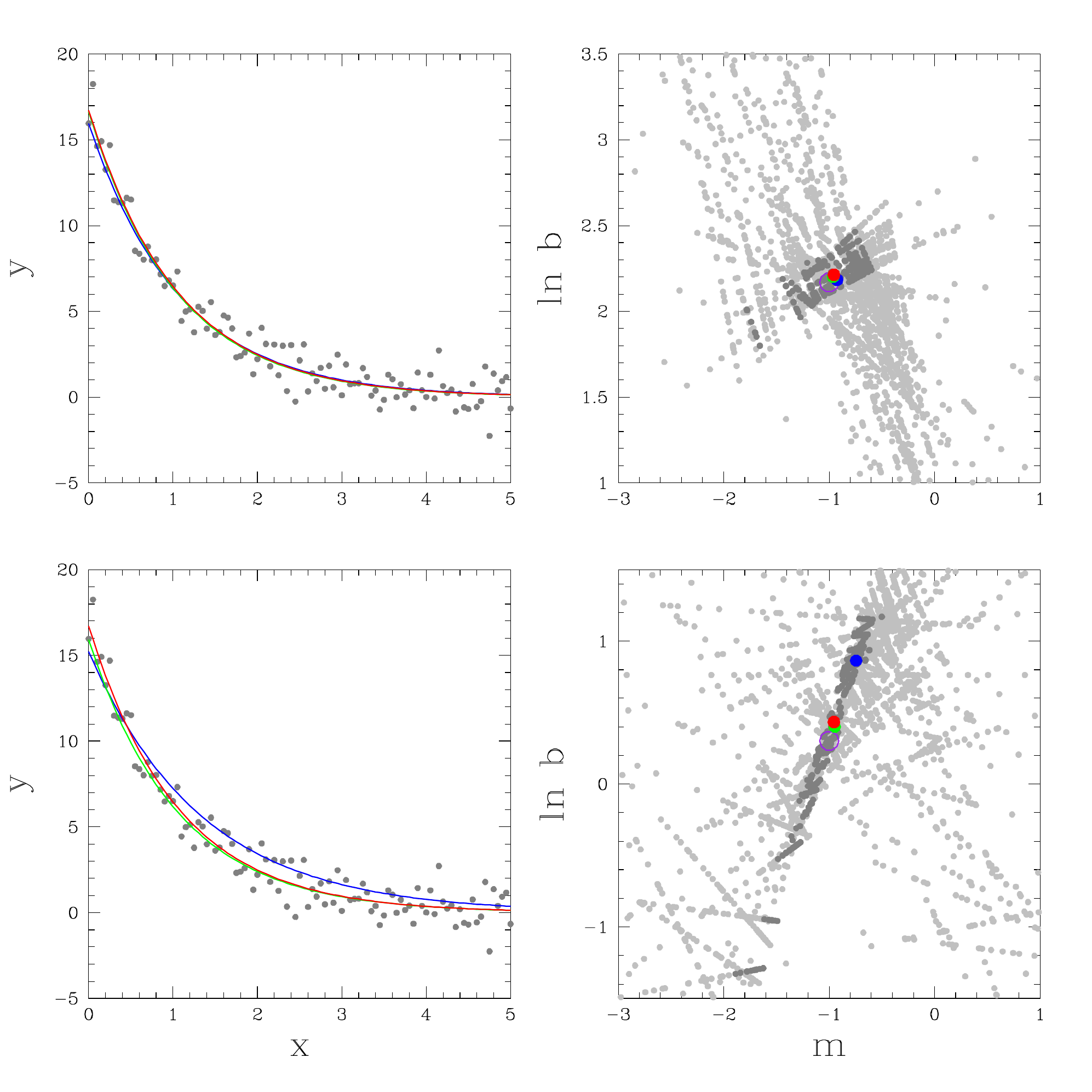}
			\caption{\textbf{Left column:}  101 uncontaminated measurements, drawn from a Gaussian distribution of mean $y(x)=10e^{-\left(x-0.5\right)}$ and standard deviation 1.  \textbf{Right column:}  {\color{black}Model solutions}, $\left\{\theta\right\}_j$, calculated from each pair of measurements in the panel to the left, using $y(x)=be^{m\left(x-\overline{x}\right)}$, {\color{black}with $\overline{x}$ calculated as described below}, and {\color{black}for} model parameters $\ln{b}$ and $m$ (see {\color{black}\textsection8.3.6}).  Each calculated parameter value is weighted {\color{black}(\textsection8.2.2),} and darker points correspond to models where the product of these weights is in the top 50\%.  {\color{black}The purple circle} corresponds to the original, underlying model, and in both columns, blue corresponds to the weighted mode of $\left\{\left\{\theta\right\}_j\right\}$, green corresponds to the weighted median of $\left\{\left\{\theta\right\}_j\right\}$, and red corresponds to maximum-likelihood model fitting.  \textbf{Top row:}  Here, we additionally weight each $x_i$ by $\sigma_{\ln{y}}^{-2} \left(x_i\right)\propto y^2 \left(x_i\right)$ when calculating {\color{black}$\overline{x}$, which} results in a fairly uncorrelated/fairly concentrated distribution for the highest-weight $\ln{b}$ vs\@. $m$ values, and consequently, the weighted mode of $\left\{\left\{\theta\right\}_j\right\}$, in particular, is less susceptible to imprecision.  \textbf{Bottom row:}  Here, we do not additionally weight each $x_i$ when calculating $\overline{x}$, which results in a strongly correlated/dispersed distribution for the highest-weight $\ln{b}$ vs.\@ $m$ values, and consequently, the weighted mode of $\left\{\left\{\theta\right\}_j\right\}$, in particular, is more susceptible to imprecision.}
		\end{figure*}
		
		Sometimes, however, these} correlations cannot be avoided.  For example, if presented with quadratic $y_i$ vs.\@ $x_i$ data, one can design away correlations between two {\color{black}of the three pairings} of the model's three parameters, but not between all three {\color{black}pairings} simultaneously:  If one models {\color{black}these} data with $y(x)=b+m_1\left(x-\overline{x}\right)+m_2\left(x-\overline{x}\right)^2$, {\color{black}with $\overline{x}=\sum_iw_ix_i/\sum_iw_i$,} both (1)~the highest-weight $b$ vs.\@ $m_1$ values and (2)~the highest-weight $m_1$ vs.\@ $m_2$ values will, for the most part, be uncorrelated, but the highest-weight $b$ vs.\@ $m_2$ values will be marginally (negatively) correlated {\color{black}(see Figure~41)}.  Despite this, {\color{black}RCR} is still effective through fairly high contamination fractions, which we demonstrate in Figure~42.
	
	\begin{figure*}
		\plotone{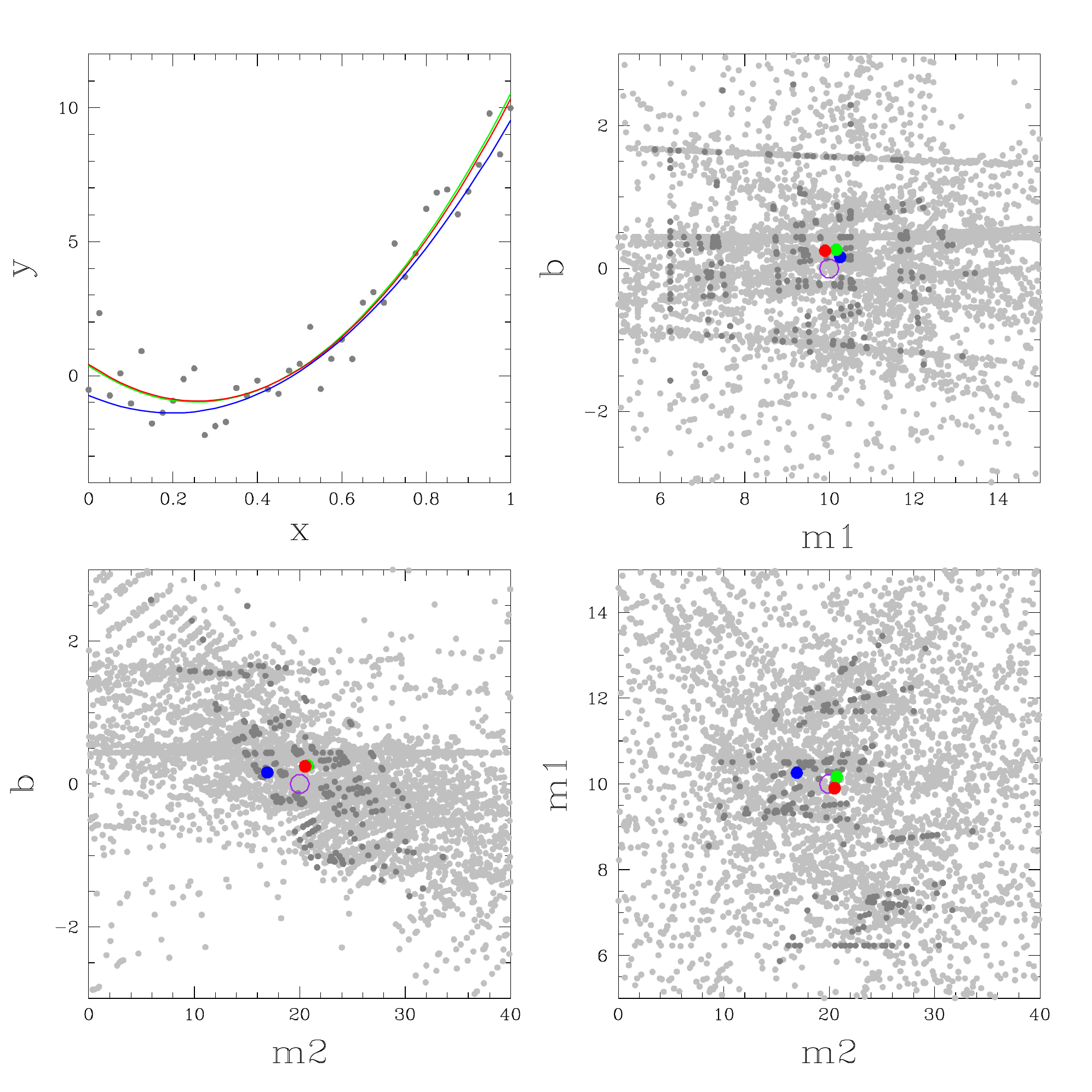}
		\caption{\textbf{Upper left:}  43 uncontaminated measurements, drawn from a Gaussian distribution of mean $y(x)=10\left(x-0.5\right)+20\left(x-0.5\right)^2$ and standard deviation 1.  \textbf{Upper right and bottom row:}  {\color{black}Model solutions}, $\left\{\theta\right\}_j$, calculated from each triplet of measurements in the upper-left panel, using $y(x)=b+m_1\left(x-\overline{x}\right)+m_2\left(x-\overline{x}\right)^2${\color{black}, with $\overline{x}=\sum_iw_ix_i/\sum_iw_i$,} and model parameters $b$, $m_1$, and $m_2$.  Each calculated parameter value is weighted {\color{black}(\textsection8.2.2),} and darker points correspond to models where the product of these weights is in the top 50\%.  {\color{black}The purple circle} corresponds to the original, underlying model, and in all panels, blue corresponds to the weighted mode of $\left\{\left\{\theta\right\}_j\right\}$, green corresponds to the weighted median of $\left\{\left\{\theta\right\}_j\right\}$, and red corresponds to maximum-likelihood model fitting.  The highest-weight $b$ vs.\@ $m_1$ values, corresponding to linear $y_{eff}(x)\equiv y(x)-m_2\left(x-\overline{x}\right)^2= b+m_1\left(x-\overline{x}\right)$, and the highest-weight $m_1$ vs.\@ $m_2$ values, corresponding to linear $y_{eff}(x)\equiv\left[y(x)-b\right]/(x-\overline{x})= m_1+m_2\left(x-\overline{x}\right)$, are largely uncorrelated, but the highest-weight $b$ vs.\@ $m_2$ values, corresponding to non-linear $y_{eff}(x)\equiv y(x)-m_1\left(x-\overline{x}\right)= b+m_2\left(x-\overline{x}\right)^2$, are marginally, negatively correlated:  Since $\left(x-\overline{x}\right)^2$ is always positive, if $m_2$ is high, $b$ tends to be low, to compensate.}
	\end{figure*}
	
	\begin{figure*}
		\centering
		\includegraphics[height=0.9\textheight]{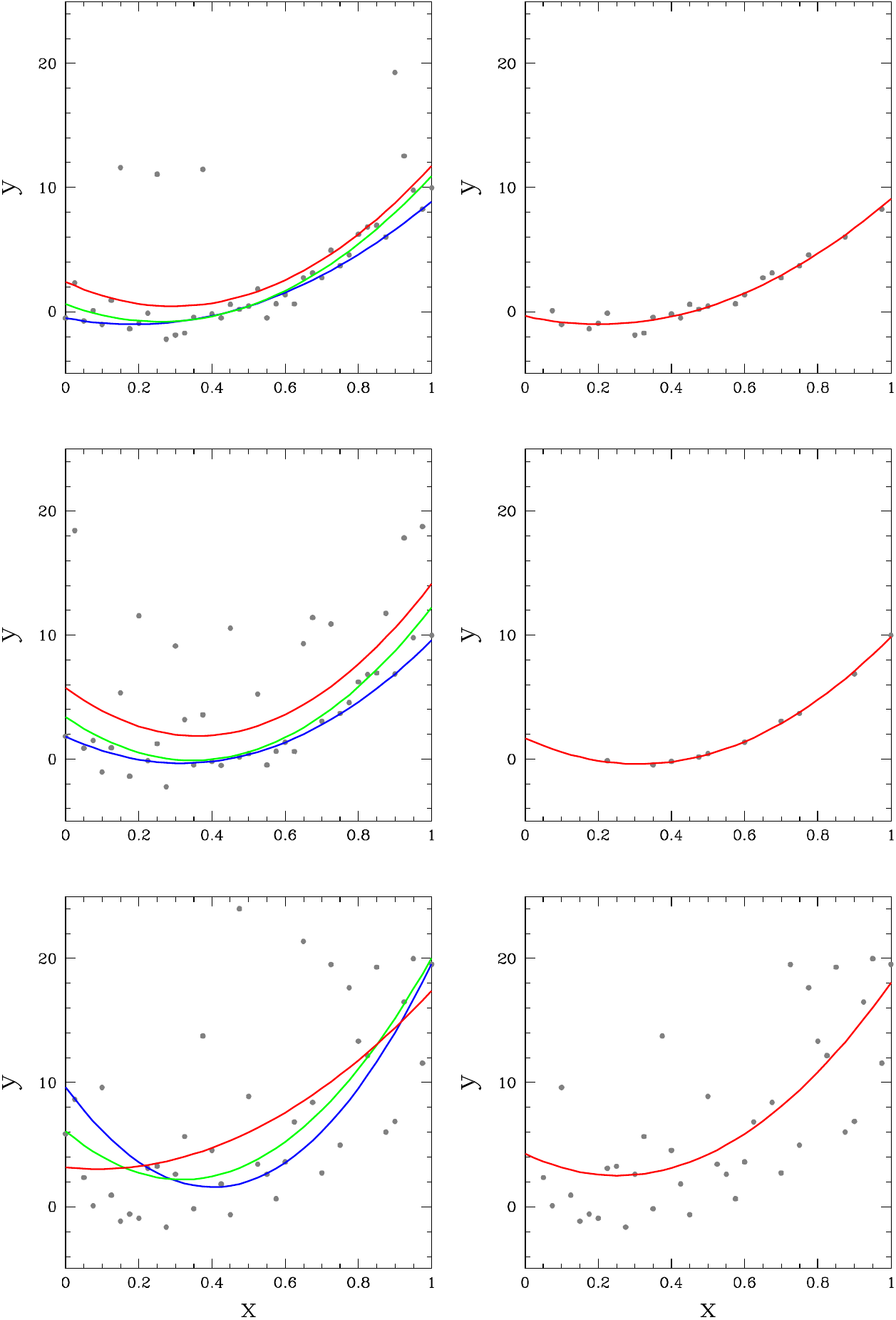}
		\caption{\textbf{Left column:}  43 measurements, with fraction $f_1=1-f_2$ drawn from a Gaussian distribution of mean $y(x)=10\left(x-0.5\right)+20\left(x-0.5\right)^2$ and standard deviation 1, and fraction $f_2=0.15$ (top row), 0.5 (middle row), and 0.85 (bottom row), representing contaminated measurements, drawn from the positive side of a Gaussian distribution of mean zero and standard deviation 10, and added to uncontaminated measurements, drawn as above.  Blue corresponds to the weighted mode of $\left\{\left\{\theta\right\}_j\right\}$, green corresponds to the weighted median of $\left\{\left\{\theta\right\}_j\right\}$, and red corresponds to maximum-likelihood model fitting.  \textbf{Right column:}  After {\color{black}RCR}.  Here, we have again performed bulk rejection {\color{black}as in \textsection5, but} using our generalization of the mode {\color{black}instead of the mode}, followed by {\color{black}individual rejection as in \textsection4, using (1)}~our most{\color{black}-}general robust technique for symmetrically distributed uncontaminated measurements -- {\color{black}now consisting of} our generalization of the mode $+$ technique~3 (the broken-line fit) -- followed by {\color{black}(2)}~our most{\color{black}-}precise robust technique -- {\color{black}now consisting of} our generalization of the median $+$ technique~1 (the 68.3\% value) -- followed by {\color{black}(3)~traditional} Chauvenet rejection{\color{black}, but} using our generalization of the mean {\color{black}instead of the mean (e.g., Figures 30, 28, and 19)}.  {\color{black}RCR} proves effective in the face of fairly heavy contamination, but is unable to overcome the greater fraction of contaminated models as $f_2\rightarrow0.85$ ({\color{black}\textsection8.3.4}).}
	\end{figure*}
	
	{\color{black}Of course, more sophisticated implementations can be imagined, in which one would not have to consider these correlations at all.  For example, instead of determining (1)~the weighted median of $\left\{\left\{\theta\right\}_j\right\}$ and especially (2)~the weighted mode of $\left\{\left\{\theta\right\}_j\right\}$ along the given parameter-space coordinate system, as we do in this paper (\textsection8.1), one could imagine doing this (or perhaps something else a bit more sophisticated) in a rotated, or even non-linearly transformed, coordinate system, with a principal axis determined (robustly) from the calculated parameter values and weights.  This is beyond the scope of the current work, but would be a natural next investigation.\\
		
		\subsubsection{Make Good Basis Decisions}}
	
	Another example of good model design is proper choice of basis.  For example, when {\color{black}fitting an exponential model, e.g., $y(x)=be^{m(x-\overline{x})}$, or a power-law model, e.g., $y(x)=b\left(x/e^{\overline{\ln{x}}}\right)^m$, to measurements, one usually calculates $\ln{b}$ and $m$, instead of $b$ and $m$ (e.g., Figures 36, 37, and 40).}  This is called choice of basis.  When performing maximum-likelihood model fitting, {\color{black}basis choices (like normalization choices; \textsection8.3.5) do} not affect the best fit, but {\color{black}good ones} can {\color{black}yield more} concentrated/symmetric probability {\color{black}distributions} for the {\color{black}model's parameters, and consequently,} more concentrated/symmetric error bars {\color{black}for these parameters.
		
		Similarly, good basis choices} can {\color{black}yield more} concentrated/symmetric {\color{black}distributions} of calculated parameter values, $\left\{\left\{\theta\right\}_j\right\}$.  While this does not affect the weighed median of $\left\{\left\{\theta\right\}_j\right\}$, {\color{black}as in \textsection8.3.5,} it can affect the weighted mode of {\color{black}$\left\{\left\{\theta\right\}_j\right\}$.  
		
		That said, as long as one's uncontaminated measurements are not scatter-dominated}, this is usually a very small effect, and {\color{black}multiple, equivalent parameterizations are perfectly acceptable.\footnote{\color{black}Another example is using $\tan{m}$ instead of $m$ for slopes when they are very large.}}
	
	Application of {\color{black}RCR} to parameterized models is potentially a very broad topic, with applications spanning not only science, but all {\color{black}quantitative} disciplines.  Here, we have but scratched the surface with a few, simple examples.\\
	
	\section{Peirce Rejection}
	
	{\color{black}Traditional} Chauvenet rejection is sigma clipping plus a rule for selecting a reasonable number of sigma for the threshold, given $N$ measurements (\textsection1).  It is straightforward to use, and as such has been adopted as standard by many government and industry laboratories, and is commonly taught at universities (Ross 2003).  However, it is not the only approach one might take to reject outliers.  For example, even Chauvenet (1863) deferred to Peirce's approach (1852; Gould 1855),\footnote{It is interesting to note that this topic, although statistical in nature, originates in the field of astronomy.  Two of these publications are in the early volumes of the Astronomical Journal, and the third, Chauvenet's ``A Manual of Spherical and Practical Astronomy", was a standard reference for decades.} which has recently seen new life with its own implementation in the R programming language (Dardis 2012).  Instead of assuming 0.5 in Equation 1, Peirce derives this value from probability theory, and finds (1)~that it is weakly dependent on $N$, asymptoting to 0.5 as $N$ increases, and (2)~that it decreases with subsequent rejections.  
	
	However, unlike Peirce's approach, Chauvenet rejection is amenable to (1)~$N$, (2)~the mean, and (3)~the standard deviation being updated after each rejection (\textsection1).  Peirce's approach requires all three of these quantities to remain fixed until all rejections have been completed, and as such Peirce rejection is less robust than Chauvenet rejection, at least when the latter is implemented iteratively, as we have done.
	
	Furthermore, our correction factors (\textsection2.3, \textsection4 -- \textsection6) empirically account for the above, weak dependence on $N$, as well as for differences in implementation (e.g., our use of one-sided deviation measurements, our use of robust quantities, etc.)  
	
	In Figures 43 -- 46, we compare Peirce rejection {\color{black}(1)~to traditional} Chauvenet rejection and (2)~{\color{black}to RCR}, for both two-sided and one-sided contaminants.  Given our iterative implementation, and our correction factors, we find {\color{black}traditional} Chauvenet rejection to be comparable to Peirce rejection when the contaminants are two-sided and $N$ is low, and better than Peirce rejection otherwise.  {\color{black}RCR} is significantly better than both of these approaches.\\
	
	\begin{sidewaysfigure*}
		\centering
		\vspace{3.75in}
		\includegraphics[width=0.9\textwidth]{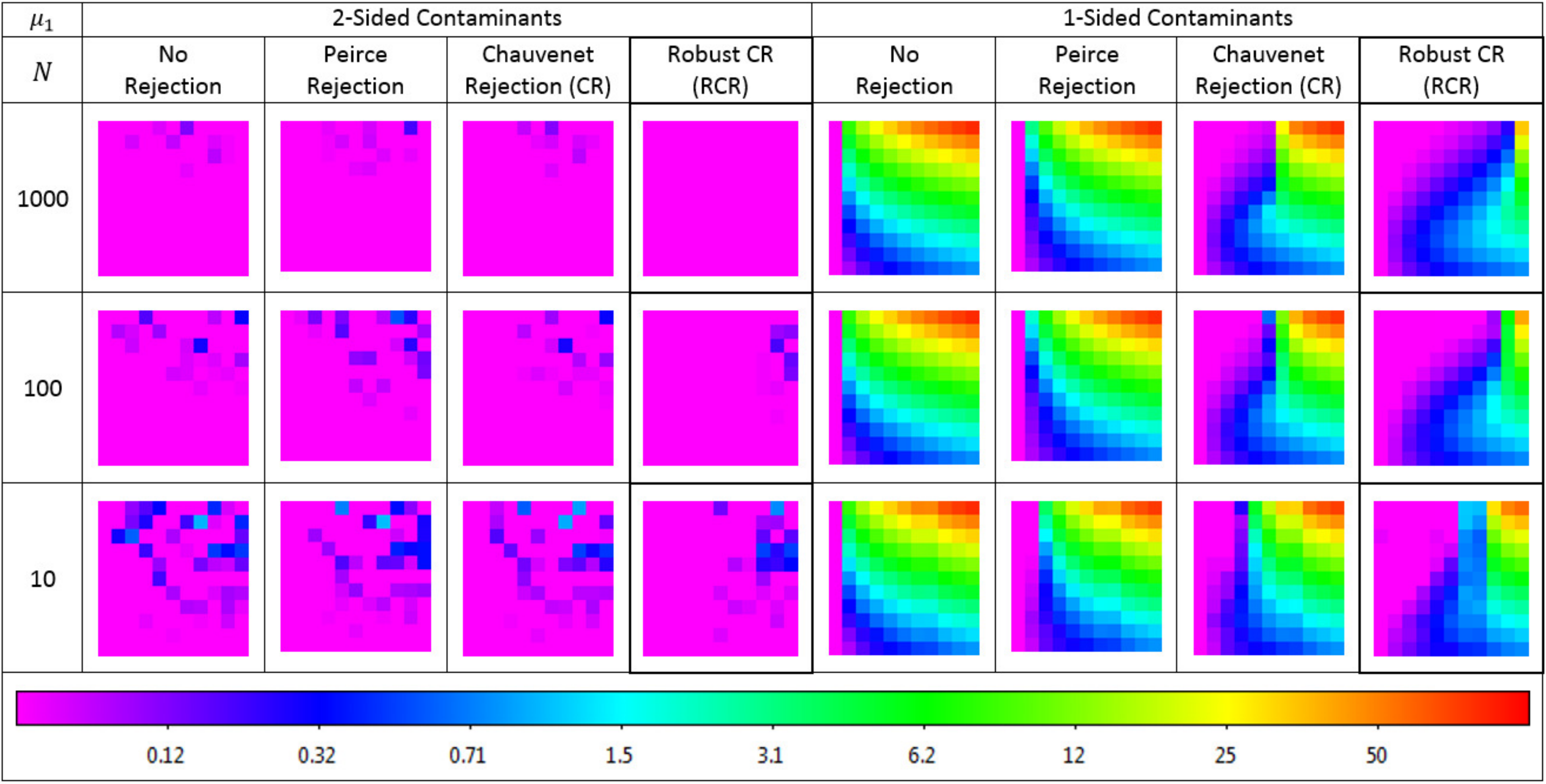}
		\caption{Average recovered $\mu_1$ for (1)~no rejection, (2)~Peirce rejection, (3)~{\color{black}traditional} Chauvenet rejection, and (4)~{\color{black}RCR}, for two-sided contaminants (left) and one-sided contaminants (right).  See Figure~5 for contaminant strength ($\sigma_2$) vs.\@ fraction of sample ($f_2$) axis information.  In the case of two-sided contaminants, all techniques recover $\mu_1\approx0$ (Figure~6).  In the case of one-sided contaminants, {\color{black}traditional} Chauvenet rejection, as implemented in this paper, is superior to Peirce rejection, and {\color{black}RCR (highlighted with a bold outline)} is superior to {\color{black}traditional} Chauvenet rejection.  The colors are scaled logarithmically, between 0.02 and 100.}
	\end{sidewaysfigure*}
	
	\begin{sidewaysfigure*}
		\centering
		\vspace{3.75in}
		\includegraphics[width=0.9\textwidth]{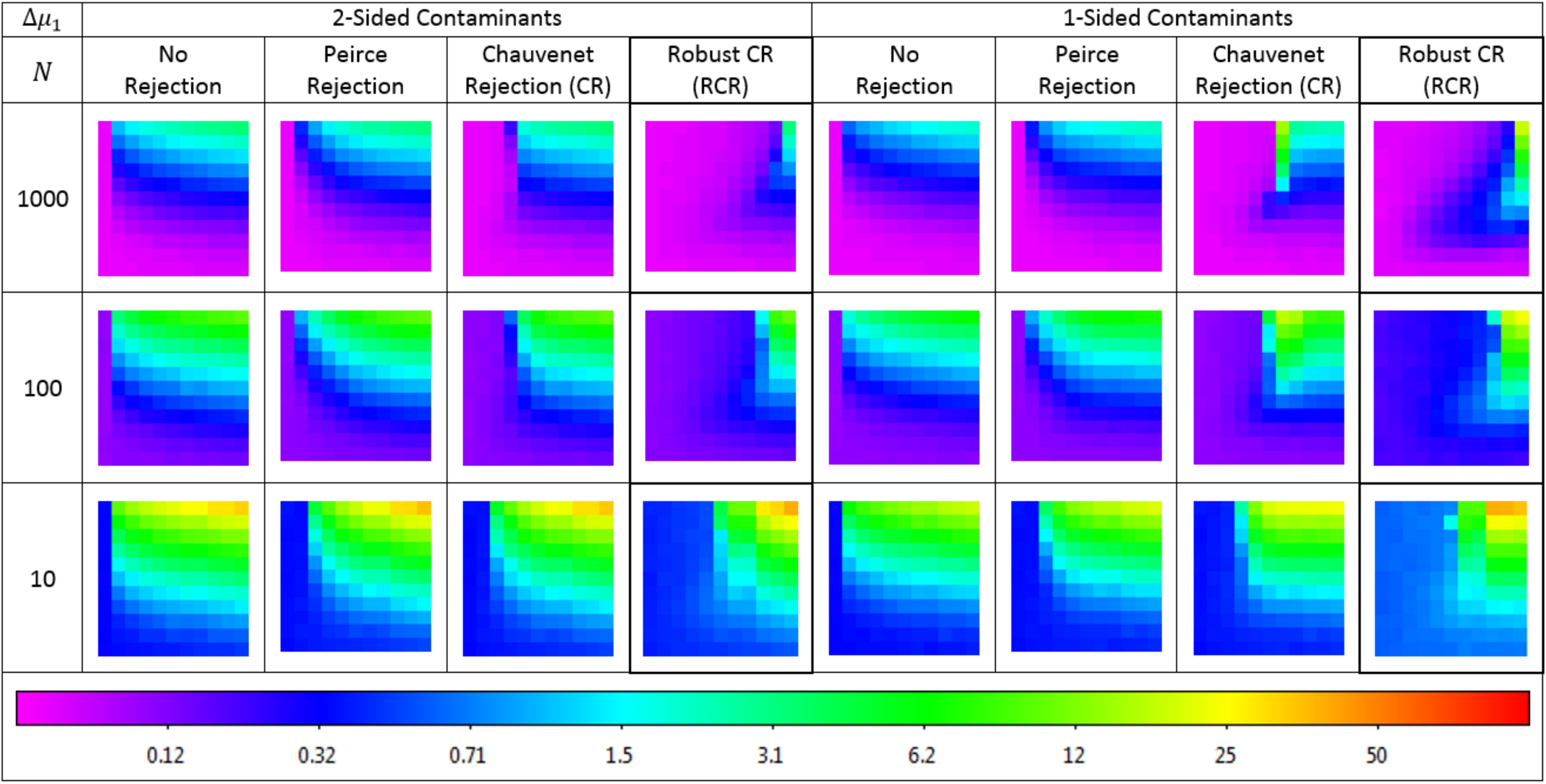}
		\caption{Average recovered $\Delta\mu_1$ for (1)~no rejection, (2)~Peirce rejection, (3)~{\color{black}traditional} Chauvenet rejection, and (4)~{\color{black}RCR}, for two-sided contaminants (left) and one-sided contaminants (right).  See Figure~5 for contaminant strength ($\sigma_2$) vs.\@ fraction of sample ($f_2$) axis information.  In the case of two-sided contaminants, all techniques recover $\mu_1\approx0$ (Figure~6).  In the case of one-sided contaminants, {\color{black}traditional} Chauvenet rejection, as implemented in this paper, is superior to Peirce rejection, and {\color{black}RCR (highlighted with a bold outline)} is superior to {\color{black}traditional} Chauvenet rejection, albeit with marginally reduced precision at low $N$ (\textsection4).  The colors are scaled logarithmically, between 0.02 and 100.}
	\end{sidewaysfigure*}
	
	\begin{sidewaysfigure*}
		\centering
		\vspace{3.75in}
		\includegraphics[width=0.9\textwidth]{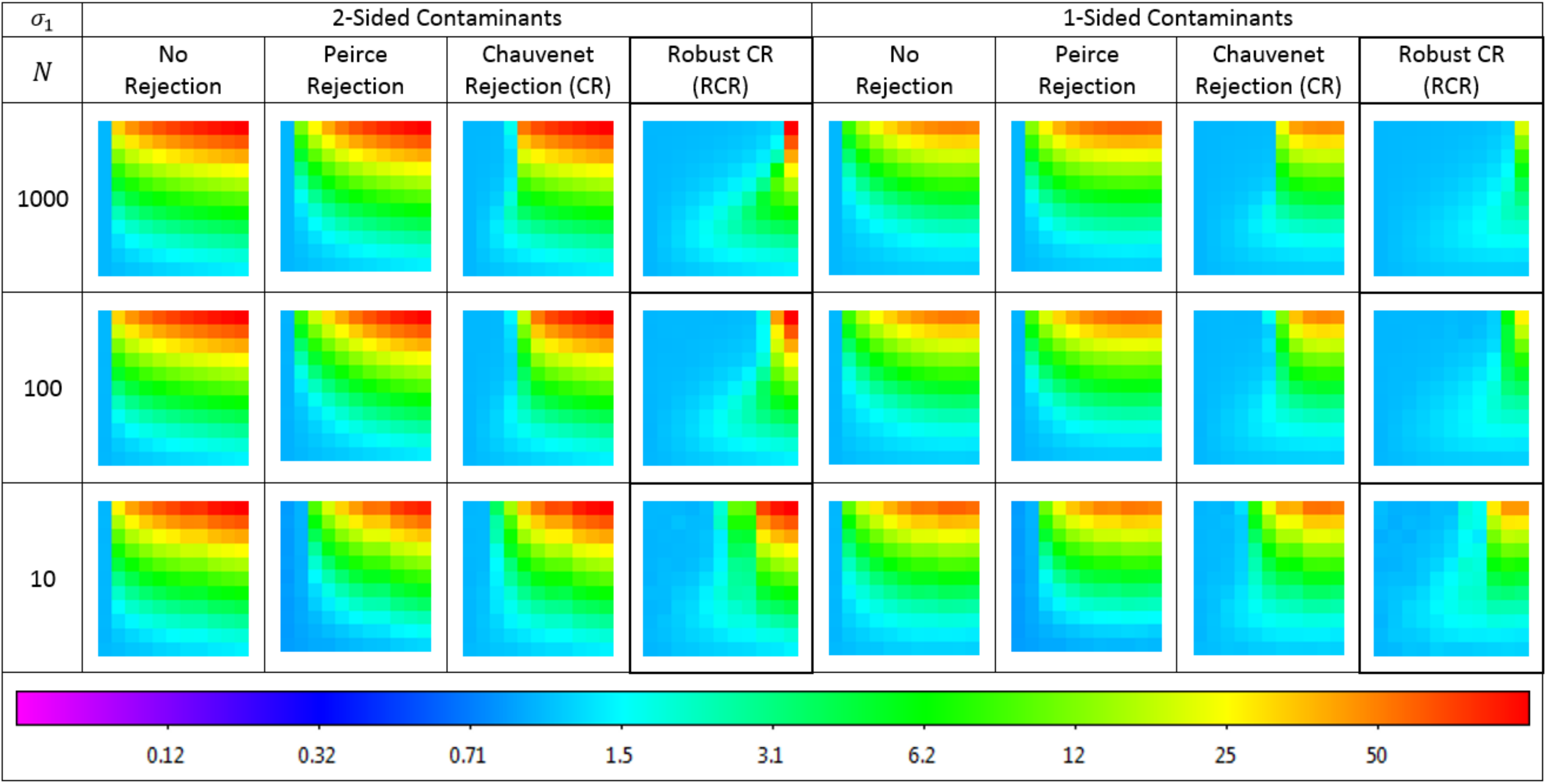}
		\caption{Average recovered $\sigma_1$ for (1)~no rejection, (2)~Peirce rejection, (3)~{\color{black}traditional} Chauvenet rejection, and (4)~{\color{black}RCR}, for two-sided contaminants (left) and one-sided contaminants (right).  See Figure~5 for contaminant strength ($\sigma_2$) vs.\@ fraction of sample ($f_2$) axis information.  In the case of two-sided contaminants, all techniques recover $\mu_1\approx0$ (Figure~6).  In the case of one-sided contaminants, {\color{black}traditional} Chauvenet rejection, as implemented in this paper, is superior to Peirce rejection, and {\color{black}RCR (highlighted with a bold outline)} is superior to {\color{black}traditional} Chauvenet rejection.  The colors are scaled logarithmically, between 0.02 and 100.}
	\end{sidewaysfigure*}
	
	\begin{sidewaysfigure*}
		\centering
		\vspace{3.75in}
		\includegraphics[width=0.9\textwidth]{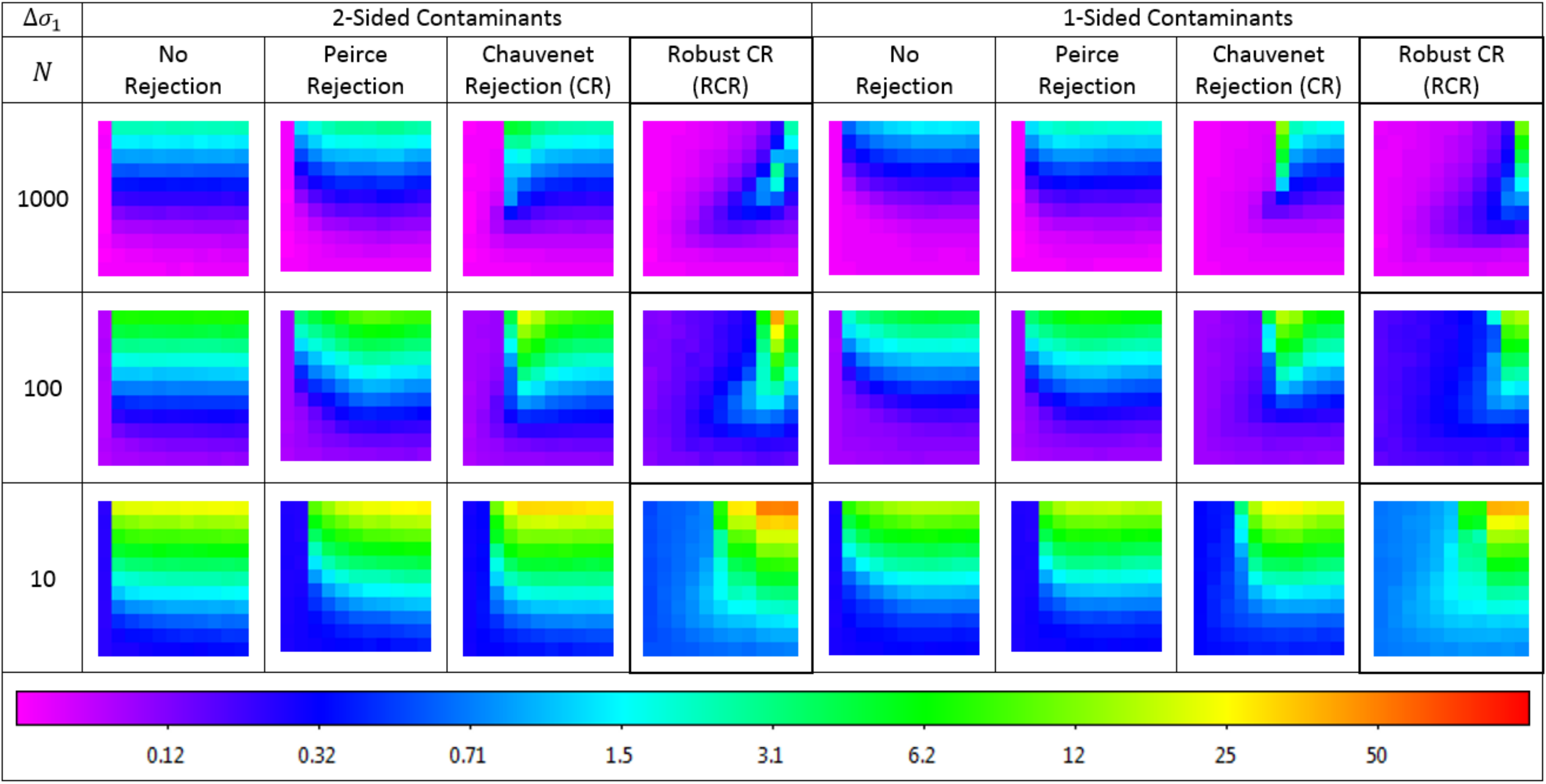}
		\caption{Average recovered $\Delta\sigma_1$ for (1)~no rejection, (2)~Peirce rejection, (3)~{\color{black}traditional} Chauvenet rejection, and (4)~{\color{black}RCR}, for two-sided contaminants (left) and one-sided contaminants (right).  See Figure~5 for contaminant strength ($\sigma_2$) vs.\@ fraction of sample ($f_2$) axis information.  In the case of two-sided contaminants, all techniques recover $\mu_1\approx0$ (Figure~6).  In the case of one-sided contaminants, {\color{black}traditional} Chauvenet rejection, as implemented in this paper, is superior to Peirce rejection, and {\color{black}RCR (highlighted with a bold outline)} is superior to {\color{black}traditional} Chauvenet rejection, albeit with marginally reduced precision at low $N$ (\textsection4).  The colors are scaled logarithmically, between 0.02 and 100.}
	\end{sidewaysfigure*}
	
	\section{Summary}
	
	The most fundamental act in science is measurement.  By combining multiple measurements, one can better constrain a quantity's true value, and its uncertainty.  However, measurements, and consequently samples of measurements, can be contaminated.  Here, we have introduced, and thoroughly tested, {\color{black}an approach} that, while not perfect, is very effective at identifying which measurements in a sample are contaminated, even if they constitute most of the sample, and especially if the contaminants are strong (making contaminated measurements easier to identify).
	
	{\color{black}In particular, we} have considered{\color{black}:}
	
	\begin{itemize}
		
		\item {\color{black}Both} symmetrically (\textsection3.1) and asymmetrically (\textsection3.2) distributed contamination of both symmetrically (\textsection3.1, \textsection3.2{\color{black}, \textsection3.3.2}) and asymmetrically (\textsection3.3.1) distributed uncontaminated measurements, {\color{black}and have developed robust outlier rejection techniques for all combinations of these cases.}
		
		\item {\color{black}The tradeoff between these techniques' accuracy and precision, and have found that by applying them in sequence, from more robust to more precise, both can be achieved} (\textsection4).  
		
		\item {\color{black}The} practical cases of bulk rejection (\textsection5), weighted data (\textsection6), and model fitting (\textsection8), {\color{black}and have generalized the RCR algorithm accordingly.}
	\end{itemize}
	
	Finally, we have developed a simple web interface {\color{black}so anyone can use the RCR algorithm}.\footnote{https://skynet.unc.edu/rcr}   Users may upload a data set, and select from the above scenarios.  They are returned their data set with outliers flagged, and with $\mu_1$ and $\sigma_1$ robustly measured.  Source code is available here as well.
	
	
	
	\acknowledgments
	
	We gratefully acknowledge the support of the National Science Foundation, through the following programs and awards:  ESP 0943305, MRI-R$^2$ 0959447, AAG 1009052, 1211782, and 1517030, ISE 1223235, HBCU-UP 1238809, TUES 1245383, and STEM$+$C 1640131.  We are also appreciative to have been supported by the Mt.\@ Cuba Astronomical Foundation, the Robert Martin Ayers Sciences Fund, and the North Carolina Space Grant Consortium.  {\color{black}We also thank the referee, and the editor, for comments that helped us to improve this paper considerably.}





\appendix

\section{A. Broken-Line Fit Through Origin}

Let $x_i = \sqrt{2}\mathrm{erf}^{-1}[(i-0.317)/N]$ and $y_i=\delta_i$.  We model these data as a broken line that passes through the origin:

\begin{equation}
y = 
\begin{cases}
\sigma_1x, & \text{if $i\leq m$} \\
\sigma_1x_m+\sigma_2(x-x_m), & \text{if $i\geq m$ and $x_i\leq1$}
\end{cases}
,
\end{equation}

\noindent where $\sigma_1$ is the slope of the line for $i\leq m$, and our modeled 68.3-percentile deviation, and $\sigma_2$ is the slope of the line for $i\geq m$ and $x_i\leq1$.  We model the break to occur at $x_m$, instead of between points, for simplicity. 

Let the fitness of this three-parameter model be measured by:

\begin{equation}
\chi_3^2 = \sum\limits_{i=1}^{N'}{[y(x_i|\sigma_1,\sigma_2,m)-y_i]^2},
\end{equation}

\noindent where $N'={\rm floor}(0.683N+0.317)$ is the number of points for which $x_i\leq1$.  Then for a given break point, $m$, the best fit is given by $d\chi_3^2/d\sigma_1=d\chi_3^2/d\sigma_2=0$, yielding:

\begin{equation}
\left[ 
\begin{array}{c}
\sigma_1 \\
\sigma_2 \\
\end{array} 
\right]
=
\left[
\begin{array}{cc}
\sum\limits_{i=1}^{m}x_i^2 + x_m^2\sum\limits_{i=m+1}^{N'}{1} & x_m\sum\limits_{i=m+1}^{N'}{(x_i-x_m)} \\
x_m\sum\limits_{i=m+1}^{N'}{(x_i-x_m)} & \sum\limits_{i=m+1}^{N'}{(x_i-x_m)^2} \\
\end{array}
\right] ^{-1}
\left[
\begin{array}{c}
\sum\limits_{i=1}^{m}{x_iy_i + x_m}\sum\limits_{i=m+1}^{N'}{y_i} \\
\sum\limits_{i=m+1}^{N'}{(x_i-x_m)y_i} \\
\end{array}
\right].
\end{equation}

\noindent We use a recursive partitioning algorithm to efficiently find the value of $m$ for which $\chi_3^2$ is minimized.  We restrict $m>1$ to avoid the following pathological case:  If $\mu$ is measured by the median and $N$ is odd, one of the measured values will always equal the median value, and consequently $y_1$ will always be zero; $m=1$ would then imply $\sigma_1=0$, but without meaning.  

Statistical equivalence to an unbroken-line fit through the origin (\textsection2.2) can similarly result in spurious values of $\sigma_1$:  In this case, any fitted value of $m$ is possible, and the lower it happens to be, the less well constrained $\sigma_1$ will be.  Consequently, if statistically equivalent, we instead use the value of $\sigma$ from the unbroken-line fit through the origin.  We measure statistical equivalency by:

\begin{equation}
\frac{\chi_1^2-\chi_3^2}{\chi_3^2}<f(N),
\end{equation}

\noindent where $\chi_1^2$ is the fitness of the one-parameter, unbroken-line fit, and $f(N)$ would be $\approx$2.3$N'$ if the points to which we are fitting were statistically independent of each other, but because we are fitting to sorted data, this is not the case.  Consequently, we determined $f(N)$ empirically:  For each technique and sample size $N$, we drew 100,000 uncontaminated samples, measured $(\chi_1^2-\chi_3^2)/\chi_3^2$ for each, sorted these values, and took the 68.3-percentile value (see Figure~47).  

\begin{figure*}
	\plotone{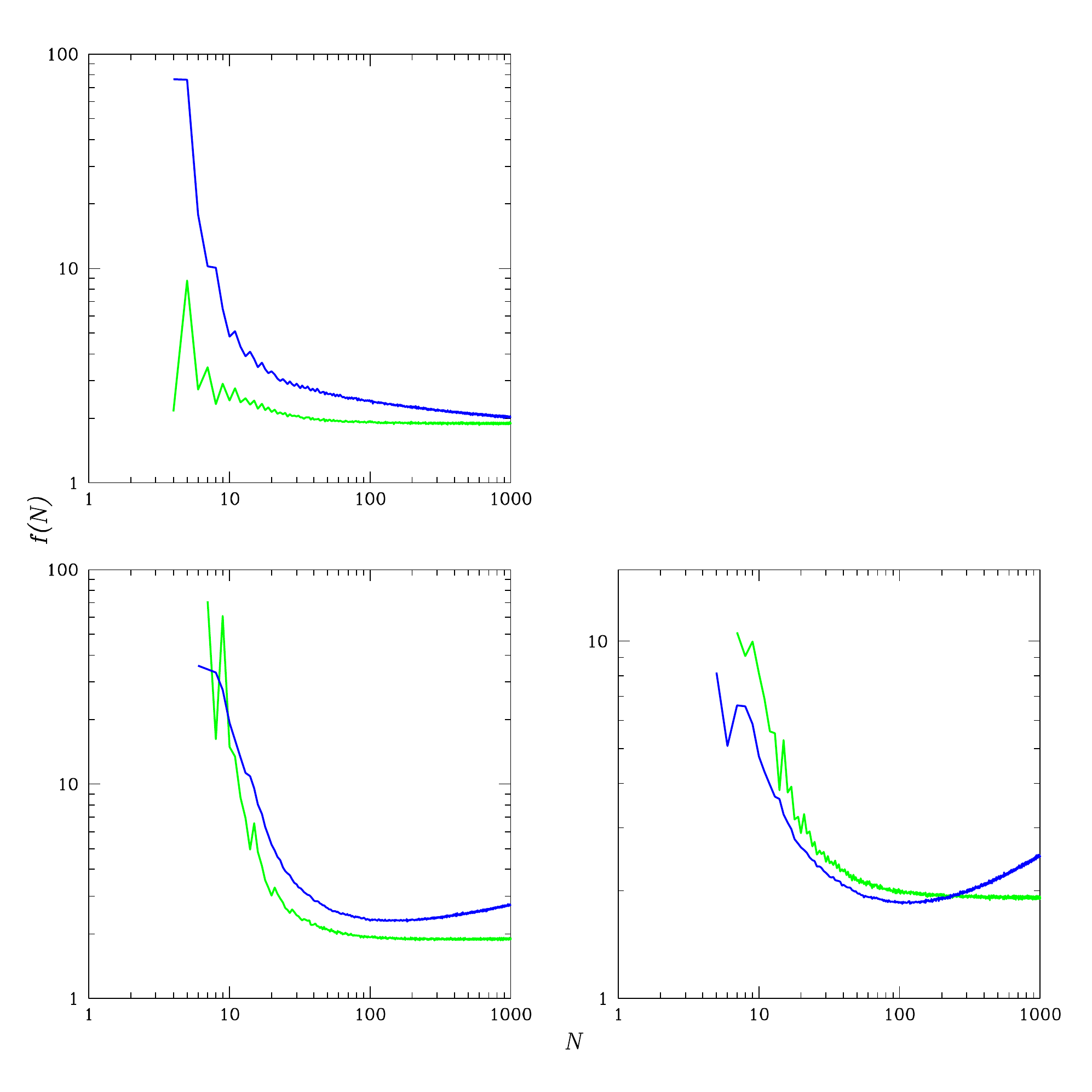}
	\caption{$f(N)$ vs.\@ $N$, for sorted, but otherwise independent data, where deviations have been measured from (1)~the median (green) and (2)~the mode (blue).  \textbf{Upper left:}  For the simplest case of computing a single $\sigma_1$, using the deviations both below and above $\mu$ (the median or the mode; \textsection3.1).  \textbf{Lower left:}  For the case of computing separate $\sigma_1$ below and above $\mu$ and computing $(\chi_1^2-\chi_3^2)/\chi_3^2$ for only the smaller of the two (\textsection3.2).  \textbf{Lower right:}  For the same case, but computing $(\chi_1^2-\chi_3^2)/\chi_3^2$ for either of the two, selected randomly (\textsection3.3.1).  Oscillations are not noise, but odd-even effects (e.g., with equally weighted data, when $N$ is odd, use of the median always results in at least one zero deviation, resulting in a larger value of $(\chi_1^2-\chi_3^2)/\chi_3^2$.  We use look-up tables for $N\leq 1000$ and empirical approximations for $f(N>1000)$ (see Appendix B).}
\end{figure*}

\begin{figure*}
	\plotone{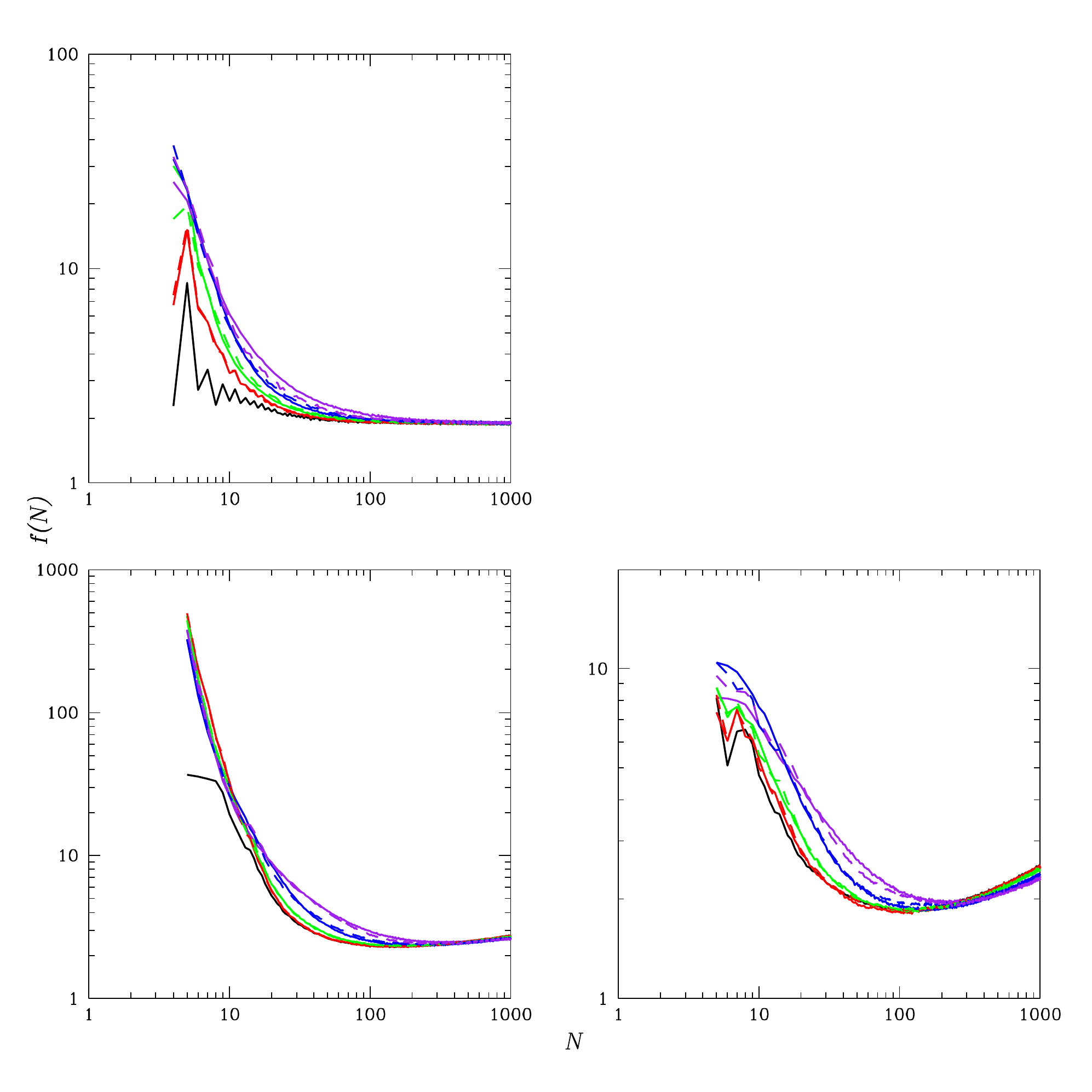}
	\caption{Same as the green, blue, and blue curves from the upper-left, lower-left, and lower-right panels of Figure~47, respectively, corresponding to what is needed for our best-option robust techniques (\textsection4), but for five representative weight distributions:  (1)~all weights equal (solid black curves -- same as the designated curves from Figure~47); (2)~weights distributed normally with standard deviation as a fraction of the mean $\sigma_w/\mu_w=0.1$ (solid red curves); (3)~weights distributed normally with $\sigma_w/\mu_w=0.3$ (solid green curves); (4)~weights distributed uniformly from zero (i.e., low-weight points as common as high-weight points; solid blue curves), corresponding to $\sigma_w/\mu_w\approx0.58$; and (5)~weights distributed inversely over one dex (i.e., low-weight points more common than high-weight points, with the sum of the weights of the low-weight points as impactful as the sum of the weights of the high-weight points; solid purple curves), corresponding to $\sigma_w/\mu_w\approx0.73$.  From these, we have produced empirical approximations, as functions of (1)~$N$ and (2)~$\sigma_w/\mu_w$ of the $x_i=\sqrt{2}\mathrm{erf}^{-1}(s_i/\sum_{i=1}^{N}w_i)<1$ points, which can be used with any sample of similarly distributed weights (dashed curves; see Appendix B).}
\end{figure*}

Otherwise, we only restrict $\sigma_1$ to be positive.  

Finally, when fitting to weighted data (\textsection6):
\begin{itemize}
	\item $x_i$ is instead given by $\sqrt{2}\mathrm{erf}^{-1}(s_i/\sum_{i=1}^{N}w_i)$, where $s_i$ is given by Equation~18, and $w_i$ is the weight of the $i$th point;  
	\item $N'$ is instead given by the largest integer such that $s_{N'}\leq 0.683\sum_{i=1}^{N}w_i$;
	\item Each term summed over $i$ in $\chi_1^2$ and $\chi_3^2$ (Equation~A2) and Equation~A3 is multiplied by $w_i$; and
	\item $f(N)$ instead depends on the weights of the data.  To this end, for the three scenarios that we consider in \textsection6 that make use of technique~3 (the broken-line fit), corresponding to all but the upper-right panel of Figure~31, we have computed $f(N)$ for the same five, representative weight distributions (see Figure~48, solid curves).  From these, we have {\color{black}similarly} produced empirical approximations, as functions of (1)~$N$ and (2)~$\sigma_w/\mu_w$ of the $x_i=\sqrt{2}\mathrm{erf}^{-1}(s_i/\sum_{i=1}^{N}w_i)<1$ points, which can be used with any sample of similarly distributed weights (Figure~48, dashed curves; see Appendix B).  We demonstrate these for the latter three weight distributions in columns 6 -- 8, respectively, of Figures 20 -- 27, and, desirably, they do not differ significantly from those of column~5, in which $\sigma_w/\mu_w=0$.  
\end{itemize}

\section{B. Empirical Approximations for $f(N)$ and Correction Factors}

For each scenario and technique presented in this paper, we calculated $f(N)$ (Appendix A) beyond $N=1000$ and correction factors ({\color{black}Figures 4, 29, and 31}) beyond $N=100$, every 0.1 dex for an additional 1 -- 2 dex, until it became computationally inefficient to continue.  

For the cases involving data of equal weight, we fitted functions of $N$ to these calculated values, yielding empirical approximations (Tables 2 -- 5).  

For the cases involving a distribution of weights, characterized by $\sigma_w/\mu_w$ of the $x_i=\sqrt{2}\mathrm{erf}^{-1}(s_i/\sum_{i=1}^{N}w_i)<1$ points (\textsection6, Appendix A), we fitted functions of both (1)~$N$ and (2)~$\sigma_w/\mu_w$ to the calculated values, yielding empirical approximations that can be used with any sample of similarly distributed weights.

For the ``single $\sigma$'' scenario, appropriate for two-sided contaminants (\textsection3.1), $f(N)$ is given by:
\begin{equation}
f(N) = f_1(N)10^{10^{a_1(N)+b_1(N)\log_{10}\left(\frac{\sigma_w}{\mu_w}\right)}},
\end{equation}
where $f_1(N)$ is the value of $f(N)$ for data of equal weight, corresponding to $\sigma_w/\mu_w=0$, and $a_1(N\leq7)$ and $b_1(N\leq7)$ are listed in Table~6.  For $7<N\leq1000$:
\begin{equation}
\begin{aligned}
a_1(N) = 0.2313(\log_{10}N)^6 - 3.02(\log_{10}N)^5 & + 15.997(\log_{10}N)^4 - 43.713(\log_{10}N)^3 + 64.629(\log_{10}N)^2 \\
& - 49.976\log_{10}N + 15.484 + 0.1513(-1)^N N^{-0.471},
\end{aligned}
\end{equation}
and
\begin{equation}
\begin{aligned} 
b_1(N)=-0.3556(\log_{10}N)^6+3.7036(\log_{10}N)^5-14.932(\log_{10}N)^4 & +29.176(\log_{10}N)^3-28.81(\log_{10}N)^2 \\
& +14.397\log_{10}N-2.64511.
\end{aligned}
\end{equation}
For $N > 1000$, $f(N)= f_1(N)$.

For this scenario and the ``bulk rejection (median) + RCR (median-T3) + RCR (median-T1) + CR'' technique, the correction factor is given by:
\begin{equation}
CF(N) = 
\begin{cases}
CF_1(N)10^{10^{a(N)+b(N)\log_{10}\left(\frac{\sigma_w}{\mu_w}\right)}} & \text{$N=2$, $N=4$, $N>5$} \\
CF_1(N)10^{-10^{a(N)+b(N)\log_{10}\left(\frac{\sigma_w}{\mu_w}\right)}} & \text{$N=3$, $N=5$}
\end{cases}
,
\end{equation}
where $CF_1(N)$ is the value of $CF(N)$ for data of equal weight, corresponding to $\sigma_w/\mu_w=0$, and $a(N\leq5)$ and $b(N\leq5)$ are listed in Table~7.  For $N>5$:
\begin{equation}
a(N)=-0.7914\log_{10}N+0.0243
\end{equation}
and
\begin{equation}
b(N)=0.1196\log_{10}N+4.5073.
\end{equation}

For the ``smaller of $\sigma_-$ and $\sigma_+$'' scenario, appropriate for one-sided contaminants and in-between cases (\textsection3.2), $f(N\leq264)$ is given by Equation B1, $f(264<N\leq567)$ is given by:
\begin{equation}
f(N)=
\begin{cases}
f_1(N)10^{10^{a_1(N)+b_1(N)\log_{10}\left(\frac{\sigma_w}{\mu_w}\right)}} & \text{$a_1(N)+b_1(N)\log_{10}\left(\frac{\sigma_w}{\mu_w}\right)>a_2(N)+b_2(N)\log_{10}\left(\frac{\sigma_w}{\mu_w}\right)$} \\
f_1(N) & \text{$a_1(N)+b_1(N)\log_{10}\left(\frac{\sigma_w}{\mu_w}\right)=a_2(N)+b_2(N)\log_{10}\left(\frac{\sigma_w}{\mu_w}\right)$} \\
f_1(N)10^{-10^{a_2(N)+b_2(N)\log_{10}\left(\frac{\sigma_w}{\mu_w}\right)}} & \text{$a_1(N)+b_1(N)\log_{10}\left(\frac{\sigma_w}{\mu_w}\right)<a_2(N)+b_2(N)\log_{10}\left(\frac{\sigma_w}{\mu_w}\right)$}
\end{cases}
,
\end{equation}
and $f(N>567)$ is given by:
\begin{equation}
f(N) = f_1(N)10^{-10^{a_2(N)+b_2(N)\log_{10}\left(\frac{\sigma_w}{\mu_w}\right)}},
\end{equation}
where $a_1(N\leq8)$ and $b_1(N\leq8)$ are listed in Table~6.\footnote{When $N = 5$, technique~3 (the broken-line fit) always defaults to technique~2 (the linear fit) if the data are weghted equally (Figure~4), leaving $f_1(5)$ undefined.  However, this is not always the case if the data are not weighted equally.  Consequently, we must define $f_1(5)$ before determining $a_1(5)$ and $b_1(5)$.  To this end, we adopt $f_1(5)=36.8534$, by extrapolation.}  For $8<N\leq1000$:
\begin{equation}
a_1(N)=-0.541(\log_{10}N)^5+4.6943(\log_{10}N)^4-15.407(\log_{10}N)^3+21.875(\log_{10}N)^2-11.211\log_{10}N-0.3798,
\end{equation}
\begin{equation}
b_1(N)=0.1462(\log_{10}N)^3-4.2139(\log_{10}N)^2+14.366\log_{10}N-10.658,
\end{equation}
\begin{equation}
a_2(N)=18.149(\log_{10}N)^3-149.27(\log_{10}N)^2+410.15\log_{10}N-378.47,
\end{equation}
and
\begin{equation}
b_2(N)=26.945(\log_{10}N)^3-221.42(\log_{10}N)^2+606.91\log_{10}N-553.89.
\end{equation}
For $N > 1000$:
\begin{equation}
a_2(N)=0.3861\log_{10}N - 2.5852
\end{equation}
and
\begin{equation}
b_2(N)=0.0424\log_{10}N+1.4479.
\end{equation}
This approximation should be used with caution beyond $N \sim 10^6$. 

\begin{deluxetable*}{ccc}
	\tablewidth{0pt}
	\tablecaption{Empirical Approximations for $f(N>1000)$\tablenotemark{a}}
	\tablehead{
		\colhead{Scenario} & \colhead{Median} & \colhead{Mode}
	}
	\startdata
	Single $\sigma$ & 1.90 & $39.2519N^{-0.7969}+1.8688$ \\
	Smaller of $\sigma_-$ and $\sigma_+$ & 1.90 & $1.3399^{N^{0.1765}}$ \\
	Random of $\sigma_-$ and $\sigma_+$ & 1.90 & $1.2591^{N^{0.2052}}$
	\enddata
	\tablenotetext{a}{For data of equal weight (Appendix A).  The latter two approximations should be used with caution beyond $N \sim 10^6$.}
\end{deluxetable*}

\begin{deluxetable*}{cc}
	\tablewidth{0pt}
	\tablecaption{Correction Factors for the ``Single $\sigma$'' Scenario\tablenotemark{a}}
	\tablehead{
		\colhead{Technique} & \colhead{Correction Factor}
	}
	\startdata
	Mean-Standard Deviation (No Rejection) & $(1-0.2897N^{-1.033})^{-1}$ \\
	CR (Mean-Standard Deviation) & $(1-0.7240N^{-0.773})^{-1}$ \\
	RCR (Median-Technique~1) & $(1-1.7198N^{-1.022})^{-1}$ \\
	RCR (Median-Technique~2) & $(1-2.9442N^{-1.073})^{-1}$ \\
	RCR (Median-Technique~3) & $(1-4.2145N^{-1.153})^{-1}$ \\
	RCR (Mode-Technique~1) & $1-\frac{0.1052}{(N/41.99)^{-0.5328}+(N/41.99)^{0.4130}}$ \\
	RCR (Mode-Technique~2) & $1-\frac{0.05104}{(N/104.9)^{-3.2545}+(N/104.9)^{0.3444}}$ \\
	RCR (Mode-Technique~3) & $(1-2.1893N^{-0.803})^{-1}$ \\
	RCR (Median-T3) $+$ CR & $(1-4.2134N^{-0.971})^{-1}$ \\
	RCR (Median-T3) $+$ RCR (Median-T1) $+$ CR & $(1-4.3185N^{-0.975})^{-1}$ \\
	Bulk Rejection (Median) $+$ RCR (Median-T3) $+$ RCR (Median-T1) $+$ CR & $(1-3.5780N^{-0.942})^{-1}$
	\enddata
	\tablenotetext{a}{For data of equal weight (\textsection2.3, \textsection3.1, \textsection4, \textsection5).  Appropriate for two-sided contaminants.  Empirical approximations are for $N>100$.}
\end{deluxetable*}

\begin{deluxetable*}{cc}
	\tablewidth{0pt}
	\tablecaption{Correction Factors for the ``Smaller of $\sigma_-$ and $\sigma_+$'' Scenario\tablenotemark{a}}
	\tablehead{
		\colhead{Technique} & \colhead{Correction Factor}
	}
	\startdata
	Mean-Standard Deviation (No Rejection) & $(1-0.5092N^{-0.514})^{-1}$ \\
	CR (Mean-Standard Deviation) & $(1-0.6939N^{-0.522})^{-1}$ \\
	RCR (Median-Technique~1) & $(1-1.3320N^{-0.549})^{-1}$ \\
	RCR (Median-Technique~2) & $(1-1.5058N^{-0.559})^{-1}$ \\
	RCR (Median-Technique~3) & $(1-1.0426N^{-0.443})^{-1}$ \\
	RCR (Mode-Technique~1) & $(1-0.5736N^{-0.265})^{-1}$ \\
	RCR (Mode-Technique~2) & $(1-0.7285N^{-0.279})^{-1}$ \\
	RCR (Mode-Technique~3) & $(1-0.8790N^{-0.264})^{-1}$ \\
	RCR (Mode-T1) $+$ CR & $(1-1.7079N^{-0.602})^{-1}$ \\
	RCR (Mode-T3) $+$ CR & $(1-2.8415N^{-0.630})^{-1}$ \\
	RCR (Mode-T1) $+$ RCR (Median-T1) $+$ CR & $(1-1.7453N^{-0.605})^{-1}$ \\
	RCR (Mode-T3) $+$ RCR (Median-T1) $+$ CR & $(1-2.9047N^{-0.633})^{-1}$ \\
	Bulk Rejection (Mode) $+$ RCR (Mode-T1) $+$ RCR (Median-T1) $+$ CR & $(1-2.3525N^{-0.627})^{-1}$ \\
	Bulk Rejection (Mode) $+$ RCR (Mode-T3) $+$ RCR (Median-T1) $+$ CR & $(1-3.3245N^{-0.650})^{-1}$
	\enddata
	\tablenotetext{a}{For data of equal weight (\textsection2.3, \textsection3.2, \textsection4, \textsection5).  Appropriate for one-sided contaminants and in-between cases.  Empirical approximations are for $N>100$.}
\end{deluxetable*}

\begin{deluxetable*}{cc}
	\tablewidth{0pt}
	\tablecaption{Correction Factors for the ``Random of $\sigma_-$ and $\sigma_+$'' Scenario\tablenotemark{a}}
	\tablehead{
		\colhead{Technique} & \colhead{Correction Factor}
	}
	\startdata
	Mean-Standard Deviation (No Rejection) & $(1-0.4176N^{-1.293})^{-1}$ \\
	CR (Mean-Standard Deviation) & $(1-0.4482N^{-0.717})^{-1}$ \\
	RCR (Median-Technique~1) & $(1-2.0285N^{-1.021})^{-1}$ \\
	RCR (Median-Technique~2) & $(1-2.5569N^{-1.050})^{-1}$ \\
	RCR (Median-Technique~3) & $1-\frac{0.06629}{(N/41.83)^{-2.8626}+(N/41.83)^{0.9580}}$ \\
	RCR (Mode-Technique~1) &
	$\begin{cases}
		1.02187-0.00907\log_{10}N & \text{if $100 < N \leq 1000$} \\
		1-0.03946N^{-0.2895} & \text{if $N > 1000$}
	\end{cases}$ \\
	RCR (Mode-Technique~2) & 
	$\begin{cases}
		1.07422- 0.02651\log_{10}N & \text{if $100 < N \leq 1000$} \\
		1-0.01616N^{-0.2895} & \text{if $N > 1000$}
	\end{cases}$ \\
	RCR (Mode-Technique~3) & $(1-3.4414N^{-0.849})^{-1}$ \\
	RCR (Mode-T3) $+$ CR & $(1-3.2546N^{-0.840})^{-1}$ \\
	RCR (Mode-T3) $+$ RCR (Median-T1) $+$ CR & $(1-2.8989N^{-0.824})^{-1}$ \\
	Bulk Rejection (Mode) $+$ RCR (Mode-T3) $+$ RCR (Median-T1) $+$ CR & $(1-3.1666N^{-0.833})^{-1}$
	\enddata
	\tablenotetext{a}{For data of equal weight (\textsection2.3, \textsection3.3.1, \textsection4, \textsection5).  Appropriate for (mildy) asymmetric uncontaminated distributions.  Empirical approximations are for $N>100$.}
\end{deluxetable*}

For this scenario and the ``bulk rejection (mode) + RCR (mode-T1) + RCR (median-T1) + CR'' technique, the correction factor is given by:
\begin{equation}
CF(N) = 
\begin{cases}
CF_1(N)10^{10^{a(N)+b(N)\log_{10}\left(\frac{\sigma_w}{\mu_w}\right)}} & \text{$N=2$, $N=3$, $N>4$} \\
CF_1(N) & \text{$N=4$}
\end{cases}
,
\end{equation}
where $a(N\leq5)$ and $b(N\leq5)$ are listed in Table~7.  For $5<N\leq100$:
\begin{equation}
a(N)=-1.1937(\log_{10}N)^4+6.5268(\log_{10}N)^3-13.308(\log_{10}N)^2+11.432\log_{10}N-4.4769
\end{equation}
and
\begin{equation}
b(N)= -1.4528(\log_{10}N)^3+5.3519(\log_{10}N)^2-5.33\log_{10}N+2.2902+0.1879(-1)^N(\log_{10}N)^{0.9521}.
\end{equation}
For $N>100$:
\begin{equation}
a(N)=-0.5408\log_{10}N-0.6482
\end{equation}
and
\begin{equation}
b(N)=1.4154+0.3635(-1)^N.
\end{equation}

For this scenario and the ``bulk rejection (mode) + RCR (mode-T3) + RCR (median-T1) + CR'' technique, the correction factor is given by Equation B15, where $a(N\leq5)$ and $b(N\leq5)$ are listed in Table~7.  For $5<N\leq20$:
\begin{equation}
a(N)=-0.2683(\log_{10}N )^4+1.9174(\log_{10}N )^3-5.062(\log_{10}N )^2 
+5.452 \log_{10}N-2.9999
\end{equation}
and
\begin{equation}
\begin{aligned}
b(N)=43.179(\log_{10}N )^6-331.85(\log_{10}N )^5 & 
+968.25(\log_{10}N )^4-1399.1(\log_{10}N )^3+1070.7(\log_{10}N )^2 \\
&-415.81 \log_{10}N+65.002+0.1365(-1)^N (\log_{10}N)^{2.4716}.
\end{aligned}
\end{equation}
For $20<N\leq100$, $a(N)$ is given by Equation B20, and:
\begin{equation}
b(N)=1.5144 \log_{10}N-0.0448+0.1365(-1)^N (\log_{10}N)^{2.4716}.
\end{equation}
For $N>100$:
\begin{equation}
a(N)=-0.4282 \log_{10}N-0.4412
\end{equation}
and
\begin{equation}
b(N)=2.9881+0.7530(-1)^N.
\end{equation}

\begin{deluxetable}{ccccccc}
\tablewidth{0pt}
\tablecaption{Empirical Approximation Parameter Values for $f(N\leq8)$}
\tablehead{
	\colhead{Scenario:} & \multicolumn{2}{c}{Single $\sigma$} & \multicolumn{2}{c}{Smaller of $\sigma_-$ and $\sigma_+$} & \multicolumn{2}{c}{Random of $\sigma_-$ and $\sigma_+$} \\ 
	\colhead{$N$} & \colhead{$a_1(N)$} & \colhead{$b_1(N)$} & \colhead{$a_1(N)$} & \colhead{$b_1(N)$} & \colhead{$a_1(N)$} & \colhead{$b_1(N)$} 
}
\startdata
4 &	0.2024 &	0.4642 & \nodata & \nodata & \nodata & \nodata\\
5 &	-0.2916 &	0.2603 &	-0.0828 &	-0.3003 & -0.4664 & 2.1342\\
6 &	-0.0332 &	0.3638 &	-0.2675 &	-0.2443 &	-0.3124 &	1.0197\\
7 &	-0.1818 &	0.4547 &	-0.5588 &	-0.4097 &	-0.7502 &	0.5797\\
8 & Equation B2 & Equation B3 &	-0.8893 &	-0.4884 & Equation B25 & Equation B26
\enddata
\end{deluxetable}

\begin{deluxetable}{ccccccccc}
\tablewidth{0pt}
\tablecaption{Empirical Approximation Parameter Values for $CF(N\leq5)$}
\tablehead{
	\colhead{Scenario:} & \multicolumn{2}{c}{Single $\sigma$} & \multicolumn{4}{c}{Smaller of $\sigma_-$ and $\sigma_+$} & \multicolumn{2}{c}{Random of $\sigma_-$ and $\sigma_+$} \\ 
	\colhead{Technique:} & \multicolumn{2}{c}{Bulk Rejection (Median)} & \multicolumn{2}{c}{Bulk Rejection (Mode)} & \multicolumn{2}{c}{Bulk Rejection (Mode)} & \multicolumn{2}{c}{Bulk Rejection (Mode)} \\ 
	\colhead{ } & \multicolumn{2}{c}{$+$ RCR (Median-T3)} & \multicolumn{2}{c}{$+$ RCR (Mode-T1)} & \multicolumn{2}{c}{$+$ RCR (Mode-T3)} & \multicolumn{2}{c}{$+$ RCR (Mode-T3)} \\
	\colhead{ } & \multicolumn{2}{c}{$+$ RCR (Median-T1)} & \multicolumn{2}{c}{$+$ RCR (Median-T1)} & \multicolumn{2}{c}{$+$ RCR (Median-T1)} & \multicolumn{2}{c}{$+$ RCR (Median-T1)} \\ 
	\colhead{ } & \multicolumn{2}{c}{$+$ CR} & \multicolumn{2}{c}{$+$ CR} & \multicolumn{2}{c}{$+$ CR} & \multicolumn{2}{c}{$+$ CR} \\ 
	\colhead{$N$} & \colhead{$a(N)$} & \colhead{$b(N)$} & \colhead{$a(N)$} & \colhead{$b(N)$} & \colhead{$a(N)$} & \colhead{$b(N)$} & \colhead{$a(N)$} & \colhead{$b(N)$} 
}
\startdata
2 & -3.1528 &	0.2739 &	-0.3984 &	1.0815 &	-0.3984 &	1.0815 &	-2.5951 &	0.7336\\
3 &	-1.1913 &	0.4487 &	-1.1462 &	1.0699 &	-1.1094 &	1.5143 &	-0.8546 &	0.8160\\
4 &	-1.0509 &	3.3825	& \nodata & \nodata & \nodata & \nodata &	-0.9543 &	 1.1605\\
5 &	-1.4145 &	0.1185 &	-1.1196 &	0.4597 &	-1.1446 &	0.3394 &	Equation B32 &	Equation B33
\enddata
\end{deluxetable}

For the ``random of $\sigma_-$ and $\sigma_+$'' scenario, appropriate for (mildly) asymmetric uncontaminated distributions (\textsection3.3.1), $f(N\leq190)$ is given by Equation B1, $f(190<N\leq305)$ is given by Equation B7, and $f(N>305)$ is given by Equation B8, where $a_1(N\leq7)$ and $b_1(N\leq7)$ are listed in Table~6.  For $7<N\leq1000$:
\begin{equation}
\begin{aligned}
a_1(N)=3.1767(\log_{10}N)^6-34.561(\log_{10}N)^5+152.16(\log_{10}N)^4 & -347.96(\log_{10}N)^3+435.59(\log_{10}N)^2\\
&-282.57\log_{10}N+73.696,
\end{aligned}
\end{equation}
\begin{equation}
b_1(N)=5.8718(\log_{10}N)^4-47.049(\log_{10}N)^3+131.12(\log_{10}N)^2-150.24\log_{10}N+61.727,
\end{equation}
\begin{equation}
a_2(N)=-1.8953(\log_{10}N)^2+11.745\log_{10}N-19.36,
\end{equation}
and
\begin{equation}
b_2(N)=-2.7584(\log_{10}N)^2+17.078\log_{10}N-24.602.
\end{equation}
For $N > 1000$:
\begin{equation}
a_2(N)=-1.1827
\end{equation}
and
\begin{equation}
b_2(N)=1.8064.
\end{equation}
This approximation should also be used with caution beyond $N \sim 10^6$.  

For this scenario and the ``bulk rejection (mode) + RCR (mode-T3) + RCR (median-T1) + CR'' technique, the correction factor is given by:
\begin{equation}
CF(N) = CF_1(N)10^{10^{a(N)+b(N)\log_{10}\left(\frac{\sigma_w}{\mu_w}\right)}},
\end{equation}
where $a(N\leq4)$ and $b(N\leq4)$ are listed in Table~7.  For $4<N\leq19$:
\begin{equation}
a(N)=-1.3993(\log_{10}N )^3+6.5746(\log_{10}N )^2-9.8844 \log_{10}N+2.8572
\end{equation}
and
\begin{equation}
b(N)=4.0458(\log_{10}N )^2-6.4354 \log_{10}N+2.7667.
\end{equation}
For $19<N\leq100$, $a(N)$ is given by Equation B32, and:
\begin{equation}
b(N)=1.7394 \log_{10}N  - 1.0435.
\end{equation}
For $N>100$:
\begin{equation}
a(N)=-0.5989 \log_{10}N  - 0.6097
\end{equation}
and
\begin{equation}
b(N)=1.4123 \log_{10}N-0.3893.
\end{equation}




\end{document}